\documentclass[
reprint,
 amsmath,amssymb,
 aps,
 prc,
floatfix,
]{revtex4-2}
\usepackage{comment}
\usepackage{graphicx}
\usepackage{dcolumn}
\usepackage{bm}
\usepackage{color}
\usepackage{xcolor}
\usepackage{mathtools}
\usepackage{url}        % enables \url{...} command
\usepackage{hyperref}
\usepackage{lineno}
\usepackage{amsmath}

\begin{document}
%\linenumbers
\preprint{APS/123-QED}
\title{Tilted geometry of the pion emission source in Au+Au collisions in the 
RHIC Beam Energy Scan}

\author{STAR Collaboration}

\date{\today}

\begin{abstract}

We present the first systematic measurement of the tilt of the pion emission source in relativistic Au+Au collisions at center-of-mass energies per nucleon pair, $\sqrt{s_{NN}}$ = 7.7, 14.5, 17.3 and 27 GeV, using data from the STAR experiment. The tilt angle and final freeze-out eccentricity are extracted through azimuthally sensitive femtoscopy of identical pion pairs. Our results reveal a strong dependence of the tilt parameter on the pair transverse momentum, indicating that the apparent source geometry is strongly coupled to expansion dynamics. Moreover, we observe a rapid decrease of the tilt magnitude with increasing collision energy, consistent with the emission source approaching longitudinal boost invariance at higher energies. These findings demonstrate that the commonly assumed boost-invariant geometry is insufficient and highlight the necessity of exploring the spatial structure of a tilted source, which is required in hydrodynamic models to reproduce features of the longitudinally expanding system, such as the slope of the directed flow. Comparisons with the UrQMD transport model show that it reproduces the overall energy dependence of the tilt magnitude qualitatively, but not quantitatively.

\end{abstract}

\pacs{25.75.-q, 25.75.Gz}

\maketitle

%%%%%%%%%%%%%%%%%%%%%%%%%%%%%%%%%%%%%%%%%%%%%%%%%%%%%%%%%%%%%%%%%%%%%%%%%%%%%%%%%%%%%%%%%
\section{Introduction}

\begin{figure}[h!]
    \centering
    \includegraphics[width=1.0\linewidth]{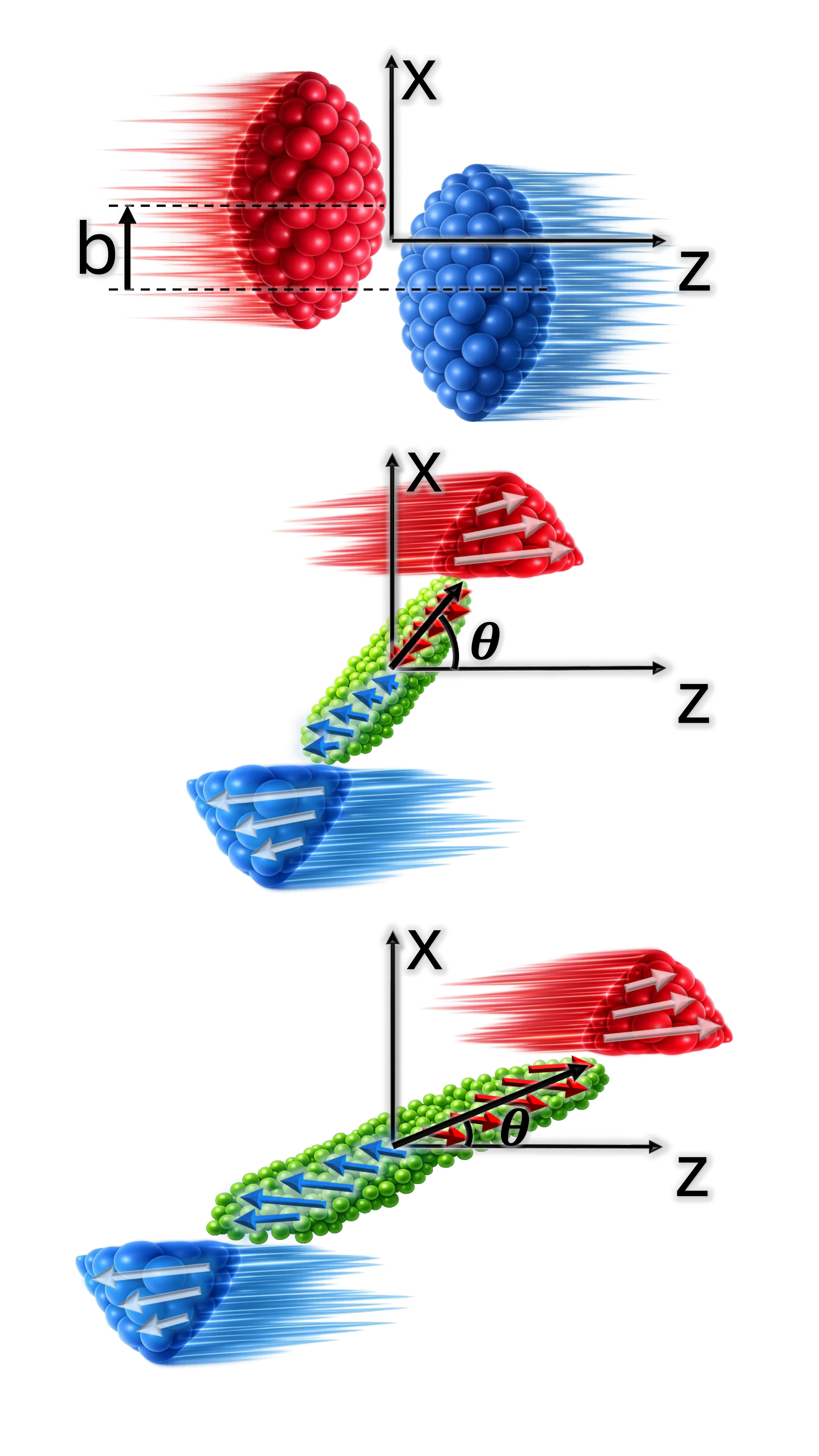}
\caption{Schematic illustration of a non-central collision between two nuclei.
Top panel: Initial geometry of the collision in the reaction plane, defined by the beam direction $z$ and the impact-parameter direction $x$. The projectile (red) and target (blue) nuclei pass each other with a finite impact parameter $b$.
Middle and bottom panels: After the collision, asymmetric energy and momentum deposition in the participant region (green) induces a spatial shift of the matter distribution in the $x$--$z$ plane, leading to the formation of a tilted fireball. The spectators are deflected outward, opposite to the transverse flow of the participant matter, consistent with overall momentum conservation. The resulting emitting source is characterized by a rotation of its principal axis by an angle $\theta$ relative to the beam direction, reflecting longitudinal--transverse correlations in the initial state and providing the geometric origin of directed flow and other rapidity-odd observables.}
    \label{fig:collision}
\end{figure}

Understanding the still incompletely constrained three-dimensional structure of the initial state has become one of the central goals of relativistic heavy-ion collisions~\cite{Arslandok:2023utm,Achenbach:2023pba}. The spatial and momentum anisotropies established in the earliest moments of the collision drive much of the subsequent collective behavior of the expanding system~\cite{Poskanzer:1998yz}. They drive the development of anisotropic flow and the emergence of complex space--time patterns during freeze-out~\cite{Bozek:2010vz,Jia:2014ysa,Pang:2015zrq,Schenke:2016ksl,Monnai:2015sca}. Among the geometric features that have attracted increasing attention in recent years is the tilt of the particle-emitting source -- a rotation of the principal axis of the fireball away from the beam direction in the reaction plane as shown in Fig.~\ref{fig:collision}.

This tilt is particularly pronounced at collision energies of a few GeV, where the nuclear passing time is significantly longer than at higher energies, so energy and momentum are deposited over an extended time interval during the early overlap stage of the collision. In non-central collisions, the finite duration of energy–momentum deposition from the projectile and target nuclei into the produced medium preserves a left–right imbalance of locally deposited longitudinal momentum across the impact-parameter direction. This results in a tilted initial configuration and leaves clear imprints~\cite{Bozek:2010bi, Chatterjee:2018lsx, Parida:2022lmt} on final-state observables, most notably the directed flow $v_1(y)$. 

The tilted fireball provides a natural mechanism for redistributing part of the large orbital angular momentum (OAM) of a non-central A{+}A collision into local vorticity and angular-momentum density in the medium. Because the initial deposition of matter and longitudinal momentum is asymmetric across the transverse (impact-parameter) direction in the reaction plane, fluid elements at different transverse positions $x$ acquire different longitudinal velocities, generating a rapidity-odd shear $\partial_x v_z$ and hence a vortical structure with nonzero kinematic vorticity. In general, the component of the kinematic vorticity perpendicular to the reaction plane is $\omega_y = \tfrac{1}{2}\left(\partial_z v_x - \partial_x v_z\right) \approx -\tfrac{1}{2}\,\partial_x v_z$,
where the approximation holds when $\partial_z v_x$ is subleading.
In hydrodynamic and OAM-based approaches, this shear redistributes part of the global OAM $L_y$ into local angular-momentum density and thermal vorticity on the freeze-out surface, which, through spin--vorticity coupling, gives rise to the experimentally observed global hyperon polarization~\cite{STAR:2017ckg}. This connection is quantified, for example, in the thermal-vorticity framework~\cite{Becattini:2016gvu}.

Figure~\ref{fig:collision} illustrates, in a schematic way, how a tilted emitting source gives rise to directed flow and global polarization. The collision is shown in the reaction plane, spanned by the beam direction ($z$ axis) and the impact parameter $b$, which in this case is oriented along the $x$ direction. In the top panel, the two nuclei are shown before the collision, approaching each other intact, together with the coordinate system. After the collision, part of the nuclear matter interacts and forms a hot and dense participant zone (shown in green in the middle and bottom panels), while the remaining nucleons become spectators and continue their motion.

As shown in the middle and bottom panels, the spectators (blue and red) are schematically depicted as being deflected outward, on average, opposite to the net transverse momentum of the participant region and, by momentum conservation, opposite to the transverse flow of the participants, i.e. away from the overlap region~\cite{Voloshin:2016ppr}. The arrows in the participant zone illustrate the average collective momentum of the expanding matter and help visualize the mechanism responsible for the generation of directed flow. In a non-central collision, the energy density is deposited in a spatially asymmetric manner, leading to an initial density profile that is effectively tilted in the $x$--$z$ plane. As a result, the position of the maximum energy density shifts from one transverse side at backward rapidity ($z<0$) to the opposite side at forward rapidity ($z>0$). Since pressure gradients point away from regions of highest density, fluid elements located at different $z$ experience transverse pressure gradients of opposite sign. During the subsequent hydrodynamic expansion, these gradients accelerate the participant matter in opposite transverse directions (with respect to the beam) at forward and backward rapidities, producing a rapidity-odd collective motion. This longitudinal variation of the transverse flow is imprinted on the final-state particle emission and manifests itself as directed flow, quantified by the first harmonic coefficient $v_1$.

Although the underlying mechanism of directed flow is the same, the magnitude of $v_1$ depends on the strength of the initial tilt of the participant source. A larger tilt produces a stronger longitudinal dependence of the energy density generating a sizable rapidity-odd directed flow. When the tilt is small, as at higher collision energies, these gradients are weaker and partially cancel, resulting in a reduced rapidity-odd $v_1$.

Several theoretical studies~\cite{Bozek:2010bi, Chatterjee:2018lsx, Parida:2022lmt} have demonstrated that introducing an initial tilt in the energy density significantly affects the development of both directed and elliptic flow, for light hadrons as well as for heavy-flavor particles. For light hadrons, hydrodynamic simulations can often reproduce the observed $v_1$ without an explicit tilt, relying instead on baryon transport and local momentum conservation. For heavier species, however, this picture changes dramatically. In particular, calculations in~\cite{Chatterjee:2018lsx} show that the rapidity-odd directed flow of open-charm mesons originates, at least in part, from a fireball whose energy density is initially tilted in the reaction plane. The resulting flow of the medium then transfers momentum to heavy quarks through drag interactions, generating the observed rapidity-dependent directed-flow pattern of open-charm D mesons~\cite{Chatterjee:2018lsx}. These findings highlight the intimate link between the geometric orientation of the source and the final-state collectivity of particles with different masses and quark content.

Experimentally, it is therefore of great importance to determine the degree and orientation of this tilt. Yet, measuring it directly remains a formidable challenge. The initial tilt discussed in~\cite{Bozek:2010bi, Chatterjee:2018lsx, Parida:2022lmt} characterizes the very early energy-density distribution---a quantity that cannot be accessed experimentally. Detectors record only those particles that survive until kinetic freeze-out, long after the system has undergone hydrodynamic expansion, cooling, and hadronization. What can be reconstructed from experimental data is, consequently, the final tilt of the particle-emitting source, rather than the tilt of the initial quark--gluon plasma (QGP) fireball.

This distinction is crucial. The initial tilt describes the orientation of the energy-density profile at formation time, before pressure gradients and collective flow have had time to act. By the time particles decouple, the system has evolved significantly, and its shape has been altered by expansion, anisotropic pressure, and rescattering. The final tilt that we observe carries the imprint of the initial geometry but also reflects the full dynamical evolution of the medium. In this sense, the measured tilt acts as an integrated signal of the system’s history---linking the early-state geometry to the late-stage freeze-out configuration.

Given the tilt is estimated, another quantity of interest is the eccentricity of the freeze-out surface. By “freeze-out eccentricity” we mean how elliptic (almond-shaped) the particle-emitting region is at kinetic freeze-out in the transverse plane. Tilt and eccentricity provide complementary information: the tilt specifies the orientation of the emission ellipsoid, while the eccentricity sets the magnitude of its transverse deformation.

%%%%%%%%%%%%%%%%%%%%%%%%%%%%%%%%%%%%%%%%%%%%%%%%%%%%%%%%%%%%%%%%%%%%%%%%%%%%%%%%%%%%%%%%%
\section{Method of azimuthally-sensitive femtoscopy}

At present, only one experimental method exists that allows for the extraction of the tilt of the particle-emitting source. This method is known as azimuthally sensitive femtoscopy (asHBT)~\cite{Lisa:2000ip,E895:2000opr,Mount:2010ey,Lisa:2011na}—that is, Hanbury Brown and Twiss (HBT) interferometry performed with respect to the reaction plane using pairs of identical particles. By analyzing two-particle momentum correlations as a function of the pair’s azimuthal angle, one can reconstruct the three-dimensional geometry and orientation of the emission region at kinetic freeze-out.

A key aspect of femtoscopy is that it does not probe the entire freeze-out source simultaneously. Instead, it accesses the so-called regions of homogeneity—the localized space–time regions that emit particles with similar velocities~\cite{Akkelin:1995gh}. The size and position of these regions depend on the kinematic properties of the emitted particles. By varying the average transverse momentum of the pair, $k_{\mathrm{T}} = \frac{p_{\mathrm{T},1} + p_{\mathrm{T},2}}{2}$, or by selecting different azimuthal angles with respect to the reaction plane, one effectively probes different portions of the source. It has been shown that pairs with higher $k_{\mathrm{T}}$ originate from smaller regions of homogeneity and are typically emitted at earlier times~\cite{Akkelin:1996sg, Lisa:2005dd}. Consequently, the $k_{\mathrm{T}}$-dependence of femtoscopic observables—including the extracted tilt angle—serves as a valuable tool for disentangling the space–time evolution of the emitting source.

\begin{figure}[h!]
    \centering
    \includegraphics[width=1.0\linewidth]{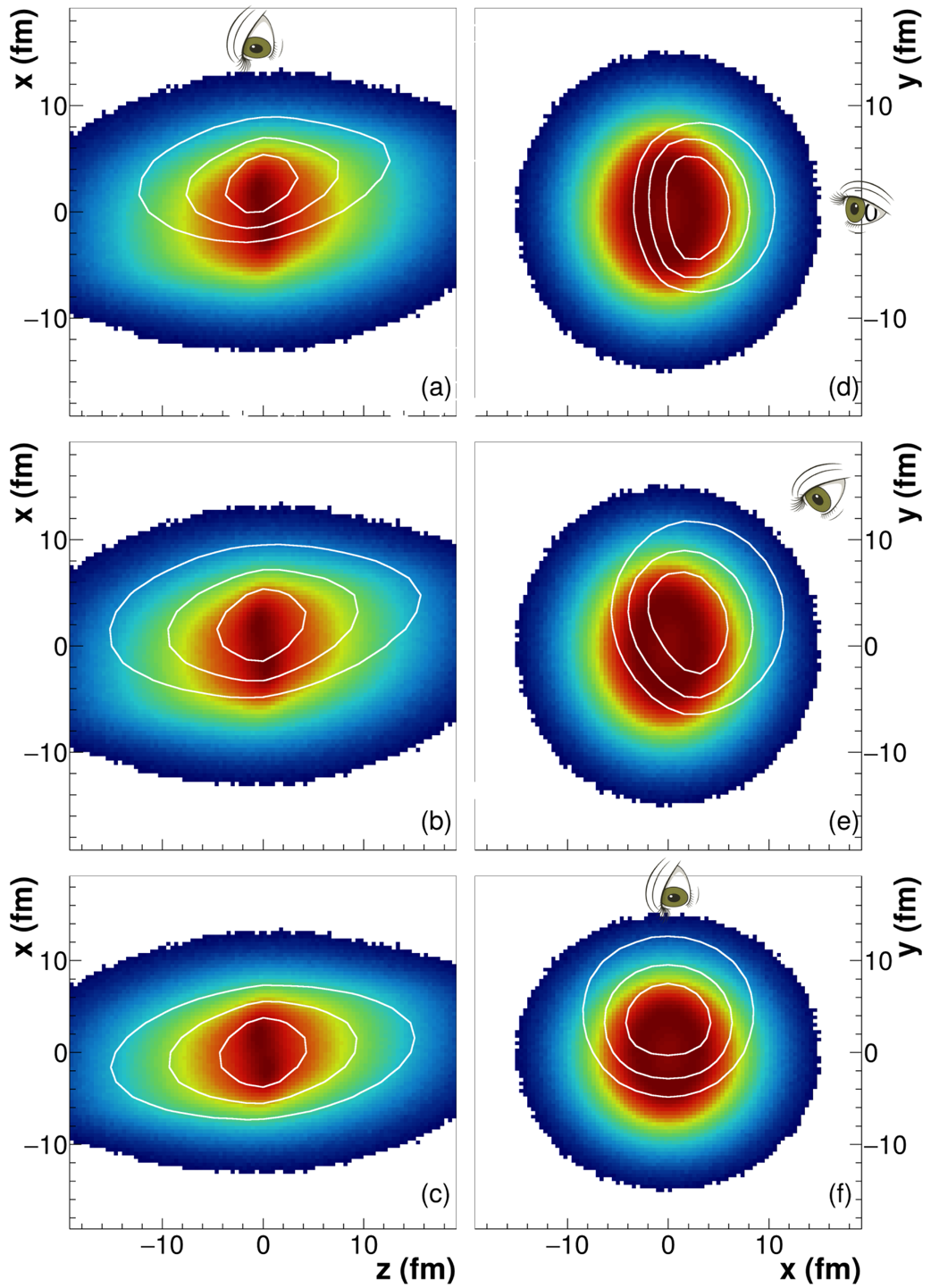}
    \caption{Two-dimensional projections of the three-dimensional spatial distribution of pion kinetical freeze-out points in Au+Au collisions at $\sqrt{s_{\mathrm{NN}}} = 7.7$~GeV, obtained from the UrQMD model~\cite{Bass:1998ca,Bleicher:1999xi}. Panels (a--c) show projections onto the $x$--$z$ plane, while panels (d--f) show projections of the same distribution onto the $y$--$x$ plane. The white contours show freeze-out coordinate distributions for pion pairs selected in \(30^\circ\)-wide azimuthal bins relative to the reaction plane. The three cases shown are centered at \(0^\circ\) (panels a, d), \(45^\circ\) (panels b, e), and \(90^\circ\) (panels c, f). The eye symbol schematically indicates the corresponding average emission direction. For the $45^{\circ}$ and $90^{\circ}$ cases, this direction cannot be clearly represented in the $x$–$z$ projection because it lies partially or entirely out of the plane of the figure.
}
    \label{fig:freeze-out}
\end{figure}

The asHBT method is based on a fundamental assumption: the reconstructed tilt of the emission source is independent of the azimuthal angle of the particle pairs selected for the analysis. In other words, irrespective of the direction from which the source is probed in the transverse plane, its intrinsic spatial orientation is expected to be recovered. Figure~\ref{fig:freeze-out} illustrates this idea. The figure is organized into two columns: the left column shows the $x-z$ projection of the three-dimensional distribution of pion freeze-out coordinates from Au+Au collisions at \(\sqrt{s_{\mathrm{NN}}}=7.7\)~GeV, while the right column presents the corresponding $y-x$ projection. The coordinates are obtained using the Ultrarelativistic Quantum Molecular Dynamics model (UrQMD~3.4 Cascade)~\cite{Bass:1998ca,Bleicher:1999xi}. From the $x-z$ projections in the left column, one can clearly see that the freeze-out distribution is tilted away from the beam ($z$) direction.

To further illustrate this behavior, we consider particle pairs emitted at different average azimuthal angles, \(\langle \phi \rangle = 0^\circ\), \(45^\circ\), and \(90^\circ\). The corresponding freeze-out coordinate distributions are indicated by overlapping white contours in the first, second, and third rows of Fig.~\ref{fig:freeze-out}, respectively, with the azimuthal angle indicated by eye. By comparing the contours constructed for different azimuthal angles, one observes that while certain features vary—leading to different magnitudes of the extracted femtoscopic radii—the tilt of the source away from the \(z\) direction remains essentially unchanged. Consequently, a consistent source tilt is expected to be reconstructed for all azimuthal orientations.

While the experimental goal is to determine the spatial tilt of the source, the only directly accessible quantities in the detector are the momenta of the emitted particles. Femtoscopy provides the crucial bridge between momentum and coordinate space: it allows us to infer space–time characteristics of the emission region from measured momentum correlations. The key idea is that identical bosons—such as charged pions—exhibit enhanced correlations at small relative momentum due to quantum-statistical (Bose–Einstein) interference. The magnitude and shape of this correlation enhancement depend on the relative spatial separation of the particle pairs at freeze-out. As a result, femtoscopy translates measurable structures in momentum space into information about the size, shape, and orientation of the homogeneity regions of the source.

To perform a quantitative analysis, two-particle momentum correlations are studied as a function of the azimuthal angle of the pair relative to the reaction plane. The selected angles are optimized to capture the extrema of the expected azimuthal oscillations of the femtoscopic radii, arising from the anisotropic geometry of the emission region in non-central collisions, while ensuring sufficient statistical precision in each bin. For every azimuthal interval, the correlation function is analyzed in the Bertsch–Pratt coordinate system~\cite{Pratt:1986cc, Bertsch:1988db}, which decomposes the relative momentum vector into the \emph{out~(o)}, \emph{long (l)} and \emph{side (s)} components corresponding, respectively, to the directions of the pair transverse momentum, the beam axis and the perpendicular direction to those two.

Experimentally, the two-particle correlation function is constructed as the ratio
\begin{equation}
C(\mathbf{q},\mathbf{K},\phi)=\frac{A(\mathbf{q},\mathbf{K},\phi)}{B(\mathbf{q},\mathbf{K},\phi)},
\label{eq:expCF}
\end{equation}
where \(A\) and \(B\) denote the relative-momentum distributions of pion pairs from the same event and from mixed events, respectively. Here, \(\mathbf{q}=\mathbf{p}_1-\mathbf{p}_2\) is the relative momentum, expressed in the longitudinal co-moving system (LCMS)~\cite{Lisa:2005dd}, in which the longitudinal component of the pair velocity vanishes; \(\mathbf{K}=(\mathbf{p}_1+\mathbf{p}_2)/2\) is the average pair momentum; and \(\phi\) is the angle of the transverse pair momentum \(\mathbf{K}_T=(K_x,K_y)\) with respect to the reaction plane. The measured correlation function is then fitted with the parametrization described below.

The fitting function used to describe the three-dimensional correlation function in the Bertsch--Pratt system is given by the Bowler--Sinyukov function~\cite{Bowler:1991vx,Sinyukov:1998fc}:
\begin{equation}
C(\mathbf{q}, \mathbf{K}, \phi)
= N \times BS(\mathbf{q}, \mathbf{K}, \phi),
\label{eq:cfFit}
\end{equation}
where
\begin{equation}
\begin{split}
BS(\mathbf{q}, \mathbf{K}, \phi)
= 1 - \lambda
  + \lambda\, K_{\mathrm{Coul}}(\mathbf{q}, \mathbf{K}, \phi)\times \\
  \times \left(1 + e^{-\sum_{i,j=o,s,l} q_i q_j R_{ij}^2}\right),
\end{split}
\label{eq:BS}
\end{equation}
Here, \(N\) is a normalization constant, \(\lambda\) represents the strength of the two-particle correlation, and \(K_{\mathrm{Coul}}(\mathbf{q}, \mathbf{K}, \phi)\) is the Coulomb correction factor accounting for the final-state electromagnetic interaction between two charged particles. In this work, it is obtained from the squared Coulomb wave function of identical pion pairs integrated over a spherical Gaussian source~\cite{Bowler:1991vx,Sinyukov:1998fc}. The Gaussian term in the exponent encodes the size and shape of the emission region through the squared radii \(R_{ij}^2\). According to~\cite{Akkelin:1995gh}, \(R_{\mathrm{side}}\) primarily reflects the geometry of the source, \(R_{\mathrm{out}}\) contains information on both the geometry and the emission duration, and \(R_{\mathrm{long}}\) is sensitive to the system lifetime. The cross terms (\(i \neq j\)) represent correlations between different spatial directions of the source and reflect possible tilts of the emission ellipsoid in \(\mathbf{q}\) space. In an azimuthally integrated analysis, these cross terms vanish by symmetry, provided that the system exhibits longitudinal boost invariance~\cite{Heinz:2002au,Lisa:2005dd}.

The fit parameters are obtained by minimizing a log-likelihood function of the form~\cite{E802:2002fwe}:
\begin{equation}
P = -2 \sum_{q_i}\left(A\ln\left[\frac{C}{A}\frac{A+B}{C+1}\right] + B\ln\left[\frac{1}{B}\frac{A+B}{C+1}\right]\right),
\label{eq:minimization}
\end{equation}
where the sum over \(q_i\) runs over all \(q\) bins of the numerator and denominator histograms. Here, \(A\) and \(B\) denote the bin contents of the same-event and mixed-event relative-momentum distributions, respectively, while \(C\) is the value of the theoretical correlation function from Eq.~\ref{eq:cfFit} evaluated at the same \(q\) points. For a given \(\mathbf{K}\) and \(\phi\) bin, all three quantities are functions of \(\mathbf{q}\).

The full implementation of the correlation function fitting procedure described above is provided in the authors’ publicly available analysis code~\footnote[1]{The code was developed by Yevheniia Khyzhniak and is available at \url{https://gitlab.com/sam_teyla/femtofitter}.}.

Equation~\ref{eq:cfFit} assumes that the freeze-out distribution follows a Gaussian profile in coordinate space, which, in non-central collisions, results in the extracted femtoscopic radii oscillations depending on the azimuthal angle of the pion pair~\cite{Lisa:2000ip}.

It was shown in~\cite{Heinz:2002au} that at midrapidity, the measured femtoscopic radii dependence on $\phi$ can be described by the following Fourier expansions:

\begin{align}
& R^2_{\text{out}}(\phi)
= R^2_{\text{out},0}
+ 2\sum_{n=2,4,6,\dots} R^2_{\text{out},n}\cos(n\phi) \nonumber \\
& R^2_{\text{side}}(\phi)
= R^2_{\text{side},0}
+ 2\sum_{n=2,4,6,\dots} R^2_{\text{side},n}\cos(n\phi) \nonumber \\
& R^2_{\text{long}}(\phi)
= R^2_{\text{long},0}
+ 2\sum_{n=2,4,6,\dots} R^2_{\text{long},n}\cos(n\phi)
\label{eq:oscFit} \\
& R^2_{\text{out-side}}(\phi)
= 2\sum_{n=2,4,6,\dots} R^2_{\text{out-side},n}\sin(n\phi) \nonumber \\
& R^2_{\text{out-long}}(\phi)
= 2\sum_{n=1,3,5,\dots} R^2_{\text{out-long},n}\cos(n\phi) \nonumber \\
& R^2_{\text{side-long}}(\phi)
= 2\sum_{n=1,3,5,\dots} R^2_{\text{side-long},n}\sin(n\phi) \nonumber
\end{align}

The zero-order Fourier coefficients correspond to the extracted femtoscopic parameters in the azimuthally integrated analysis. Additionally, it is common for Fourier coefficients above the second order to be consistent with zero, a finding that has experimental confirmation~\cite{STAR:2004qya}.

The extracted Fourier coefficients can be used to calculate the tilt angle of the homogeneity region in coordinate space~\cite{Lisa:2000ip, Mount:2010ey}:

\begin{equation}
\theta_{\text{side-long}} 
= \frac{1}{2}\arctan\left(
\frac{-4\,R^{2}_{\text{side-long},1}}
     {R^{2}_{\text{long},0}
     - R^{2}_{\text{side},0}
     + 2\,R^{2}_{\text{side},2}}
\right)
\label{eq:tiltSl}
\end{equation}
where the subscript following each radius term denotes the order of the Fourier coefficient. It is important to note that the applicability of this formula relies on assumptions about the source geometry and will be revisited in the Section~\ref{sec:energy}.

%%%%%%%%%%%%%%%%%%%%%%%%%%%%%%%%%%%%%%%%%%%%%%%%%%%%%%%%%%%%%%%%%%%%%%%%%%%%%%%%%%%%%%%%%
\section{Details of the experiment}
\label{sec:exp}

The analysis presented in this work is based on data from the Beam Energy Scan (BES) program at the Relativistic Heavy Ion Collider (RHIC). The data were recorded by the STAR experiment during the 2018, 2019, and 2021 running periods, corresponding to center-of-mass energies per nucleon pair of $\sqrt{s_{\mathrm{NN}}} = 7.7$, 14.5, 17.3 and 27~GeV, respectively. These data sets were collected as part of the second phase of the BES program (BES-II), designed to explore the properties of strongly interacting matter at high baryon density and to search for possible signatures of the QCD critical point~\cite{STAR:2025zdq}.

Events were selected using minimum-bias trigger conditions. Pile-up events and bad runs were removed following the standard STAR quality-assurance procedure, similar to that described in~\cite{STAR:2025zdq,STAR:2021mii}. The primary vertex position along the beam direction was required to satisfy $|V_{\mathrm{z}}| < 70~\mathrm{cm}$, ensuring uniform detector acceptance, while the radial vertex position relative to the beam axis was constrained to $V_{\mathrm{r}} < 2~\mathrm{cm}$ to reject beam–pipe interactions.

Charged particle reconstruction was performed using the Time Projection Chamber (TPC)~\cite{STAR:1997sav,Anderson:2003ur}, which provides precise tracking and particle identification through ionization energy loss ($dE/dx$) measurements. Tracks were required to have transverse momentum within the range $0.15 < p_{\mathrm{T}} < 0.8~\mathrm{GeV}/c$ and pseudorapidity $|\eta| < 1.0$. To ensure reliable momentum reconstruction and particle identification, each track was required to have at least 15 measured hits in the TPC ($N_{\mathrm{hits}} \geq 15$), and to originate from the primary vertex within a distance of closest approach (DCA) of less than 3 cm.

Particle identification (PID) for charged pions was performed using the $dE/dx$ information from the TPC, expressed in terms of the normalized deviation
\begin{equation}
    n_{\sigma,i} = \frac{1}{\sigma_{i}}\ln\!\left(\frac{(dE/dx)_{\mathrm{measured}}}{(dE/dx)_{\mathrm{expected},i}}\right),
\end{equation}
where $i$ denotes the particle species hypothesis~\cite{Bichsel:2006cs}. Tracks were identified as pions by requiring $|n\sigma_{\pi}| < 2$, while tracks consistent with other species were excluded by demanding $|n\sigma_{\mathrm{other}}| > 3$. This combination of cuts ensures a high-purity pion sample with little contamination from kaons, protons, and electrons in the studied momentum range.

Collision centrality was determined from the charged-particle multiplicity measured in the TPC within $|\eta|~<~0.5$, using a Monte Carlo Glauber model to relate multiplicity percentiles to collision geometry~\cite{STAR:2008med,Miller:2007ri}. The centrality calibration and event categorization are consistent with those used in previous BES analyses~\cite{STAR:2017sal}.

In femtoscopic analyses, a careful treatment of two-track reconstruction effects is essential to ensure the integrity of the measured correlation functions. Although single-track inefficiencies are common to both the same-event and mixed-event samples and thus cancel in the ratio that defines the correlation function~\cite{Lisa:2005dd}, two-track artifacts can distort the numerator and denominator differently, particularly at small relative momentum $q = | \mathbf{p}_{1} - \mathbf{p}_{2} |$~\cite{Lisa:2005dd}. To minimize such distortions, standard anti-splitting and anti-merging pair cuts were applied~\cite{STAR:2004qya}.

Track splitting occurs when the TPC reconstruction algorithm mistakenly interprets the ionization-cluster pattern of a single charged particle as two distinct tracks. This effect artificially enhances the same-event pair distribution at very low $q$, producing a spurious peak in the correlation function. To suppress such artifacts, a topological ``splitting-level'' (SL) variable was evaluated for each pair. The splitting level characterizes the pattern of hit sharing between the two reconstructed tracks across TPC pad rows; following the standard STAR definition, it is not restricted to positive values, so negative values can occur for well-separated pairs. Candidate pairs were required to satisfy $-0.5 \leq \mathrm{SL} \leq 0.6$. This selection effectively removes pairs that could arise from track duplication or spurious splitting~\cite{STAR:2004qya}.

The complementary effect, track merging, occurs when two close tracks—which originate mostly from lower relative momentum pairs—are reconstructed as a single trajectory. Merging suppresses the number of same-event pairs at small $q$, leading to an artificial depletion in the correlation function. Since lost pairs cannot be recovered directly, the corresponding phase-space region must also be removed from the mixed-event distribution to maintain consistency. For this purpose, an anti-merging criterion was applied based on the fraction of merged hits (FMH), defined as the percentage of TPC padrows on which both tracks in a pair share a reconstructed cluster~\cite{STAR:2004qya}. Only pairs with $\mathrm{FMH} = 0\%$
were accepted in the final analysis. This ensures that both tracks in a pair are separated by the reconstruction algorithm over all padrows traversed in the TPC.

A detailed discussion of these two-particle selection criteria and their impact on the correlation function can be found in~\cite{STAR:2004qya}.

\section{Event-plane reconstruction}

In order to perform an azimuthally sensitive femtoscopic analysis, it is necessary to determine the orientation of the reaction plane. Because the reaction plane is determined by the unobservable initial-state geometry, it cannot be measured directly in experiment. Instead, experiments reconstruct an event plane, which serves as an experimental estimate of the true reaction plane, using the azimuthal anisotropy of the final-state particle distributions. The reconstructed event plane reflects the collective response of the medium to the initial spatial asymmetry and provides a practical proxy for studying anisotropic flow~\cite{Voloshin:2008dg}.

Event planes can be defined for different harmonic orders, each corresponding to a specific type of azimuthal anisotropy. The first-order event plane ($\Psi_{1}$) is associated with directed flow ($v_{1}$), representing a dipole-like emission pattern with enhanced particle yield along the impact-parameter direction. The second-order event plane ($\Psi_{2}$) corresponds to elliptic flow ($v_{2}$), arising from the elliptic overlap geometry of non-central collisions. Higher-order event planes ($\Psi_{n}$ with $n \geq 3$) are sensitive to finer spatial structures and fluctuations in the initial energy density, giving rise to triangular and higher-order flow components. While the true reaction plane is fixed by the collision geometry, different reconstructed event planes (\(\Psi_n\)) provide experimentally accessible approximations based on different harmonic anisotropies of the final-state particle emission.

\subsection{First-order event plane from the EPD}

In the present analysis, the plane of interest is the first-order event plane ($\Psi_{1}$), which is used to reconstruct the forward–backward geometry of the particle-emitting source after the collision. The first-order event plane was determined using the Event Plane Detector (EPD), a segmented scintillator array positioned symmetrically on both sides of the interaction point and covering the pseudorapidity range of $2.14 < |\eta| < 5.09$. The full description of the EPD design, calibration, and performance is given in~\cite{Adams:2019fpo}. The reconstruction procedure follows the standard formalism developed in~\cite{Poskanzer:1998yz, Voloshin:2008dg}, in which the $n$th-order event plane is calculated from the event-wise flow vector components,
\begin{equation}
\Psi_{n} = \frac{1}{n}\tan^{-1}\!\left(\frac{Q_{n,y}}{Q_{n,x}}\right),
\label{eq:psi}
\end{equation}
where the components of the $n$-th order $Q$-vectors are defined as 
\begin{align}
Q_{n,x} &= \sum_{k=1}^{N_{\mathrm{tiles}}} w_{k}\cos(n\phi_{k}), \label{eq:Qx}\\
Q_{n,y} &= \sum_{k=1}^{N_{\mathrm{tiles}}} w_{k}\sin(n\phi_{k}). \label{eq:Qy}
\end{align}
here, $\phi_{\mathrm{k}}$ is the azimuthal angle of the $k$-th EPD tile, $w_{\mathrm{k}}$ is its weight (proportional to its signal amplitude), and the summation extends over all tiles with valid signals in the event.

A unique feature of the Beam Energy Scan energies ($\sqrt{s_{\mathrm{NN}}}=7.7$, 14.5, 17.3 and 27~GeV) is that the beam rapidities fall within the acceptance of the EPD. Consequently, the EPD receives contributions from both ``spectator" fragments and produced particles from the ``participant" region, which are characterized by opposite directed flow patterns~\cite{STAR:2022ahj}. The inner EPD rings are dominated by spectator protons that define the projectile and target deflection directions, while the outer rings contain a mixture of participants and secondaries. This dual sensitivity must be taken into account when constructing $\Psi_{1}$ to ensure the correct determination of the event-plane orientation.

To optimize the event-plane resolution, larger weights are assigned to pseudorapidity regions where the magnitude of $v_{1}(\eta)$ is larger, and negative weights are applied where $v_{1}$ changes sign. A natural and self-consistent choice for the weighting function is therefore the measured $v_{1}(\eta)$ distribution in the EPD acceptance itself. In practice, first an approximate $v_{1}(\eta)$ is obtained in the EPD by evaluating truncated normalized-MIP (nMIP -- minimum-ionizing particle) signals as a function of $\eta$, relative to a reference event plane reconstructed in the TPC~\cite{STAR:2024ujm}. The resulting $v_{1}(\eta)$ points are then fitted with a smooth polynomial function, which serves as the $\eta$-dependent weight $w_{\mathrm{\eta}}$ in Eqs.~(\ref{eq:Qx}) and~(\ref{eq:Qy}).

Ideally, the reconstructed event-plane angle distribution should be uniform in azimuth, indicating that all orientations are equally probable. In practice, residual detector asymmetries, segmentation effects, and nonuniform response can introduce artificial modulations in the event-plane distribution. These are corrected through a standard event-plane flattening procedure~\cite{Poskanzer:1998yz, Voloshin:2008dg}, in which harmonic corrections are applied to shift the reconstructed angles toward a uniform distribution. Typically, sixth- and higher-order Fourier components are subtracted to remove residual acceptance effects and achieve a flat event-plane distribution.

The event-plane reconstruction described above yields a high-resolution estimate of $\Psi_{1}$, suitable for azimuthally differential femtoscopic analyses.

\subsection{Event-plane resolution}

The finite precision with which the event plane is reconstructed must be accounted for in azimuthally differential analyses. The event-plane resolution quantifies how well the reconstructed event-plane angle $\Psi_{n}$ approximates the true reaction-plane angle $\Psi_{\mathrm{RP}}$. It is defined as~\cite{Poskanzer:1998yz, Voloshin:2008dg}
\begin{equation}
\mathcal{R}_{nk} = \left\langle \cos\!\left[k\left(\Psi_{n} - \Psi_{\mathrm{RP}}\right)\right] \right\rangle,
\label{eq:resolution}
\end{equation}
where $n$ is the order of the reconstructed event plane, $k$ is the
harmonic of the observable being corrected, and the brackets $\langle \cdots \rangle$ denote an
average over all events.

Since the true reaction-plane angle $\Psi_{\mathrm{RP}}$ is not experimentally accessible, Eq.~\ref{eq:resolution}
cannot be measured directly. Instead, the resolution is estimated using
the subevent method: each event is divided into two independent
subevents (east and west halves of the EPD), which are used to reconstruct
two separate subevent-plane angles $\Psi_n^{E}$ and $\Psi_n^{W}$.
For symmetric, statistically independent subevents with equal resolution,
the measurable east--west correlation factorizes as
\begin{equation}
\begin{split}
  \bigl\langle
    \cos\bigl[k\,(\Psi_n^{E} - \Psi_n^{W})\bigr]
  \bigr\rangle
  = \bigl\langle\cos\bigl[k(\Psi_n^{E}-\Psi_{\mathrm{RP}})\bigr]\bigr\rangle
  \times \\
  \times \bigl\langle\cos\bigl[k(\Psi_n^{W}-\Psi_{\mathrm{RP}})\bigr]\bigr\rangle 
  \equiv \bigl(\mathcal{R}_{nk}^{\mathrm{sub}}\bigr)^{2},
\end{split}
\label{eq:factor}
\end{equation}
where $\mathcal{R}_{nk}^{\mathrm{sub}} \equiv
\langle \cos[k(\Psi_{n,\mathrm{sub}}-\Psi_{\mathrm{RP}})]\rangle$ is the
single-subevent resolution relative to the true reaction plane for the
harmonic $k$. Equation~\ref{eq:factor} makes explicit that the measurable east--west
correlation is not a resolution by itself; it equals the square
of the single-subevent resolution. Therefore, the quantity needed as input
to the standard resolution formalism is obtained as
\begin{equation}
\mathcal{R}_{nk}^{\mathrm{sub}}
=
\sqrt{
\bigl\langle\cos\bigl[k\,(\Psi_n^{E}-\Psi_n^{W})\bigr]\bigr\rangle }.
\label{eq:resolution_sub}
\end{equation}

To convert between resolution and the underlying precision parameter
$\chi$, we use the Bessel-function expression~\cite{Poskanzer:1998yz, Voloshin:2008dg}
\begin{equation}
\mathcal{R}_{nk} = 
\frac{\sqrt{\pi}}{2\sqrt{2}}\,
\chi_{n}\,
e^{-\chi_{n}^{2}/4}
\left[
I_{\frac{k-1}{2}}\!\left(\frac{\chi_{n}^{2}}{4}\right)
+ 
I_{\frac{k+1}{2}}\!\left(\frac{\chi_{n}^{2}}{4}\right)
\right],
\label{eq:resolution_bessel}
\end{equation}
where $I_{\nu}$ is the modified Bessel function of the first kind of order
$\nu$, and $n$ and $k$ correspond to the order of the event plane and the
harmonic coefficient of interest, respectively. The parameter $\chi$
quantifies the event-plane precision based on the anisotropic flow magnitude
and particle multiplicity. The notation $\mathcal{R}_{nk}=\mathcal{R}_k(\chi_n)$
means: evaluate the right-hand side of Eq.~\ref{eq:resolution_bessel} at $\chi_n$ for the chosen
harmonic $k$.

Numerical inversion is needed because Eq.~\ref{eq:resolution_bessel} is nonlinear in $\chi$ and has
no closed-form inverse. In practice, $\chi_n^{\mathrm{sub}}$ is obtained by
numerically inverting Eq.~\ref{eq:resolution_bessel} such that
$\mathcal{R}_k(\chi_n^{\mathrm{sub}})=\mathcal{R}_{nk}^{\mathrm{sub}}$.
For two equal subevents combined into a full event plane,
$\chi_n=\sqrt{2}\,\chi_n^{\mathrm{sub}}$. The full-event resolution required
to correct an observable of harmonic $k$ is then obtained by evaluating
Eq.~\ref{eq:resolution_bessel} at $\chi_n$, i.e.\ $\mathcal{R}_{nk}=\mathcal{R}_k(\chi_n)$.
In this analysis $n=1$ throughout, and the resolutions used for corrections
are computed for the relevant values of $k$.

\begin{figure}[h!]
    \centering
    \includegraphics[width=1.0\linewidth]{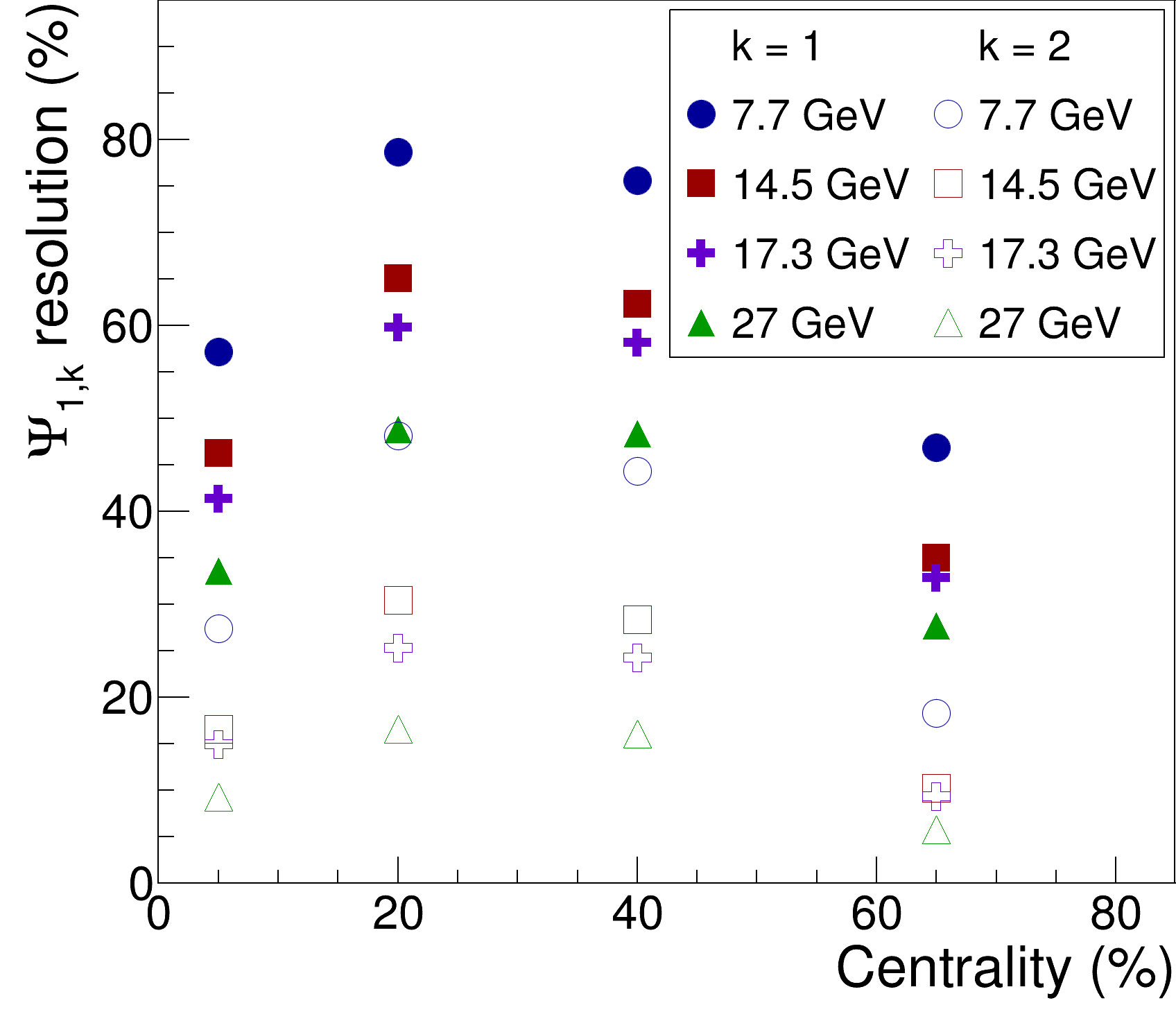}
    \caption{Resolution of the first and second-order event plane reconstructed with the EPD
detector. Blue circles correspond to Au+Au collisions at $\sqrt{s_{NN}} = 7.7$~GeV, red squares to 14.5~GeV, violet crosses to 17.3~GeV and green triangles to 27~GeV. Closed markers show the resolution of the first-order event plane using the first-harmonic coefficients (n=1, k=1). Open markers show the effective resolution of the same first-order event plane when applied to second-harmonic coefficients (n=1, k=2).}
    \label{fig:resolution}
\end{figure}
The event-plane resolution reflects the combined influence of particle
multiplicity and flow anisotropy. In very central collisions, the large
particle multiplicity improves statistical precision, but the near-azimuthal
symmetry of the overlap region reduces the anisotropic flow signal, leading
to a smaller event-plane resolution. Conversely, in very peripheral collisions, the spatial anisotropy is large, but the reduced multiplicity
degrades the statistical accuracy of the event-plane reconstruction. As a
result, the resolution reaches its maximum for mid-central collisions, where
a balance between these two competing effects is achieved. The dependence of
the first-order event-plane resolution on centrality and beam energy, as
obtained in this analysis, is shown in Fig.~\ref{fig:resolution}.

\section{Corrections}
\subsection{Purity and momentum-resolution corrections}

The raw correlation function, constructed as the ratio of same-event to mixed-event distributions, is affected by the finite sample purity and the limited momentum resolution of the detector~\cite{STAR:2004qya}. These effects modify the measured correlation strength and can bias the extracted femtoscopic parameters if not properly accounted for. For clarity, the standard correction factors are introduced first, followed by a description of how they are incorporated directly into the fit function in the present analysis, rather than applied to the measured correlation function itself.

The procedure for correcting the measured correlation function for finite sample purity is well established in the femtoscopy literature~\cite{STAR:2004qya, STAR:2009fks, ALICE:2017iga} and is given by
\begin{equation}
C^{\mathrm{pur.\,corr}}(\mathbf{q}, \mathbf{K}, \phi) 
= \frac{C^{\mathrm{raw}}(\mathbf{q}, \mathbf{K}, \phi) - 1}{PP(\mathbf{q},\mathbf{K})} + 1,
\label{eq:purityCorr}
\end{equation}
where $C^{\mathrm{raw}}(\mathbf{q}, \mathbf{K}, \phi)$ is the raw measured correlation function and
$PP(\mathbf{q},\mathbf{K}) = \mathrm{Purity}(p_{1}) \times \mathrm{Purity}(p_{2})$.
Here, $\mathrm{Purity}(p_{i})$ denotes the single-particle purity of the pion sample, evaluated as a function of the particle momentum $p_{i}$. The product $PP(\mathbf{q},\mathbf{K})$ therefore represents the pair purity for each bin of relative momentum $\mathbf{q}$.

The single-particle purity quantifies the fraction of correctly identified pions in the selected sample.
Misidentified tracks-primarily electrons, kaons, or protons-are generally assumed to be uncorrelated with the femtoscopic signal and therefore reduce the apparent correlation strength~\cite{Lisa:2005dd, WA97:2001aby}. Applying the correction of Eq.~(\ref{eq:purityCorr}) compensates for this dilution by effectively rescaling the correlation amplitude to correspond to a pure pion sample.

In addition to the purity correction, the finite momentum resolution of the TPC introduces a systematic smearing of the measured relative momentum components $(q_{o}, q_{s}, q_{l})$. If not accounted for, this smearing leads to a slight underestimation of the extracted HBT radii and correlation strength when fitting the raw correlation functions~\cite{STAR:2004qya}. The effect of finite momentum resolution is accounted for using the prescription introduced in~\cite{STAR:2004qya}:

\begin{equation}
C^{\mathrm{MR.\,corr}}(\mathbf{q}, \mathbf{K}, \phi) 
= C^{\mathrm{raw}}(\mathbf{q}, \mathbf{K}, \phi)\times MR(\mathbf{q}, \mathbf{K}, \phi),
\label{eq:mr}
\end{equation}
where $MR(\mathbf{q}, \mathbf{K}, \phi)$ denotes the momentum-resolution correction factor.

Note that the dependence of $BS(\mathbf{q}, \mathbf{K}, \phi)$ on the femtoscopic radii can be treated partly analytically (see Eq.~\ref{eq:BS}), whereas the dependence of the momentum-resolution correction $MR(\mathbf{q}, \mathbf{K}, \phi)$ on these parameters is more involved. Consequently, an iterative procedure commonly used in femtoscopy analyses is employed~\cite{NA44:1994dmh}. In this approach, both the Coulomb correction $K_{\mathrm{Coul}}(\mathbf{q}, \mathbf{K}, \phi)$ in Eq.~\ref{eq:BS} and the momentum-resolution correction $MR(\mathbf{q}, \mathbf{K}, \phi)$ in Eq.~\ref{eq:mr} are first evaluated using an initial set of radii. A fit is then performed in which only the correlation strength parameter $\lambda$ and the radii appearing in the exponential term of Eq.~\ref{eq:BS} are allowed to vary. The radii obtained from this fit are then used to recalculate $K_{\mathrm{Coul}}(\mathbf{q}, \mathbf{K}, \phi)$ and $MR(\mathbf{q}, \mathbf{K}, \phi)$, and the procedure is iterated until convergence is reached, i.e., until the output parameters agree with the input parameters within their statistical uncertainties. In practice, convergence is reached within one or two iterations.

Both the pion purity and momentum‐resolution effects are best incorporated directly into the fit procedure rather than applied as corrections to the measured correlation function. This approach preserves the statistical integrity of the experimental data and ensures that detector effects are treated consistently within the model. Applying purity and resolution corrections at the level of the correlation function effectively distorts the raw observable, potentially introducing biases or correlations between bins that are difficult to quantify. In contrast, embedding these effects into the fit function allows the theoretical correlation model to be folded with the detector response and purity factors in a controlled way, maintaining a one-to-one comparison between model and unmodified data. This strategy also facilitates a more accurate propagation of uncertainties and avoids artificial amplification of statistical fluctuations that can arise from direct corrections to the data.

Consequently, the correlation functions shown in Figs.~\ref{fig:1DCF} and~\ref{fig:2DCF} represent the raw experimental distributions. They include only the standard single-track and two-track selection cuts described earlier in the Section~\ref{sec:exp}, with no additional corrections applied to the data points.
The combination of Eqs.~\ref{eq:cfFit},~\ref{eq:purityCorr}, and~\ref{eq:mr} leads to the final expression used for fitting the correlation function:
\begin{equation}
\begin{split}
C(\mathbf{q}, \mathbf{K}, \phi)
&= N \times \Bigl[ 1 + PP(\mathbf{q},\mathbf{K}) \times \\
&\times \Bigl( BS(\mathbf{q}, \mathbf{K}, \phi) \times MR(\mathbf{q}, \mathbf{K}, \phi) - 1 \Bigr) \Bigr]
\end{split}
\label{eq:cfmodified}
\end{equation}
where the convolution with the momentum-resolution function $MR(\mathbf{q}, \mathbf{K}, \phi)$ and the purity factor $PP(\mathbf{q},\mathbf{K})$ are applied to the theoretical template $BS(\mathbf{q}, \mathbf{K}, \phi)$.

\subsection{Event-plane resolution correction}

Finite resolution in the reconstructed event-plane angle leads to an artificial suppression of the observed oscillation amplitudes of the femtoscopic radii. Since the true reaction-plane angle cannot be determined perfectly, the measured modulation of the radii with respect to the reconstructed plane is systematically reduced compared to the true underlying value~\cite{Poskanzer:1998yz}. To account for this effect, the correction procedure described in the literature~\cite{ALICE:2018fdu} is applied, in which the correction is introduced directly to the fitted Fourier coefficients:
\begin{equation}
R^{\mathrm{true}}_{nk} = \frac{R^{\mathrm{obs}}_{nk}}{\mathcal{R}_{nk}},
\label{eq:resoCorr}
\end{equation}
where $R^{\mathrm{obs}}_{nk}$ denote the observed (uncorrected) oscillation amplitudes of the femtoscopic radii, and $\mathcal{R}_{nk}$ is the corresponding event-plane resolution for event plane of order $n$ (which is 1 in this work) and harmonic coefficient of interest $k$ (which is 1 and 2 in this work), as illustrated in Fig.~\ref{fig:resolution}.

\subsection{Bin-width correction}

An additional correction is required to account for the finite azimuthal bin width used in the measurement of the femtoscopic parameters. In practice, the dependence of the squared radii on the pair emission angle $\phi$ is measured in discrete intervals rather than as a continuous function. When the data are averaged over finite angular bins, the extracted oscillation amplitudes are slightly reduced compared to their true values, leading to a small systematic underestimation of the Fourier coefficients.

Following the procedure outlined earlier~\cite{Mount:2010ey}, the bin-width correction is applied directly to the measured oscillation amplitudes according to
\begin{equation}
R^{\mathrm{true}}_{nk} = R^{\mathrm{obs}}_{nk} \,
\frac{k\,\Delta\phi/2}{\sin(k\,\Delta\phi/2)},
\label{eq:binWidth}
\end{equation}
where $\Delta\phi$ denotes the azimuthal bin width used in the analysis. The correction factor approaches unity for infinitesimally small bins and increases monotonically with bin size. For the bin width of $\Delta\phi = 45^{\circ}$ used in this study, the effect amounts to a few percent for the lowest harmonic terms. The correction is applied to all extracted Fourier coefficients of the femtoscopic radii before performing the final fits and comparisons across different centralities and collision energies.

\section{Results and discussion}

\begin{figure}[h!]
    \centering
    \includegraphics[width=0.96\linewidth]{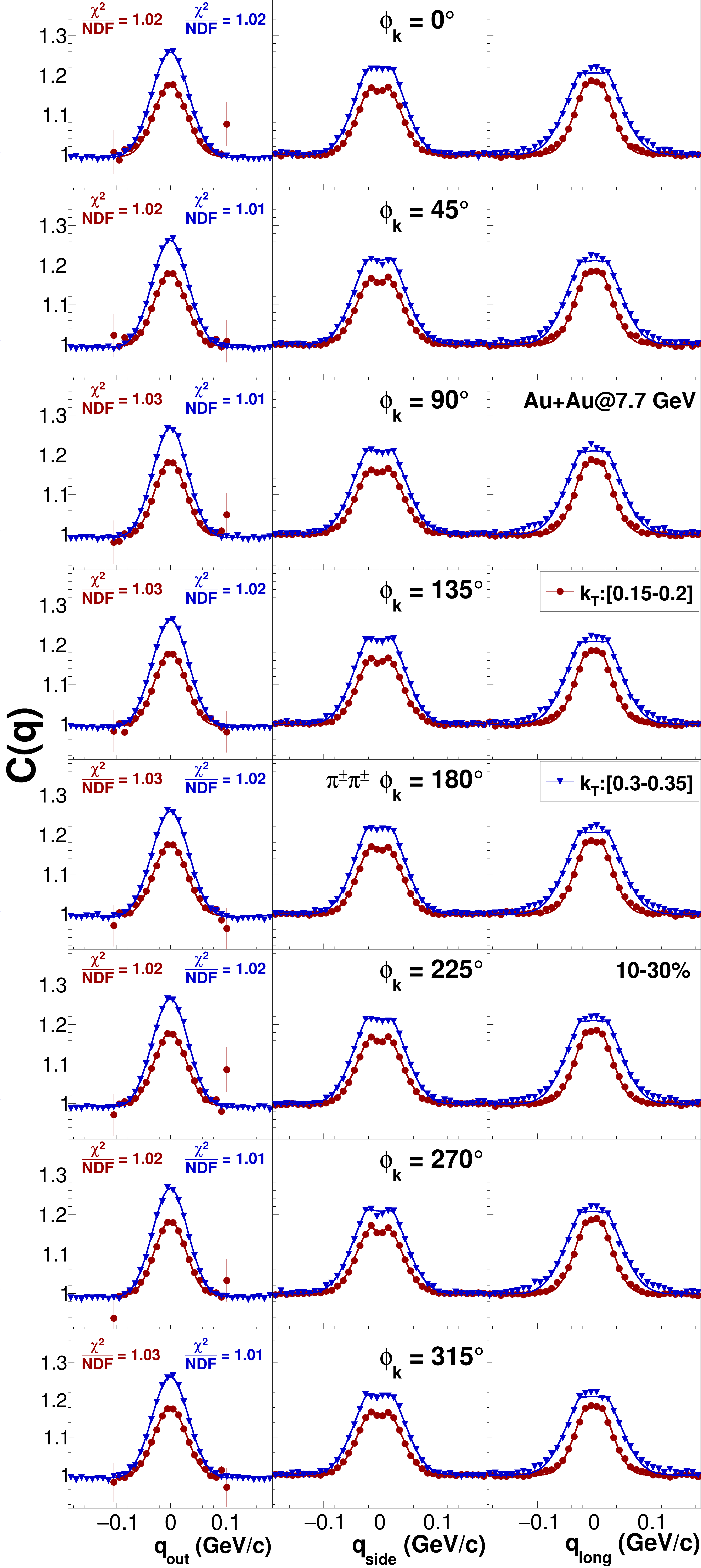}
    \caption{One-dimensional projections of the three-dimensional correlation function in the out (left column), side (central column), and long (right column) directions for eight ranges in the azimuthal angle of the pion pair relative to the first-order event plan. The lines correspond to the projections of the three-dimensional fit of the correlation functions according to Eq.~\ref{eq:cfmodified}. Results are shown for Au+Au collisions at $\sqrt{s_{NN}} = 7.7$~GeV in the 10-30\% centrality class. Projections onto a given axis are obtained with the condition $|q_{\text{others}}| < 0.05$~GeV/$c$. Red circles with red fit curves correspond to $k_{\mathrm{T}} = 0.15$--$0.20$~GeV/$c$, while blue triangles with blue fit curves correspond to $k_{\mathrm{T}} = 0.30$--$0.35$~GeV/$c$.}
    \label{fig:1DCF}
\end{figure}

Figure~\ref{fig:1DCF} shows one-dimensional projections of the three-dimensional correlation functions constructed for identical pion pairs in eight bins of the pair emission angle \(\phi_k\), defined with respect to the first-order event plane (our proxy for the reaction plane) as \(\phi_k = \phi - \Psi_1\). The correlation functions are constructed separately for \(\pi^+\pi^+\) and \(\pi^-\pi^-\) pairs and then combined, as no statistically significant difference is observed between the two charge combinations in the extracted tilt. The results are shown for two intervals of transverse pair momentum: $0.15 < k_{\mathrm{T}} < 0.20~\mathrm{GeV}/c$ (red circles) and $0.30 < k_{\mathrm{T}} < 0.35~\mathrm{GeV}/c$ (blue triangles). Projections onto each axis were obtained with the condition $|q_{\text{others}}| < 0.05~\mathrm{GeV}/c$. The data correspond to Au+Au collisions at $\sqrt{s_{\mathrm{NN}}} = 7.7~\mathrm{GeV}$ in the 10--30\% centrality class. The markers represent the measured correlation functions, while the red and blue curves show the corresponding fits performed using Eq.~\ref{eq:cfmodified}. The full set of parameters extracted from the fits is provided in the tables in Appendix~\ref{app:tables}. An absence of entries in the \(q_{\mathrm{out}}\) projection above \(\sim0.1\) GeV/\(c\) arises from the combination of the lower single-particle transverse-momentum cutoff and the selected \(k_T\) range.

By comparing the one-dimensional projections of the correlation functions with the fitted curves, small deviations from a purely Gaussian source shape are observed. Such deviations are commonly associated with the presence of resonance decays, which broaden the effective emission profile~\cite{Pratt:1986cc,Csorgo:1995bi,Heinz:1999rw,Humanic:2006ve}. These resonance contributions naturally give rise to non-Gaussian, Lévy-type emission patterns, which can be understood within the Lévy walk framework arising from anomalous diffusion and resonance decay kinematics~\cite{Csorgo:2003uv,Csanad:2007fr,Kincses:2024lnv}. Similar non-Gaussian effects have been reported previously in both model calculations~\cite{Li:2012ta,Kravchenko:2021app} and experimental measurements~\cite{STAR:2003nqk,STAR:2004qya}. Overall, the fitted functions provide a good description of the data, with $\chi^{2}/\mathrm{NDF}$ values close to unity. Importantly, the tilt angle extracted in this analysis is obtained from ratios of femtoscopic radii and is therefore expected to be less sensitive to assumptions about the detailed shape of the emission source. 
In particular, if the empirical \(\hat{R}\)-scaling behavior discussed in~\cite{Csanad:2024jpy} holds, where \(\hat{R}\) denotes an effective geometric length scale extracted from the femtoscopic radii and is largely independent of how close the emission source is to a Gaussian shape and of the overall correlation strength in the pair~\cite{PHENIX:2017ino}, then the explicit source-shape dependence cancels in the determination of the tilt angle.

\begin{figure}[h!]
    \centering
    \includegraphics[width=1.0\linewidth]{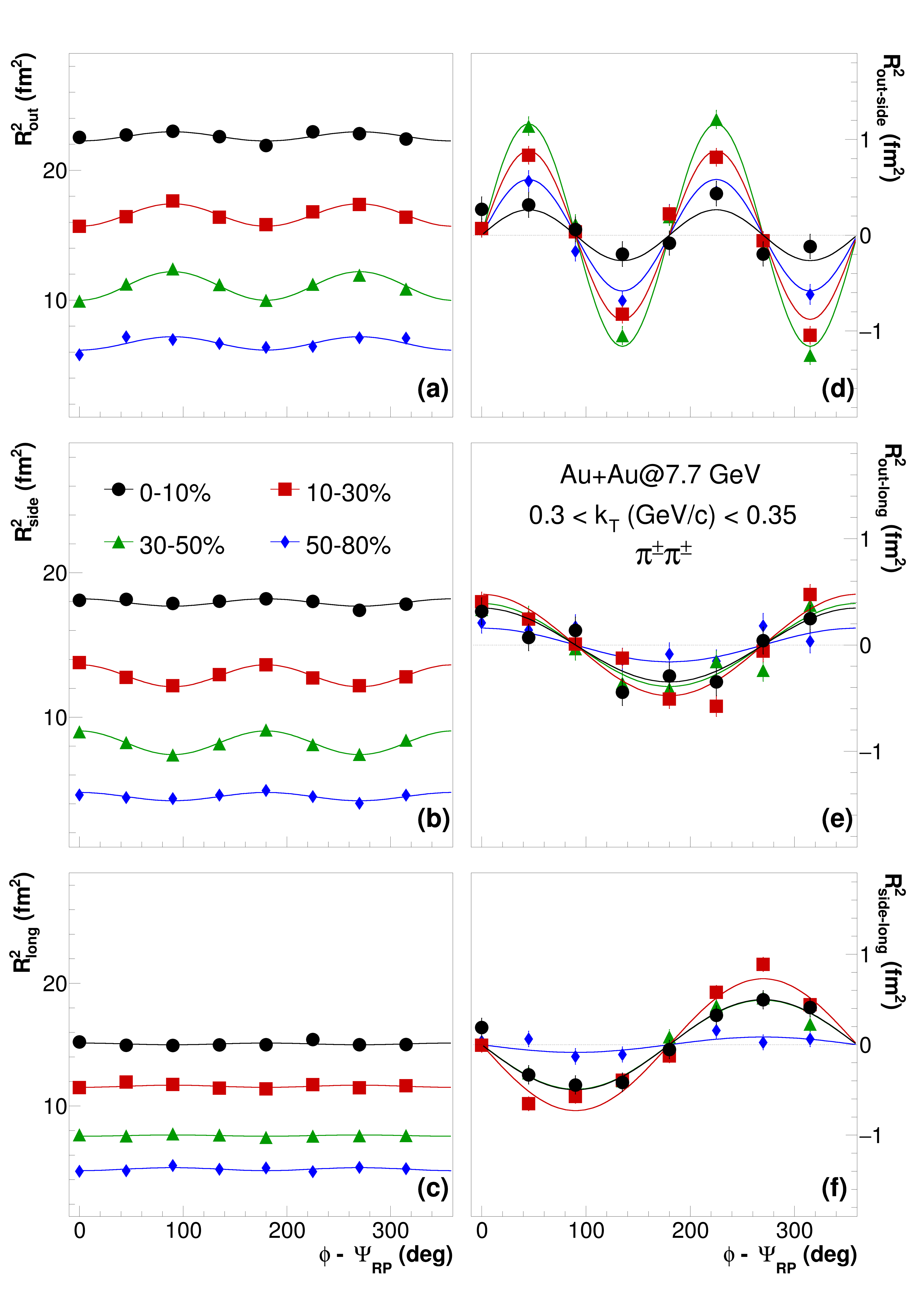}
    \caption{Extracted femtoscopic radii as a function of the azimuthal angle relative to the
event plane for four centrality classes. Black circles correspond to 0--10\%, red
squares to 10--30\%, green triangles to 30--50\%, and blue diamonds to 50--80\% in
Au+Au collisions at $\sqrt{s_{NN}} = 7.7$~GeV. Results are shown for pion pairs with transverse momentum $k_{\mathrm{T}}
= 0.3$--$0.35$ GeV/$c$. Lines of the corresponding colors represent fits to the oscillations of the femtoscopic radii using Eq.~\ref{eq:oscFit}.}
    \label{fig:radiiCent}
\end{figure}

Because the tilt angle of the emission source is primarily governed by the initial collision geometry, it is instructive to investigate how the femtoscopic radii vary with azimuthal angle for different centralities. Figure~\ref{fig:radiiCent} presents the oscillations of the extracted femtoscopic parameters for four centrality intervals: 0--10\%, 10--30\%, 30--50\%, and 50--80\%. Each data set exhibits clear harmonic modulation with respect to the first-order event plane, reflecting the spatial anisotropy and orientation of the freeze-out source. The amplitude and phase of these oscillations provide direct information about the geometric deformation and tilt of the pion-emitting region, which will be discussed in the following sections.

Panels (a--c) of Fig.~\ref{fig:radiiCent} show the oscillations of the squares of the three standard radii 
($R^{2}_{\mathrm{out}}$, $R^{2}_{\mathrm{side}}$, $R^{2}_{\mathrm{long}}$) usually
obtained in the azimuthally integrated analysis. 
Panels (d--f) display the oscillations of the cross-term components 
($R^{2}_{\mathrm{out\text{-}side}}$, $R^{2}_{\mathrm{out\text{-}long}}$, $R^{2}_{\mathrm{side\text{-}long}}$), 
which are non-zero only when the system lacks reflection symmetry 
with respect to the standard out--side--long coordinate axes. 

\begin{figure}[h!]
    \centering
    \includegraphics[width=0.83\linewidth]{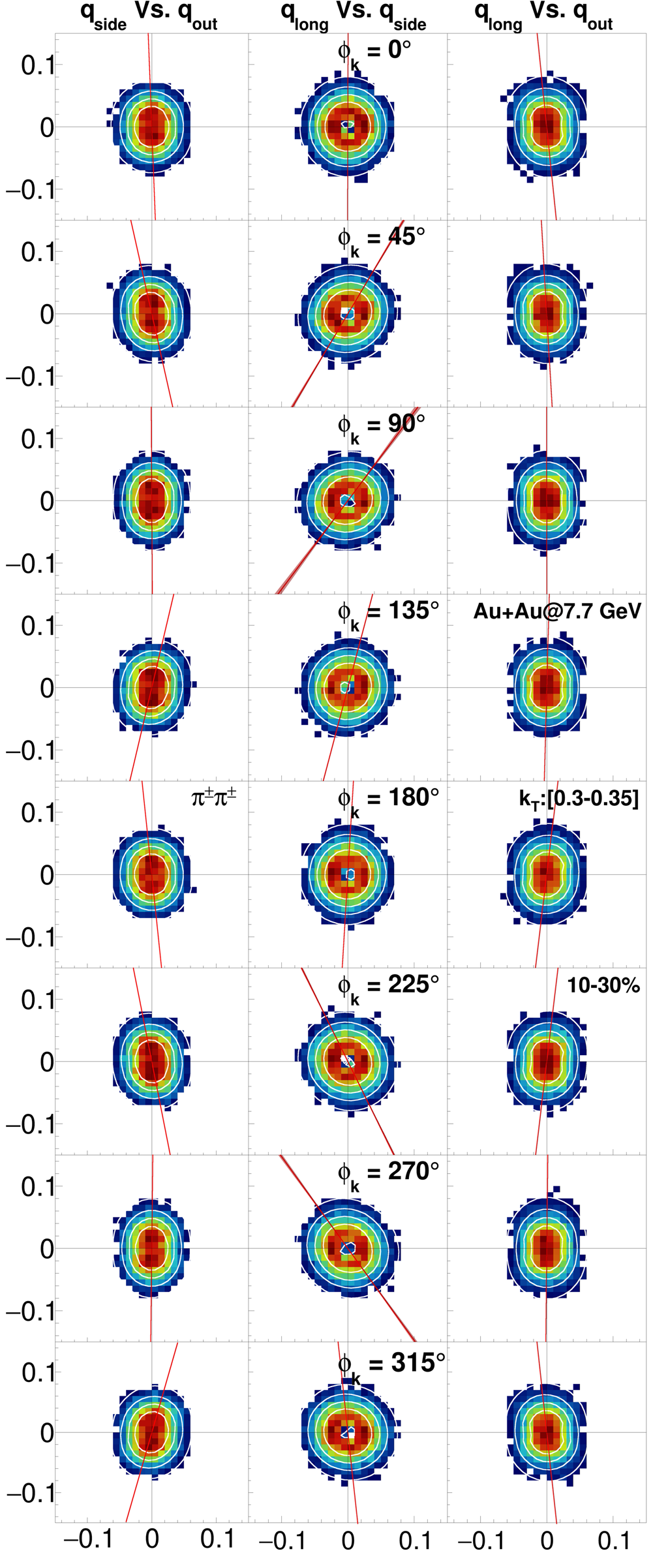}
    \caption{Two-dimensional projections of the three-dimensional correlation function in the out–side (left column), side–long (central column), and out–long (right column) directions for eight ranges in the azimuthal angle of the pion pair relative to the first-order event plane. White concentric ellipses correspond to the two-dimensional projections of the three-dimensional fit of the correlation functions according to the Eq.~\ref{eq:cfmodified}. The red shaded bands indicate the correlation angle, defined as the tilt angle of the emission ellipsoid in the corresponding two-dimensional projection. It is obtained from the off-diagonal cross term of the femtoscopic radius tensor together with the difference of the two diagonal radii in that plane. Results are presented for Au+Au collisions at $\sqrt{s_{NN}} = 7.7$~GeV in the 10--30\%
centrality class and for pion pairs with transverse momentum $k_{\mathrm{T}} = 0.30$--$0.35$~GeV/$c$. Projections
onto a given axis are obtained with the condition $|q_{\text{other}}| < 0.05$~GeV/$c$.}
    \label{fig:2DCF}
\end{figure}

The presence of non-zero cross-term components indicates correlations between the corresponding directions 
of the relative momentum $\mathbf{q}$, effectively representing ``tilts'' of the correlation function. 
These tilts can be directly observed in Fig.~\ref{fig:2DCF}, 
which shows the two-dimensional projections of the three-dimensional correlation function 
for eight ranges of the emission angle $\phi_{k}$.

For instance, the rotation of the correlation function in the $out-side$ plane, 
shown in the left column of Fig.~\ref{fig:2DCF}, 
corresponds directly to the oscillations of $R^{2}_{\mathrm{out\text{-}side}}$ 
depicted in panel (d) of Fig.~\ref{fig:radiiCent}. 
In Fig.~\ref{fig:radiiCent}, the lines represent Fourier fits 
according to Eq.~\ref{eq:oscFit}.

\begin{figure}[h!]
    \centering
    \includegraphics[width=1.0\linewidth]{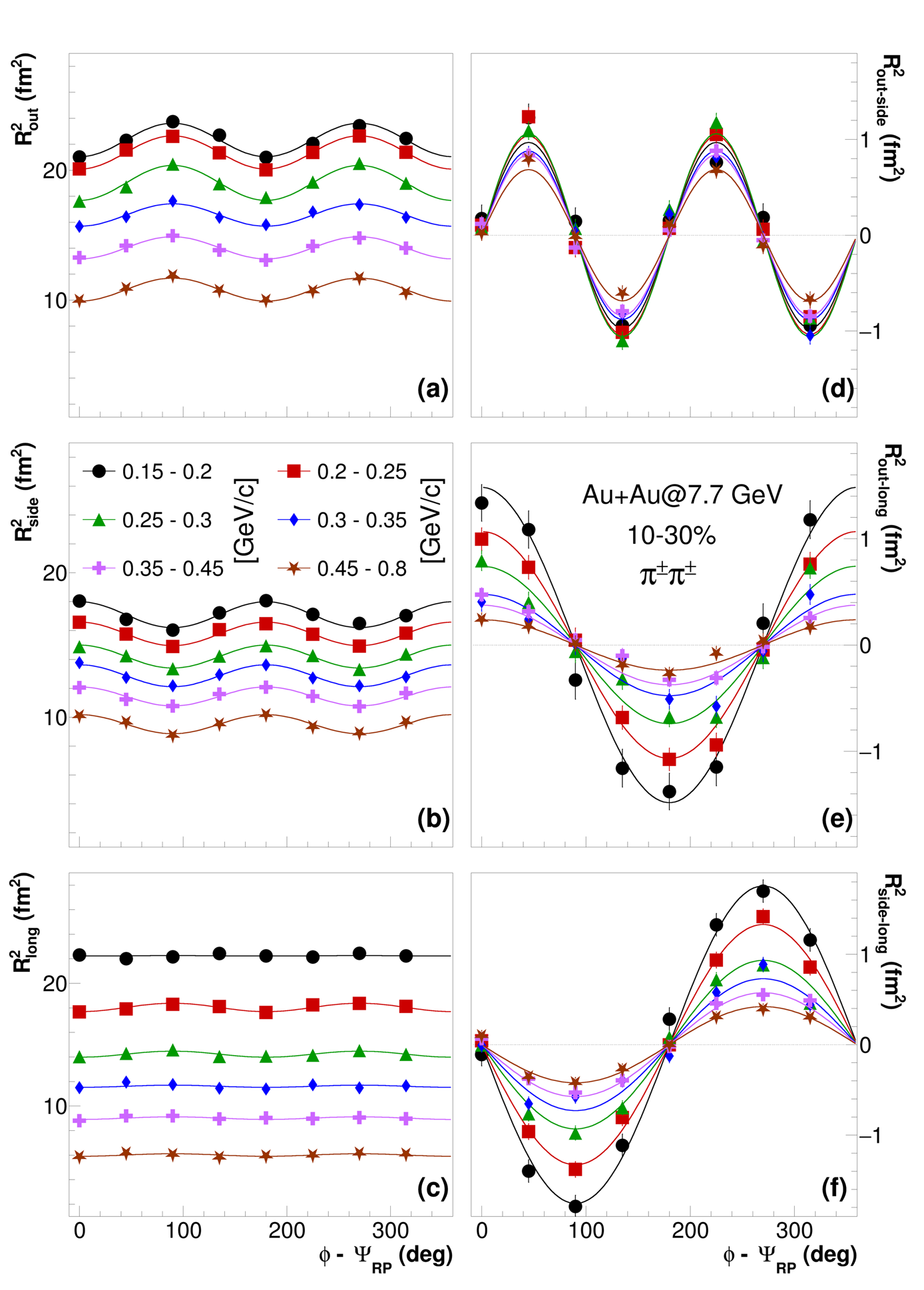}
    \caption{Extracted femtoscopic radii as a function of the azimuthal angle of the pion pair
relative to the event plane for six ranges in transverse momentum $k_{\mathrm{T}}$. Black
circles correspond to $0.15$--$0.2$~GeV/$c$, red squares to $0.2$--$0.25$~GeV/$c$, green triangles to
$0.25$--$0.3$~GeV/$c$, blue diamonds to $0.3$--$0.35$~GeV/$c$, violet crosses to $0.35$--$0.45$~GeV/$c$, and brown stars to $0.45$--$0.8$~GeV/$c$. Results are shown for Au+Au collisions at $\sqrt{s_{NN}}=7.7$~GeV in the 10--30\%
centrality class. Lines of the corresponding colors represent fits to the oscillations of the femtoscopic radii using Eq.~\ref{eq:oscFit}.}
    \label{fig:radiiKt}
\end{figure}

It is interesting to note that while a centrality dependence of the tilt angle is naturally expected from the collision geometry, a surprisingly stronger dependence of the femtoscopic parameters is observed as a function of transverse momentum, as shown in Fig.~\ref{fig:radiiKt}. The crucial component for the tilt calculation is the cross-term $R^{2}_{\text{side--long}}$ (shown in the panel (f)), which clearly differs across the $k_{\mathrm{T}}$ bins, suggesting variations in the extracted tilt angle of the emission source as well.

%%%%%%%%%%%%%%%%%%%%%%%%%%%%%%%%%%%%%%%%%%%%%%%%%%%%%%%%%%%%%%%%%%%%%%%%%%%%%%%%%%%%%%%%%
\section{Extracting the tilt parameter}

The tilt angle of the homogeneity region can be extracted by fitting the azimuthal dependence of the femtoscopic parameters with Eq.~\ref{eq:oscFit} and applying Eq.~\ref{eq:tiltSl}.

\begin{figure}[h!]
    \centering
    \includegraphics[width=1.0\linewidth]{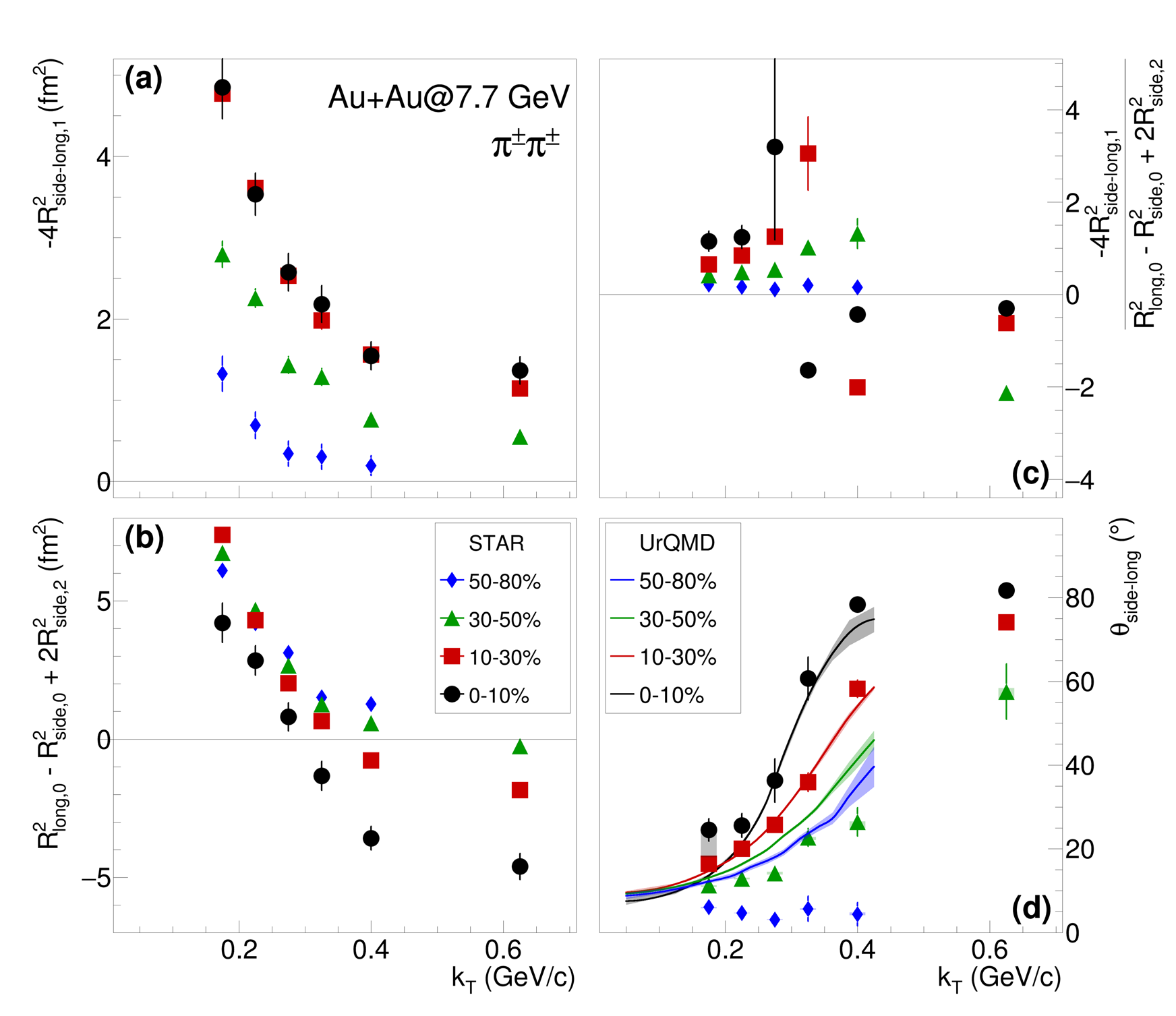}
    \caption{Illustration of the procedure for extracting the tilt parameter from the
femtoscopic radii. Panel (a) shows the transverse-momentum dependence of $-
4R^{2}_{\text{side-long},1}$. Panel (b) shows the dependence of $R^{2}_{\text{long},0} - R^{2}_{\text{side},0} + 2R^{2}_{\text{side},2}$ on transverse momentum. Panel (c) presents the ratio of the quantities from panels (a) and (b). Panel (d) shows the resulting tilt parameter, calculated as $\tfrac{1}{2}\arctan$ of the value shown in panel (c) and corresponds to the Eq.~\ref{eq:atan2}. In all panels, black circles correspond to 0--10\%, red squares to 10--30\%, green triangles to 30--50\%, and blue diamonds to 50--80\% centrality in Au+Au collisions at $\sqrt{s_{NN}} = 7.7$~GeV. Shaded bands represent the tilt obtained from UrQMD~3.4 Cascade for the corresponding centralities.  }
    \label{fig:sl_7.7}
\end{figure}

\begin{figure}[h!]
    \centering
    \includegraphics[width=1.0\linewidth]{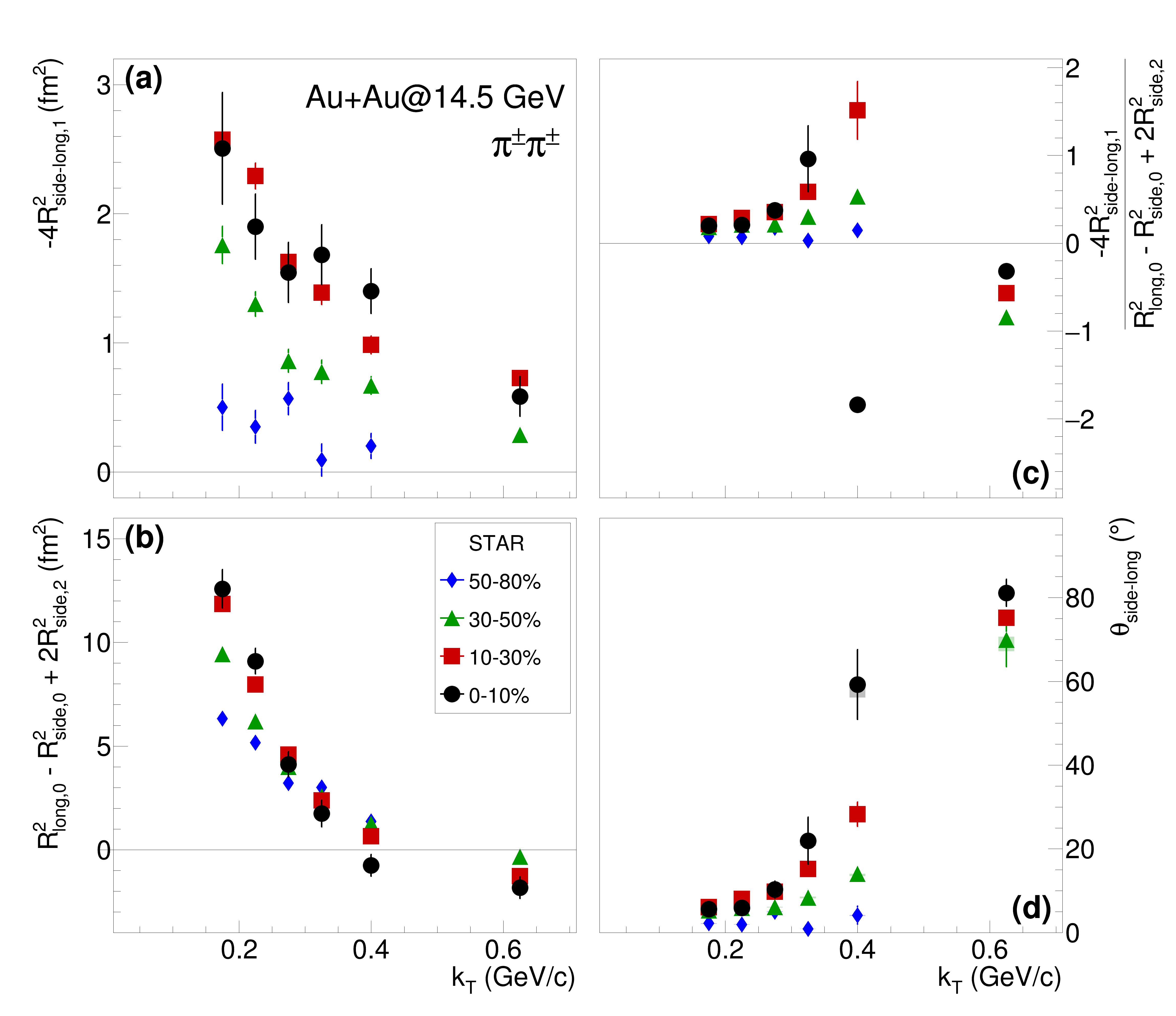}
    \caption{Same as Figure~\ref{fig:sl_7.7}, but for 14.5 GeV.}
    \label{fig:sl_14}
\end{figure}

Figures~\ref{fig:sl_7.7}, \ref{fig:sl_14}, and \ref{fig:sl_27} illustrate the procedure for extracting the tilt parameter at three different collision energies. Each figure consists of four panels. Panel (a) shows the numerator from Eq.~\ref{eq:tiltSl}, which represents the key quantity describing the correlation between the \textit{side} and \textit{long} axes of the ellipsoidal source model. Panel (b) shows the difference in the magnitudes of the longitudinal and outward radii and sets the geometric scale entering the tilt extraction. It is important to note that Eq.~\ref{eq:tiltSl} was originally derived under the assumption that the \textit{long} axis of the homogeneity region is always larger than the \textit{side} axis. While this assumption holds at low $k_{\mathrm{T}}$, experimental measurements at higher $k_{\mathrm{T}}$ reveal that it is not universally valid. For instance, at $\sqrt{s_{\mathrm{NN}}}=27$~GeV (Fig.~\ref{fig:sl_27}), the difference between $R^{2}_{\text{long}}$ and $R^{2}_{\text{side}}$ in panel (b) is almost always positive, whereas at $\sqrt{s_{\mathrm{NN}}}=7.7$~GeV (Fig.~\ref{fig:sl_7.7}), nearly half of the points fall on the negative side, indicating that the \textit{side} radius exceeds the \textit{long} one. In all cases, the parameter $2R_{\text{side,2}}^{2}$ is small and has a negligible impact on the results shown in panel (b) for any of the considered energies.

Panel (c) shows the ratio of the quantities plotted in panels (a) and (b). A sudden jump in the dependence observed in panel (c) is caused by the negative difference between $R^{2}_{\text{long}}$ and $R^{2}_{\text{side}}$ shown in panel (b). To obtain the tilt angle, one takes one half of the arctangent of the values presented in panel (c). Since the tangent function is $\pi/2$-periodic, special care must be taken to avoid an artificial sign flip, which could lead to an negative value of the extracted tilt angle in such cases. Negative values of the tilt is a subject for interpretation and will be discussed later in the section~\ref{sec:energy}. The best way to avoid any inconsistencies is to expand Eq.~\ref{eq:tiltSl}, as follows:

\begin{equation}
\theta = \frac{1}{2}
\begin{cases}
\tan^{-1}\!\left(\dfrac{Y}{K}\right), & K > 0,\\[1.2em]
\tan^{-1}\!\left(\dfrac{Y}{K}\right) + \pi, & K < 0,\, Y \ge 0,\\[1.2em]
\tan^{-1}\!\left(\dfrac{Y}{K}\right) - \pi, & K < 0,\, Y < 0,\\[0.6em]
+\dfrac{\pi}{2}, & K = 0,\, Y > 0,\\[0.6em]
-\dfrac{\pi}{2}, & K = 0,\, Y < 0,
\end{cases}
\qquad \theta \in (-\pi, \pi].
\label{eq:atan2}
\end{equation}
where $Y = -4R^2_{side-long,1}$ (or $-4R^2_{out-long,1}$ if one wants to analyze \textit{out-long} direction) and $K = R^{2}_{long,0}-R^{2}_{side,0}+2R^{2}_{side,2}$. This formulation ensures a physically meaningful and continuous determination of the tilt angle, even across sign changes in the denominator.

In practice, when performing the calculation in code (for example, using \texttt{C++}), one should use the \texttt{atan2} function, which correctly distinguishes between quadrants and thus prevents ambiguity in the determination of the tilt angle.

\begin{figure}[h!]
    \centering
    \includegraphics[width=1.0\linewidth]{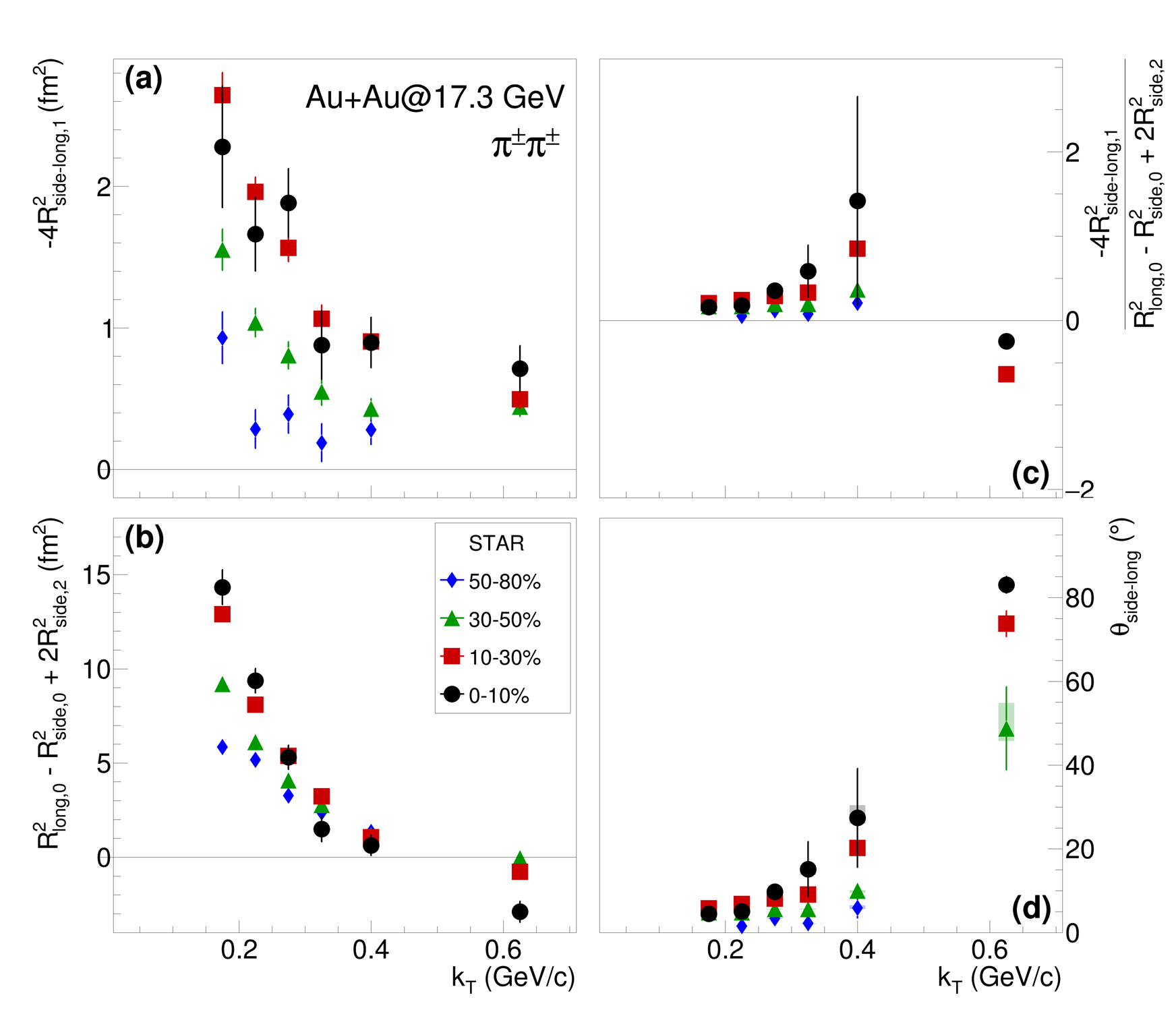}
    \caption{Same as Figure~\ref{fig:sl_7.7} and~\ref{fig:sl_14}, but for 17.3 GeV.}
    \label{fig:sl_17}
\end{figure}

\begin{figure}[h!]
    \centering
    \includegraphics[width=1.0\linewidth]{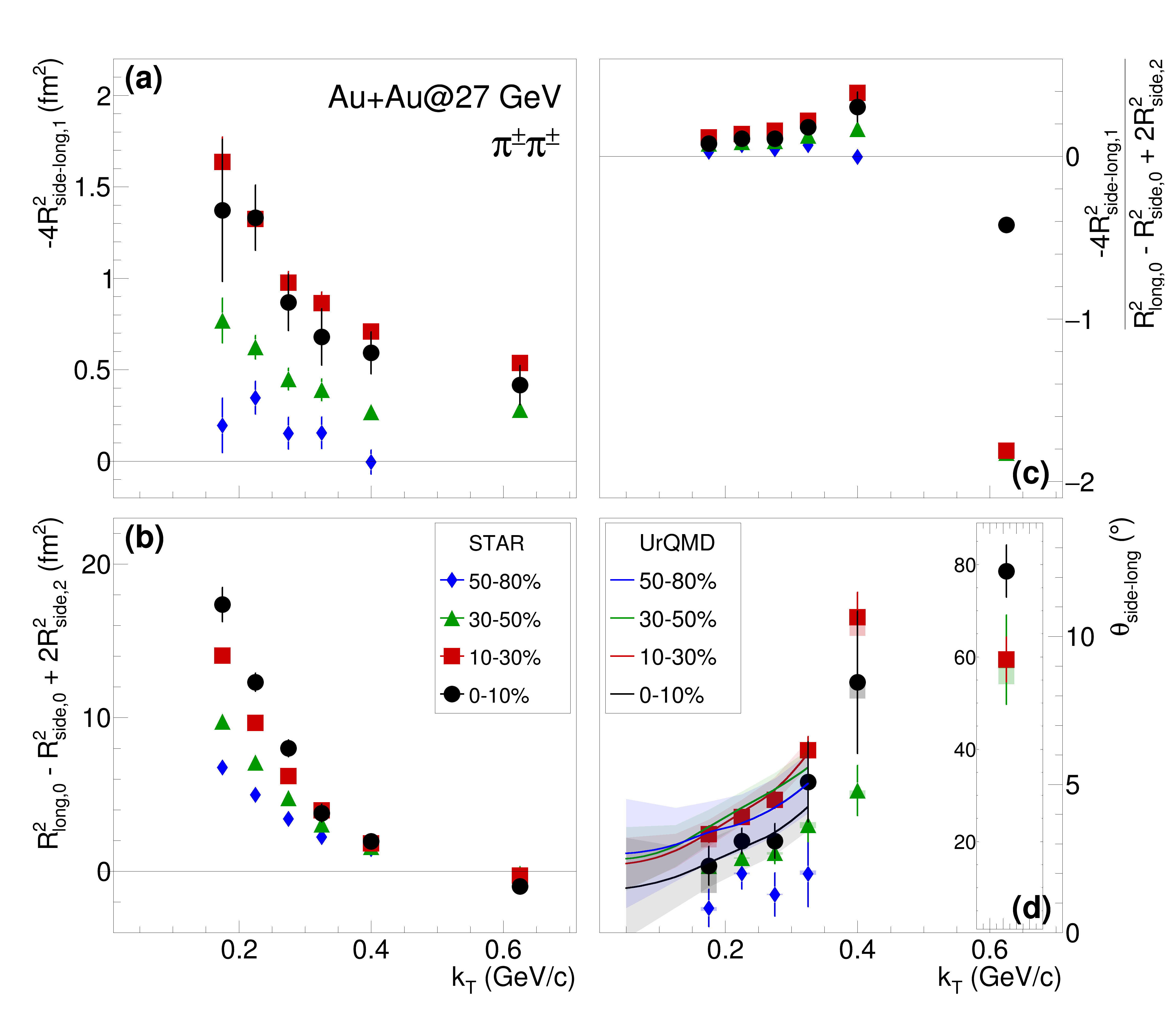}
    \caption{Same as Figure~\ref{fig:sl_7.7}, ~\ref{fig:sl_14}, and~\ref{fig:sl_17}, but for 27 GeV.}
    \label{fig:sl_27}
\end{figure}

Panel (d) of Figures~\ref{fig:sl_7.7}--~\ref{fig:sl_27} shows the extracted tilt parameter obtained using Eq.~\ref{eq:atan2} as a function of the transverse momentum of pion pairs. The experimental data are shown by the markers, while the tilt values calculated from UrQMD~3.4 Cascade are represented by the lines and have been taken from~\cite{Khyzhniak:2024chj}. Each data point includes both statistical and systematic uncertainties, shown as lines and shaded bands, respectively. The shaded regions around the model lines indicate the corresponding statistical uncertainties of the simulation.

For the most peripheral collisions in the highest \(k_T\) bin at all four considered energies, as well as for \(k_T = 0.4\) GeV/\(c\) at \(\sqrt{s_{NN}} = 27\) GeV, the extracted tilt is not shown. In these cases, the side--long correlation, visible in panel (a) of Figs.~\ref{fig:sl_7.7}--\ref{fig:sl_27}, becomes very small, while the source shape becomes increasingly round, as seen in panel (b) of Figs.~\ref{fig:sl_7.7}--\ref{fig:sl_27}. This combination, together with the available statistics, leads to a very large systematic uncertainty in the tilt determination, making the corresponding points not meaningful to display.

To evaluate the systematic uncertainty associated with the final results, four sources were considered: the choice of the fit range, the assumed source size used to account for the Coulomb interaction between particles in the pair, and the inclusion of momentum resolution and pair purity in the fit. Since these sources of uncertainty are correlated---each corresponding to an alternative fit to the same dataset---the systematic uncertainty was evaluated as  
$\sigma_{\mathrm{syst}} = \max |R_i - R_{\mathrm{def}}|$,  
where $R_i$ and $R_{\mathrm{def}}$ denote the fitted radius parameters obtained from the varied and default configurations, respectively.

The uncertainty associated with the Coulomb correction was estimated by varying the source radius within the minimal and maximal values obtained from the three-dimensional femtoscopic radii. The largest contribution to the systematic uncertainty in the tilt calculation arises from the assumed radius of the Coulomb source, as it is modeled as a spherically symmetric emitter -- an approximation that obviously does not hold for this analysis. At the time of this study, only a one-dimensional Coulomb correction was practically available due to computational limitations. This effect is visible in Fig.~\ref{fig:2DCF}. The red lines in this figure represent the tilts of the correlation function obtained using femtoscopic parameters from the three-dimensional fits. A slight misalignment between these red lines and the principal axes of the corresponding ellipses illustrates the influence of this systematic uncertainty.

One can see from panel (d) of Figures~\ref{fig:sl_7.7} and \ref{fig:sl_27} that, in general, the UrQMD model fails to reproduce the exact experimental values; however, it qualitatively captures both the centrality dependence and the overall $k_{\mathrm{T}}$ trend of the tilt parameter. It is interesting that the tilt magnitude is best captured in the 10-30\% centrality class for both
of the considered energies.

It is interesting to note that although the values of the tilt parameter are very small at low $k_{\mathrm{T}}$, the ordering of centralities appears to change in this region. More detail may be found in~\cite{Khyzhniak:2024chj}. It is also evident that the $k_{\mathrm{T}}$ dependence is much stronger than the centrality dependence. One possible interpretation of this behavior is that it may, at least in part, reflect a time evolution of the source tilt, as particles with higher $k_{\mathrm{T}}$ are typically emitted at earlier stages of the collision.

%%%%%%%%%%%%%%%%%%%%%%%%%%%%%%%%%%%%%%%%%%%%%%%%%%%%%%%%%%%%%%%%%%%%%%%%%%%%%%%%%%%%%%%%%
\section{World systematics for  the tilt parameter}
\label{sec:energy}
\begin{figure}[h!]
    \centering 
    \includegraphics[width=0.9\linewidth]{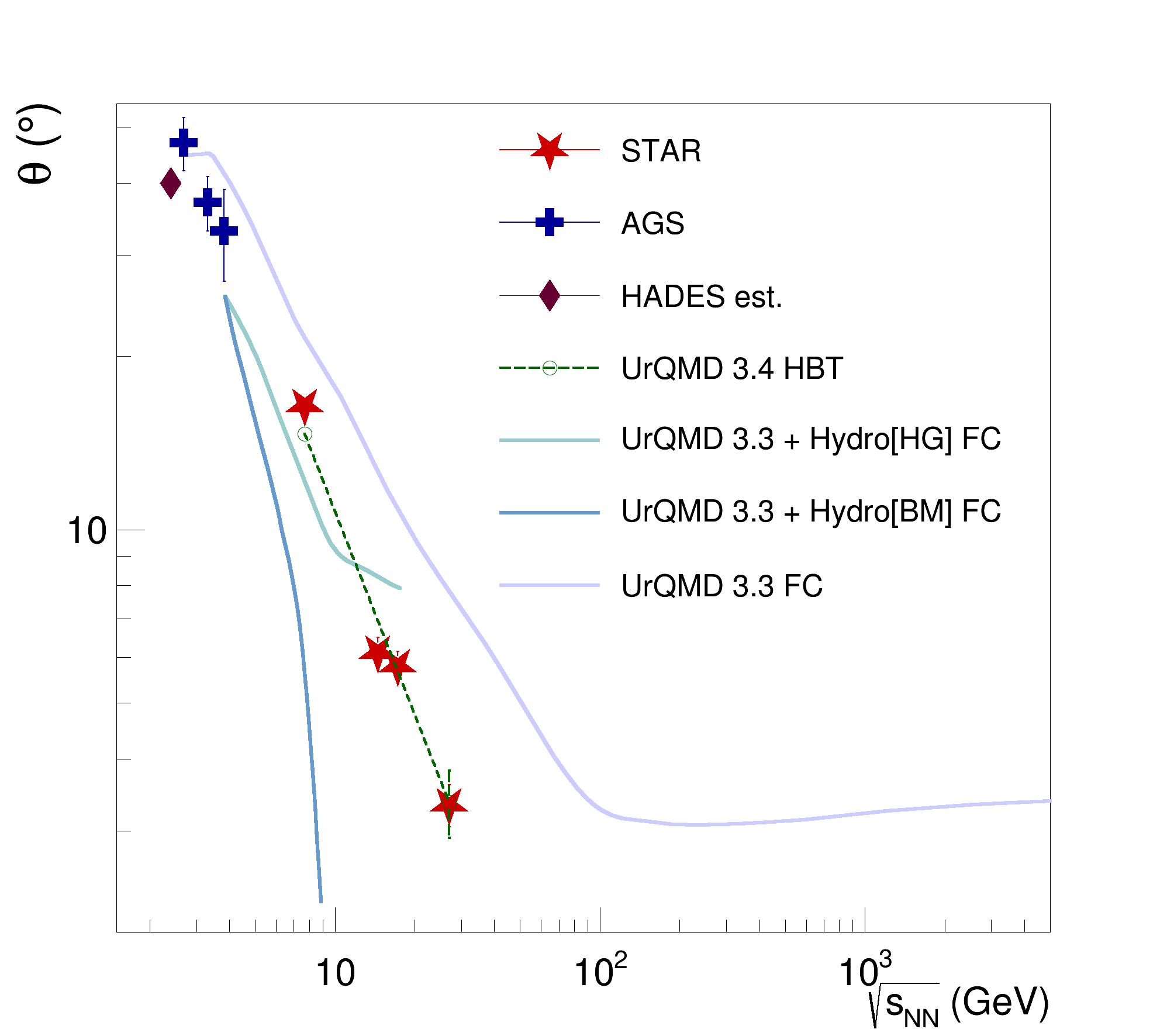}
    \caption{Energy dependence of the extracted tilt parameter. Red stars represent the STAR
results (where $\theta = \theta_{\text{side-long}}$) obtained in this work, blue crosses correspond to AGS data~\cite{E895:2000opr}, and the violet diamond denotes the estimate from HADES publication data~\cite{HADES:2019lek}. Green dashed line show the tilt values extracted from azimuthally sensitive femtoscopy using UrQMD~3.4 Cascade~\cite{Khyzhniak:2024chj}. Other filled lines indicate tilt estimates obtained by fitting the distribution of freeze-out coordinates from UrQMD~3.3 Cascade~\cite{Lisa:2011na}, UrQMD~3.3+Hydro[HG]~\cite{Lisa:2011na}, and UrQMD~3.3+Hydro[BM]~\cite{Lisa:2011na}.}
    \label{fig:energy}
\end{figure}
\begin{figure}[h!]
    \centering
    \includegraphics[width=1.0\linewidth]{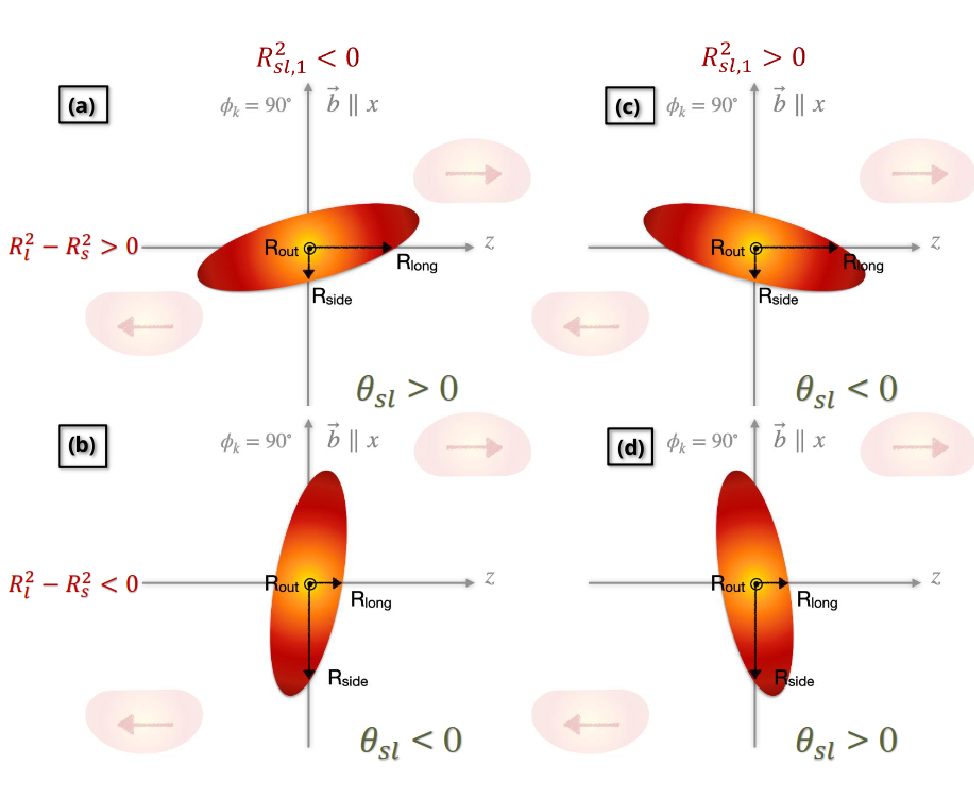}
    \caption{Schematic representation of the orientation of the homogeneity region with respect to
the fixed STAR coordinate system. One particular homogeneity region is shown for the case where the azimuthal angle of the pion pair is $\phi_{k} = 90^\circ$. In this configuration, the beam direction is along the $z$ axis, $R_{\text{long}}$ is parallel to the $z$ axis, and the impact parameter vector is aligned with the $x$ axis. The $y$ axis is perpendicular to the reaction plane, $R_{\text{out}}$ points out of the page and
$R_{\text{side}}$ is perpendicular to both $R_{\text{long}}$ and $R_{\text{out}}$. Cases (a) and (b) illustrate situations where the correlation between $R_{\text{side}}$ and $R_{\text{long}}$ is negative, yielding
$R^{2}_{\text{side-long},1} < 0$, while cases (c) and (d) show the opposite sign. Additionally, in cases (a) and (c), $R^{2}_{\text{long}}$ is larger than $R^{2}_{\text{side}}$, whereas in cases (b)
and (d)$, R^{2}_{\text{side}}$ exceeds $R^{2}_{\text{long}}$. }
    \label{fig:ellipses}
\end{figure}
To compare the results obtained in this work with previously published data, the energy dependence of the tilt parameter was constructed, as shown in Fig.~\ref{fig:energy}. The tilt parameter for the lowest $k_{\mathrm{T}}$ bin was taken from Figs.~\ref{fig:sl_7.7}-\ref{fig:sl_27}, respectively, in order to match as closely as possible the AGS acceptance.

One can see that the STAR results are consistent with the general trend observed in the AGS data: the tilt decreases with increasing collision energy, which is consistent with the expectation that the system becomes increasingly boost-invariant. The STAR data lie slightly below the UrQMD~3.4 Cascade predictions. The comparison between data and model calculations suggests that the tilt parameter is quite sensitive to the underlying equation of state (EoS).

It is interesting to compare the beam-energy dependence of the femtoscopic tilt parameter with that of directed-flow observables, in particular the midrapidity pion slope \(dv_{1}/dy\), reported, for example, in~\cite{STAR:2020dav}. Such a comparison is physically well motivated, since both observables are sensitive to the same underlying asymmetric longitudinal initial-state of the collision system. The femtoscopic tilt reflects this asymmetry in coordinate space through the orientation of the freeze-out homogeneity regions, whereas $dv_{1}/dy$ reflects it in momentum space through the sideward collective deflection of emitted particles.

This connection is also supported by hydrodynamic studies discussed in the Introduction. In addition, the experimental study~\cite{STAR:2017ykf} is particularly relevant, as the observed directed flow was interpreted there as arising, at least in part, from a tilted source in the initial state. In that work, it was argued that the measured $v_{1}$ slopes in Cu+Au and Au+Au collisions are consistent with a similar initial tilt of the created medium, and that this tilt depends more strongly on the collision energy than on the system size.

Against this background, the comparison with the present femtoscopic results is especially suggestive. Both observables exhibit a similar overall evolution with beam energy: the magnitude of $dv_{1}/dy$ decreases with increasing $\sqrt{s_{NN}}$, and the femtoscopic tilt angle extracted in this work also becomes smaller as the beam energy increases. This qualitative agreement is consistent with the expectation that the longitudinal tilt of the system becomes less pronounced at higher collision energies.

At the same time, important differences between the two observables remain. The femtoscopic tilt remains positive over the studied energy range and shows a comparatively strong beam-energy dependence, whereas the directed-flow slope is negative and varies more moderately~\cite{STAR:2020dav}. This is not surprising, since the two quantities probe different manifestations of the same underlying dynamics. It should be noted, however, that a nonzero femtoscopic tilt does not uniquely imply a purely geometric tilt of the full source, since $side-side$ correlations may also receive contributions from the collective velocity field, including longitudinal--transverse shear. In this sense, the opposite signs of the two observables are physically plausible rather than contradictory. A positive femtoscopic tilt refers to the orientation of the major axis of the emission ellipsoid, which is tilted toward positive \(z\). Since collective expansion is driven by pressure gradients, it develops preferentially along the minor axis of the emission ellipsoid. For the tilt convention illustrated in Fig.~\ref{fig:collision}, this minor axis is perpendicular to the major axis and has a component along negative \(z\). Therefore, a positive spatial tilt may contribute to a negative directed-flow slope near midrapidity, corresponding to \(v_{1}>0\) on the \(z<0\) side and \(v_{1}<0\) on the \(z>0\) side. Nevertheless, the fact that they exhibit a similar global energy dependence supports the interpretation that both observables retain sensitivity to the same longitudinal asymmetry of the system.

A note should be made regarding the HADES data. In~\cite{HADES:2019lek}, the HADES Collaboration defined tilt angle using the spatial correlation tensor $S_{\mu\nu}$. This definition of the tilt can be used only under the assumption of non-flowing source~\cite{Heinz:2002au,Mount:2010ey}, otherwise one has to switch to the definition provided by Eq.~\ref{eq:atan2}. Additionally HADES reported a negative value of the tilt parameter, which remains subject to interpretation. To clarify the sign convention and physical interpretation of the tilt parameter, a schematic illustration is presented in Fig.~\ref{fig:ellipses}. This figure shows the orientation of the homogeneity region with respect to the fixed STAR coordinate system. A particular homogeneity region is depicted for the case where the azimuthal angle of the pion pair is $\phi_{k} = 90^\circ$. In this configuration, the beam direction is along the $z$ axis, $R_{\text{long}}$ is parallel to the $z$ axis, and the impact parameter vector is aligned with the $x$ axis. The $y$ axis is perpendicular to the reaction plane, $R_{\text{out}}$ points out of the page, and $R_{\text{side}}$ is perpendicular to both $R_{\text{long}}$ and $R_{\text{out}}$.  

Cases (a) and (b) illustrate configurations where the correlation between $R_{\text{side}}$ and $R_{\text{long}}$ is negative, resulting in $R^{2}_{\text{side--long},1} < 0$, while cases (c) and (d) correspond to positive correlations. Additionally, in cases (a) and (c), $R^{2}_{\text{long}}$ is larger than $R^{2}_{\text{side}}$, whereas in cases (b) and (d), $R^{2}_{\text{side}}$ exceeds $R^{2}_{\text{long}}$.

As one can see, both the numerator and denominator are required in order to unambiguously determine which of the cases shown in Fig.~\ref{fig:ellipses} applies. Using the information provided in~\cite{HADES:2019lek} and in~\cite{Greifenhagen:2020wsz}, one can conclude that $R^{2}_{\text{side--long},1}$ is always negative for HADES. This effectively excludes the configurations illustrated in panels (c) and (d) of Fig.~\ref{fig:ellipses}. Consequently, the HADES tilt angle exceeds $45^{\circ}$ but remains below $90^{\circ}$. In order to compare the HADES result with the world data consistently, one should therefore add $\pi/2$ to the value of the tilt parameter, according to the definition introduced in Eq.~\ref{eq:atan2}.

\section{Final kinetic freeze-out eccentricity}

In addition to the tilt extraction, we have performed a measurement of the final kinetic freeze-out eccentricity of the homogeneity regions. The eccentricity with respect to the beam axis is calculated following~\cite{Retiere:2003kf} as
\begin{equation}
\varepsilon_{F} = \frac{2R^{2}_{s,2}}{R^{2}_{s,0}}
\label{eq:ecc}
\end{equation}
which is, in fact, an approximation of the transverse eccentricity in the ``natural'' frame tilted relative to the beam axis~\cite{Mount:2010ey}. This approximation is valid under the assumption that the tilt angle is negligibly small.
The full expression for the eccentricity in the tilted frame is given by
\begin{widetext}
\begin{equation}
\label{eq:eccFull}
\varepsilon_{F} =
\frac{
2 R^{2}_{s,2}\left(1 + \cos^{2}\theta\right)
+ \left(R^{2}_{s,0} - R^{2}_{l,0}\right)\sin^{2}\theta
- 2 R^{2}_{sl,1}\sin\!\left(2\theta\right)
}{
R^{2}_{s,0}\left(1 + \cos^{2}\theta\right)
+ \left(2R^{2}_{s,0} + R^{2}_{l,0}\right)\sin^{2}\theta
+ 2 R^{2}_{sl,1}\sin\!\left(2\theta\right)
}
\end{equation}
\end{widetext}
where $\theta$ is tilt angle from Eq.~\ref{eq:atan2} and $R_{i,j}$ are Fourier coefficients from Eq.~\ref{eq:oscFit}.
As can be seen from Figs.~\ref{fig:sl_7.7}--\ref{fig:sl_27}, the tilt angle is negligibly small either at high collision energies or at relatively small transverse momenta $k_{\mathrm{T}}$, with the precise $k_{\mathrm{T}}$ range depending on the collision energy.

\begin{figure}[h!]
    \centering
    \includegraphics[width=1.0\linewidth]{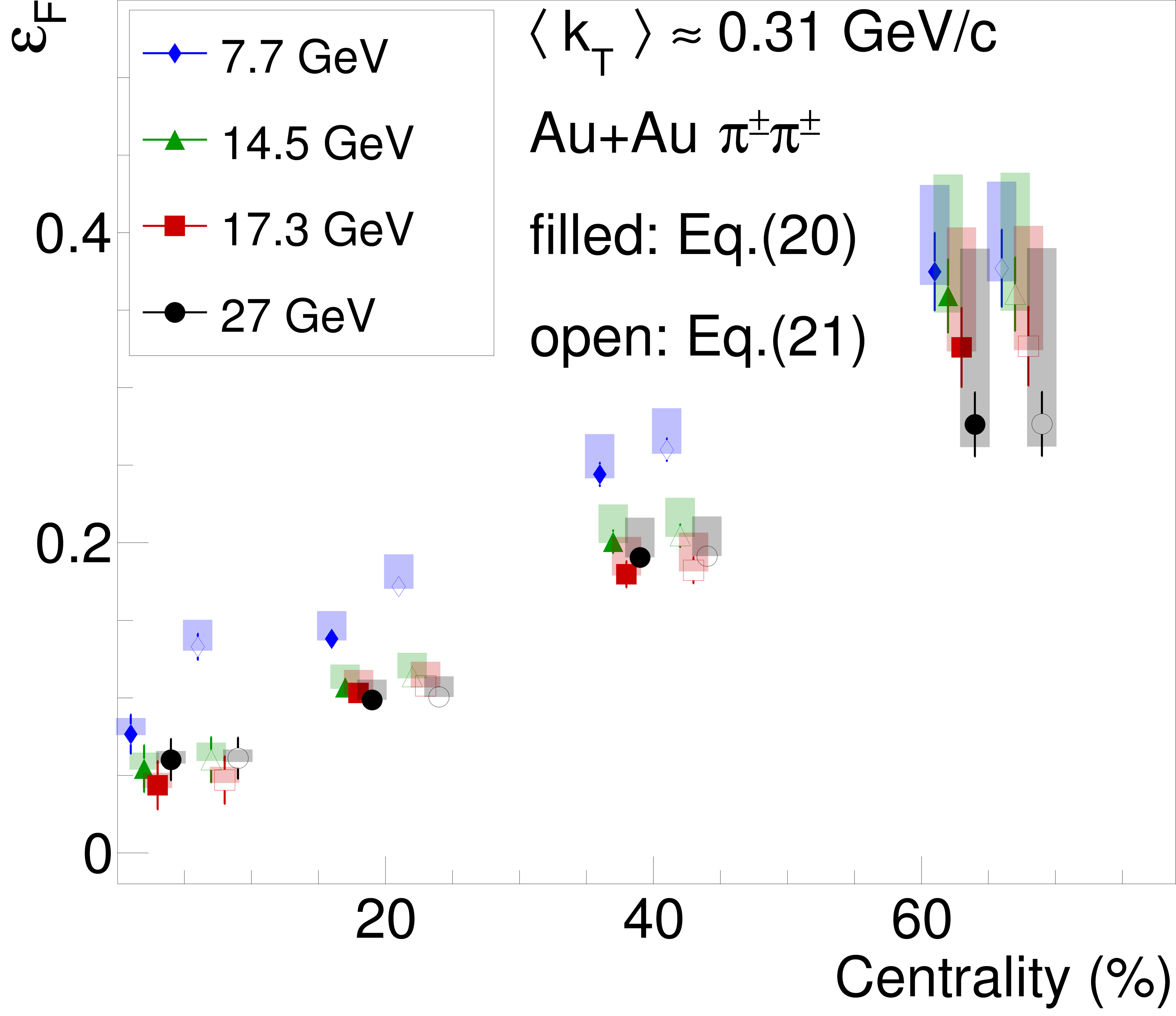}
    \caption{Final eccentricity as a function of centrality for pion pairs with combined transverse momentum
$\langle k_{\mathrm{T}} \rangle \approx 0.31~\mathrm{GeV}/c$.
Results for different collision energies are shown using distinct colors and marker styles.
Filled markers correspond to the eccentricity defined by Eq.~\ref{eq:ecc}, while open markers represent
the eccentricity calculated using Eq.~\ref{eq:eccFull}.}
    \label{fig:ecc_cent}
\end{figure}

\begin{figure}[h!]
    \centering
    \includegraphics[width=1.0\linewidth]{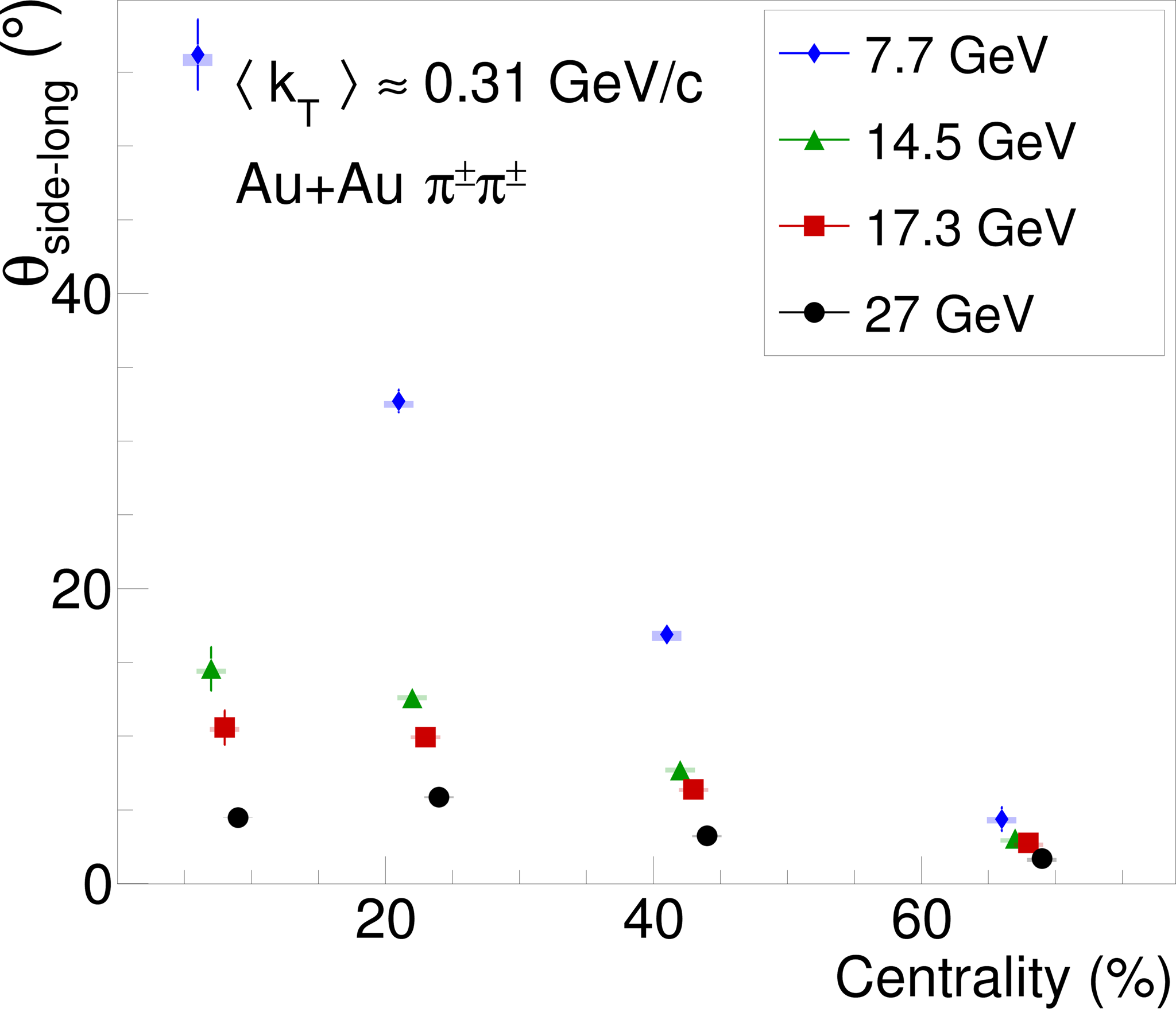}
    \caption{Tilt angle, calculated using Eq.~\ref{eq:atan2}, as a function of centrality for pion pairs with an average transverse momentum
$\langle k_{\mathrm{T}} \rangle \approx 0.31~\mathrm{GeV}/c$.
Results for different collision energies are shown using distinct colors and marker styles.}
    \label{fig:tilt_ave}
\end{figure}

Figure~\ref{fig:ecc_cent} shows the centrality dependence of the eccentricity calculated using Eqs.~\ref{eq:ecc} and~\ref{eq:eccFull} for all collision energies, for pion pairs with combined transverse momentum corresponding to an average $\langle k_{\mathrm{T}} \rangle \approx 0.31~\mathrm{GeV}/c$. The corresponding tilt angles obtained at the same centralities, $k_{\mathrm{T}}$, and energies are shown in Fig.~\ref{fig:tilt_ave}. A noticeable difference between the eccentricities evaluated with respect to the beam axis and the tilted source frame is observed only for $\sqrt{s_{NN}} = 7.7$~GeV in the 0--30\% centrality range, where the tilt angle exceeds $30^\circ$, as seen in Fig.~\ref{fig:tilt_ave}.

As expected, the homogeneity regions become progressively more azimuthally symmetric toward central collisions. This trend is consistent with the geometric evolution from an initially almond-shaped overlap region in peripheral collisions toward a nearly round configuration in central events, where collective expansion effectively washes out residual spatial anisotropy.

\begin{figure}[h!]
    \centering
    \includegraphics[width=1.0\linewidth]{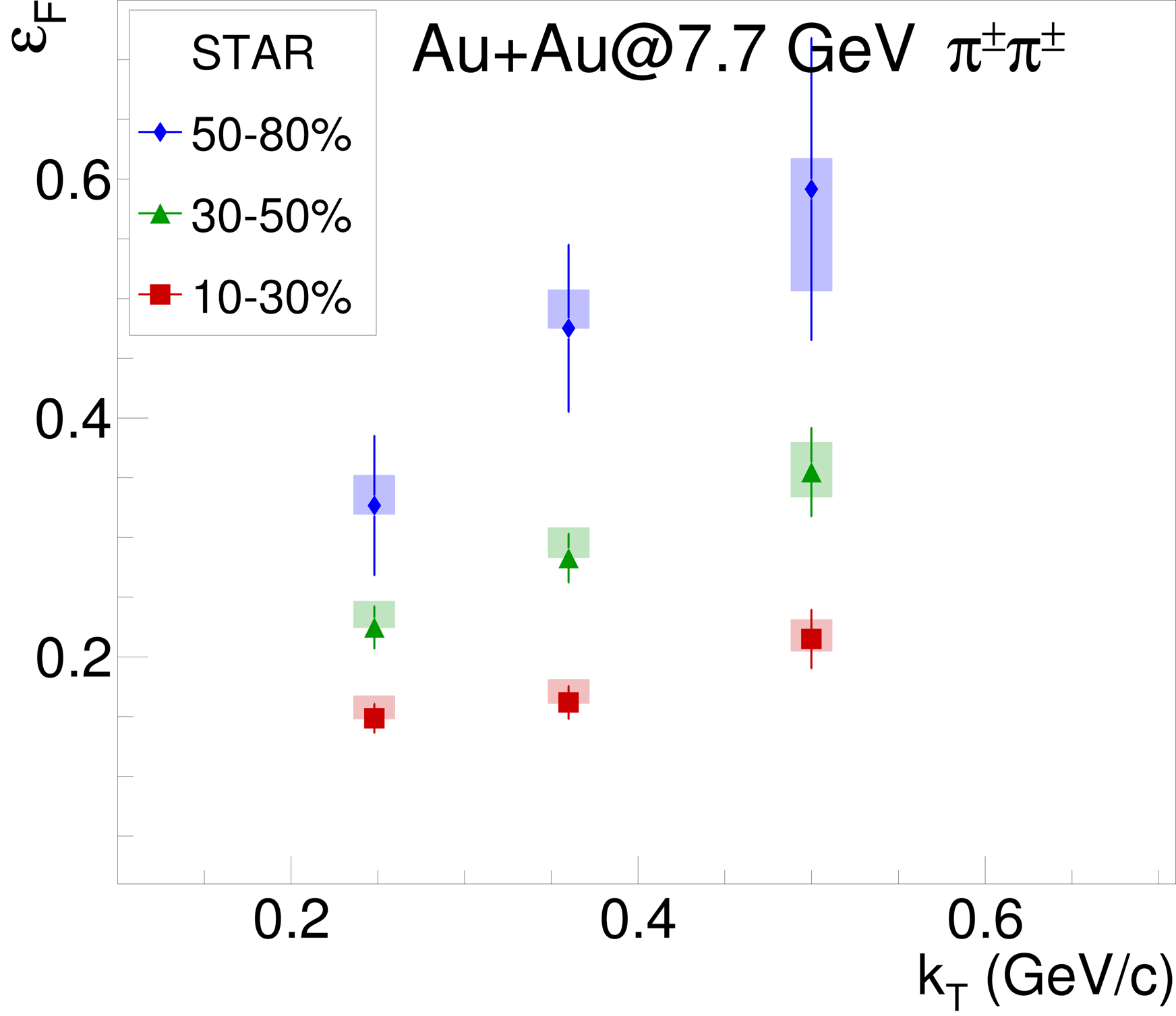}
    \caption{Transverse momentum dependence of the final eccentricity, estimated according to Eq.~\ref{eq:ecc}, for Au+Au collisions at $\sqrt{s_{\mathrm{NN}}}=7.7$~GeV. Different colors indicate different collision centralities, as specified in the legend. Vertical lines represent statistical uncertainties, while the surrounding boxes denote systematic uncertainties.}
    \label{fig:ecc_7}
\end{figure}

\begin{figure}[h!]
    \centering
    \includegraphics[width=1.0\linewidth]{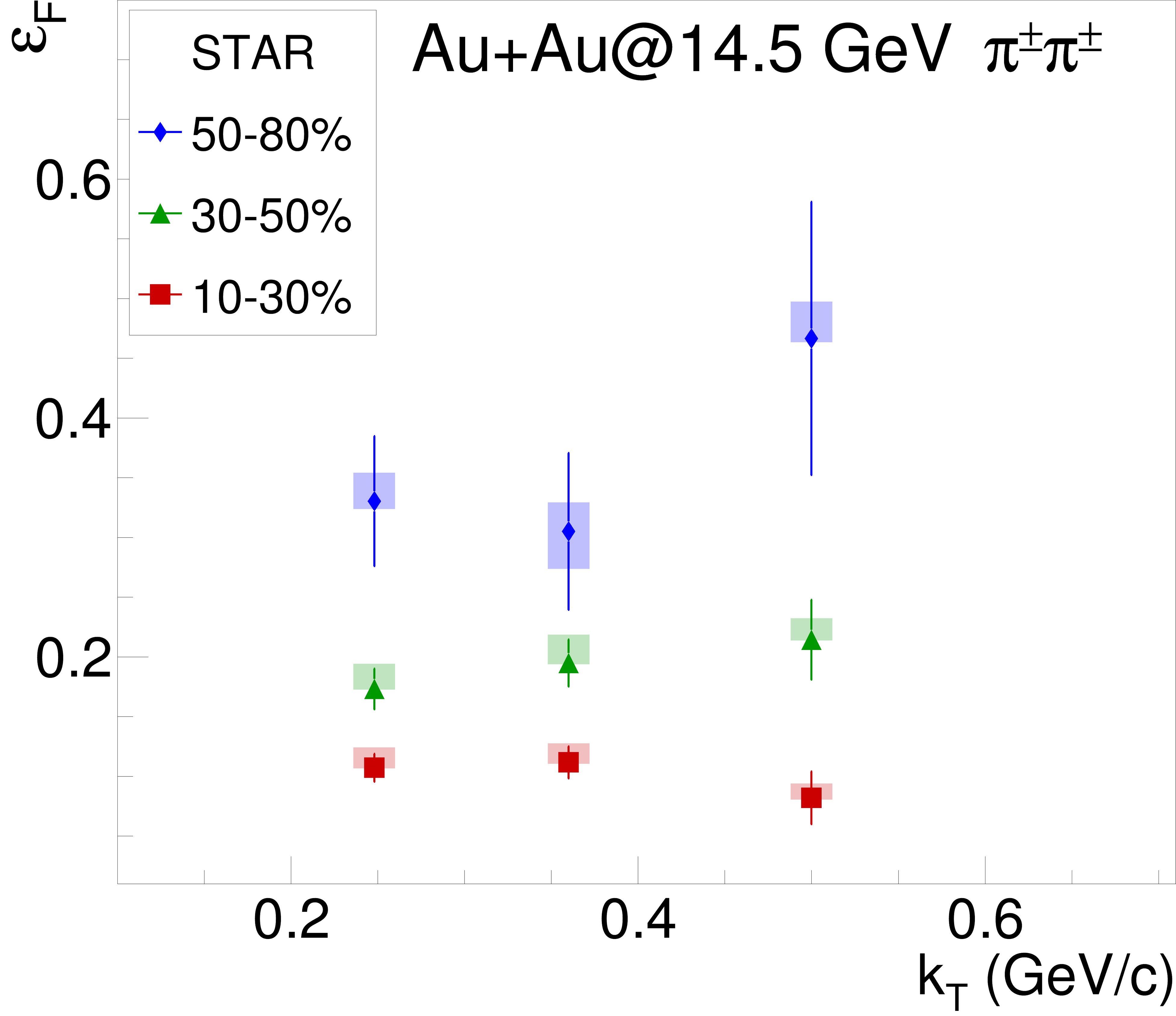}
    \caption{Transverse momentum dependence of the final eccentricity, estimated according to Eq.~\ref{eq:ecc}, for Au+Au collisions at $\sqrt{s_{\mathrm{NN}}}=14.5$~GeV. Different colors indicate different collision centralities, as specified in the legend. Vertical lines represent statistical uncertainties, while the surrounding boxes denote systematic uncertainties.}
    \label{fig:ecc_14}
\end{figure}

\begin{figure}[h!]
    \centering
    \includegraphics[width=1.0\linewidth]{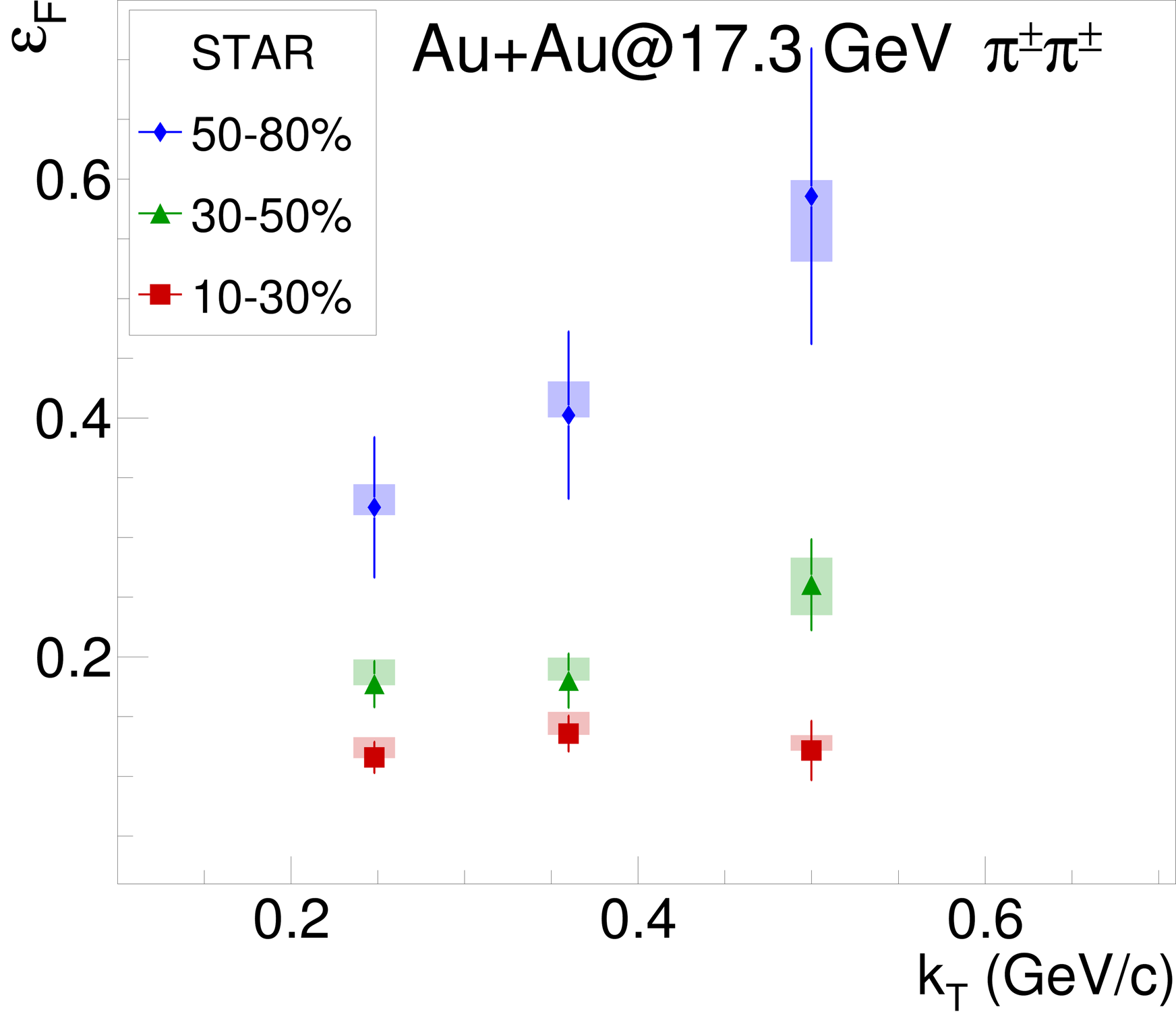}
    \caption{Transverse momentum dependence of the final eccentricity, estimated according to Eq.~\ref{eq:ecc}, for Au+Au collisions at $\sqrt{s_{\mathrm{NN}}}=17.3$~GeV. Different colors indicate different collision centralities, as specified in the legend. Vertical lines represent statistical uncertainties, while the surrounding boxes denote systematic uncertainties.}
    \label{fig:ecc_17}
\end{figure}

\begin{figure}[h!]
    \centering
    \includegraphics[width=1.0\linewidth]{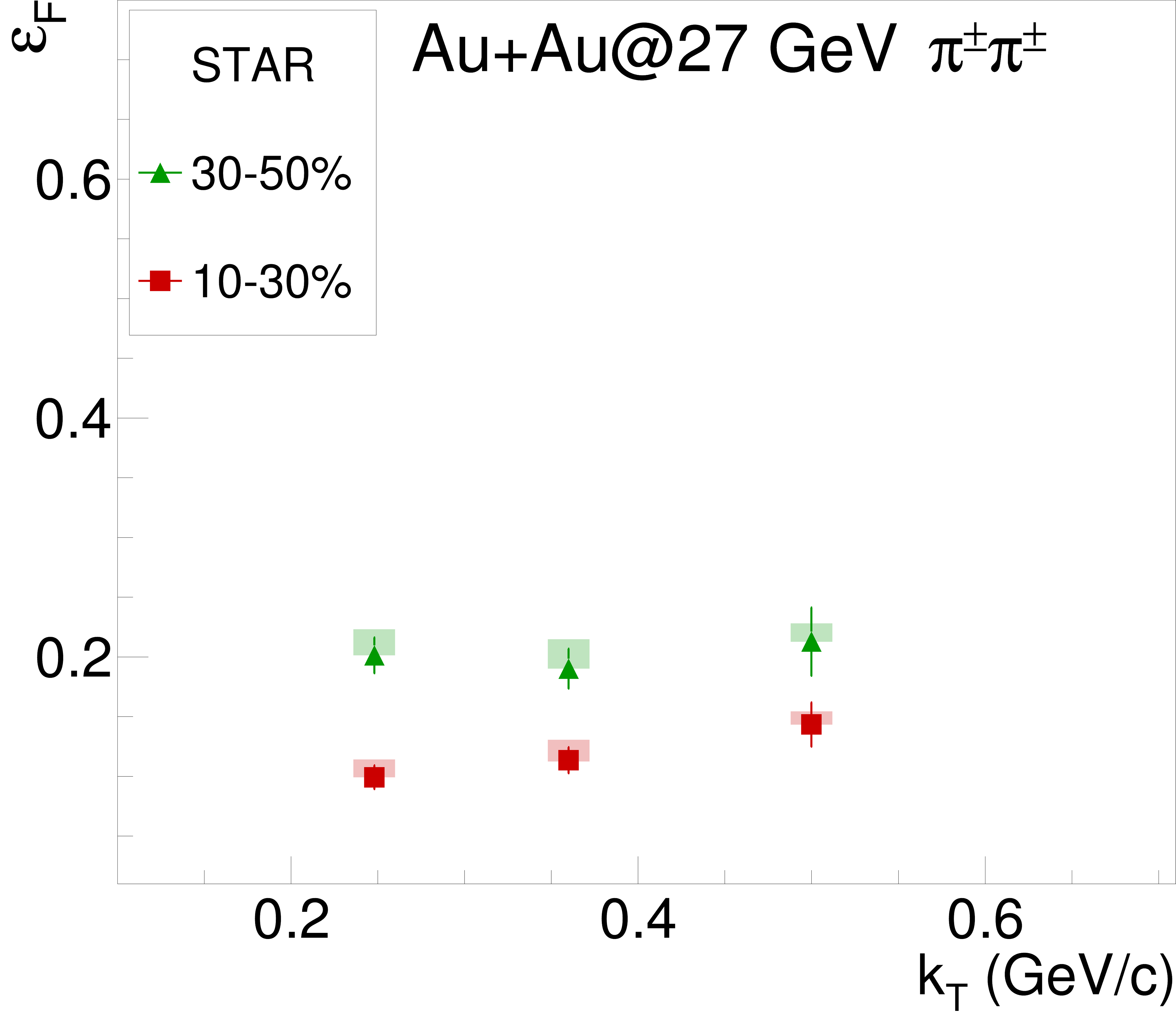}
        \caption{Transverse momentum dependence of the final eccentricity, estimated according to Eq.~\ref{eq:ecc}, for Au+Au collisions at $\sqrt{s_{\mathrm{NN}}}=27$~GeV. Different colors indicate different collision centralities, as specified in the legend. Vertical lines represent statistical uncertainties, while the surrounding boxes denote systematic uncertainties.}
    \label{fig:ecc_27}
\end{figure}

Figures~\ref{fig:ecc_7}--\ref{fig:ecc_27} show the transverse-momentum dependence of the final eccentricity, obtained according to Eq.~\ref{eq:ecc}, for all four collision energies. Despite the relatively large statistical uncertainties, the eccentricity is observed to increase with increasing $k_{\mathrm{T}}$ for all considered centralities, with the effect being more pronounced in peripheral events. The increase of the eccentricity with $k_{\mathrm{T}}$ reflects the well-established space--momentum correlations that arise in an expanding source. Higher-$k_{\mathrm{T}}$ pairs are emitted from smaller, faster-moving regions closer to the surface of the expanding fireball, where the anisotropy of the flow field is stronger, leading to a larger apparent eccentricity of the homogeneity region. 

Due to the low event-plane resolution combined with the limited signal strength, the results for the 0–10\% centrality class are not shown, as they carry large systematic uncertainties and are therefore not sufficiently reliable. The same reasoning applies to the 50–80\% centrality results at 27 GeV, where the event-plane resolution deteriorates with increasing collision energy.


\begin{figure}[h!]
    \centering
    \includegraphics[width=1.0\linewidth]{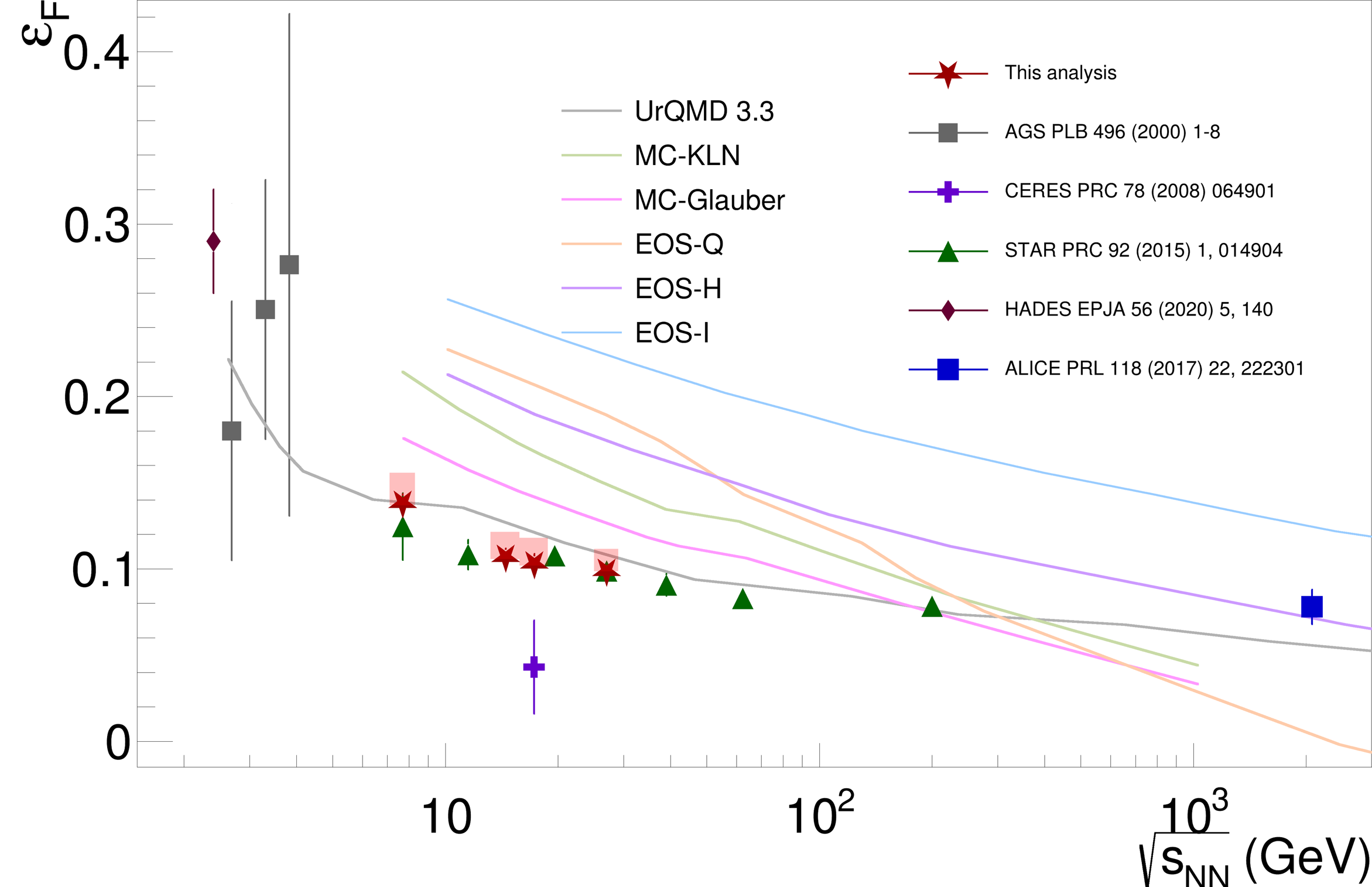}
    \caption{Final eccentricity as a function of collision energy in midcentral
Au+Au collisions (STAR, HADES, E895), Pb+Au (CERES) and Pb+Pb collisions (ALICE) for three rapidity
regions and with $\langle k_{\mathrm{T}} \rangle \approx 0.31~\mathrm{GeV}/c$. }
    \label{fig:ecc_energy}
\end{figure}
Figure~\ref{fig:ecc_energy} shows the energy dependence of the final eccentricity, combining the results of this analysis with world data. The green triangles correspond to the previously published STAR measurement based on oscillations relative to the second-order event plane, which is insensitive to forward--backward asymmetries that are the primary focus of this paper. The red stars represent the new results obtained using the first-order event plane. Very good agreement is observed between the two methods at $\sqrt{s_{NN}} = 7.7$ and $27$~GeV, despite the expected small decorrelation between the first- and second-order event planes~\cite{STAR:2022ahj}. The new measurements at $14.5$ and $17.3$~GeV align smoothly with the previously established energy dependence. In particular, the $17.3$~GeV point overlaps with the CERES measurement, effectively resolving long-standing ambiguities in the interpretation of the CERES result and closing discussions related to a possible softening of the equation of state at that energy~\cite{Lisa:2011na}.

A notable feature of Fig.~\ref{fig:ecc_energy} is the strong rise of the final eccentricity with decreasing collision energy. This trend reflects the interplay between the freeze-out geometry and collective expansion: at higher energies, stronger radial and anisotropic flow enhances space--momentum correlations, which affect the apparent anisotropy of the homogeneity regions even as the overall freeze-out geometry becomes more azimuthally symmetric. Finally, the figure includes comparisons to several model calculations, illustrating that the energy dependence of the final eccentricity is highly sensitive to the underlying equation of state. Interestingly, the model that reproduces the observed trend most closely is the UrQMD~3.3 Cascade, suggesting that hadronic transport effects capture an essential part of the relevant dynamics in this energy range.

%%%%%%%%%%%%%%%%%%%%%%%%%%%%%%%%%%%%%%%%%%%%%%%%%%%%%%%%%%%%%%%%%%%%%%%%%%%%%%%%%%%%%%%%%
\section*{Summary}

We report a measurement revealing that the particle-emitting source in relativistic heavy-ion collisions is spatially tilted. Using data from the STAR experiment at RHIC, we studied Au+Au collisions at beam energies of $\sqrt{s_{\mathrm{NN}}} = 7.7$, 14.5, 17.3 and 27~GeV. The analysis uses azimuthally sensitive femtoscopy, which examines correlations between identical pions to reconstruct the shape and orientation of the particle-emitting source at freeze-out.

We observe clear oscillations of the femtoscopic radii with respect to the event plane, showing that the freeze-out source is not symmetric but tilted relative to the beam axis. The tilt angle becomes smaller at higher collision energies, consistent with a scenario in which the geometry becomes increasingly boost-invariant with increasing $\sqrt{s_{NN}}$. The tilt parameter also depends on $k_{\mathrm{T}}$: pion pairs with higher $k_{\mathrm{T}}$, which are emitted earlier, come from regions that are less rotated, hinting at a possible dynamical evolution of the source tilt with time.

When compared with UrQMD simulations, the general trends with energy and centrality are reproduced, although the model misses the magnitude of the tilt except in mid-central collisions. Combined results from STAR, AGS, and HADES show a smooth change of the tilt angle over a wide energy range (1--30~GeV). These findings demonstrate that the fireball remains tilted even at intermediate RHIC energies and highlight the need for fully three-dimensional modeling at these energies. Comparisons with hybrid hydrodynamic--transport models also show that the observed tilt is sensitive to the equation of state of the matter created in the collision.

In addition to the tilt measurements, we have extracted the final eccentricity of the homogeneity regions across all investigated collision energies. The results show a clear and monotonic behavior: the freeze-out eccentricity is largest at the lowest $\sqrt{s_{\mathrm{NN}}}$ and gradually decreases with increasing beam energy. This trend reflects the reduced lifetime and weaker collective expansion at lower energies, where the system has insufficient time to transform the initially almond-shaped overlap geometry into a more azimuthally symmetric configuration. At higher energies, stronger flow and longer evolution naturally drive the source toward a rounder shape at freeze-out. The observed energy dependence is consistent with expectations from hydrodynamic evolution and previous world measurements, and the new STAR data help to constrain the dynamic response of the medium to its initial
geometric anisotropy.

\begin{acknowledgments}
We thank the RHIC Operations Group and SDCC at BNL, the NERSC Center at LBNL, and the Open Science Grid consortium for providing resources and support.  This work was supported in part by the Office of Nuclear Physics within the U.S. DOE Office of Science, the U.S. National Science Foundation, National Natural Science Foundation of China, Chinese Academy of Science, the Ministry of Science and Technology of China and the Chinese Ministry of Education, NSTC Taipei, the National Research Foundation of Korea, Czech Science Foundation and Ministry of Education, Youth and Sports of the Czech Republic, Hungarian National Research, Development and Innovation Office, New National Excellency Programme of the Hungarian Ministry of Human Capacities, Department of Atomic Energy and Department of Science and Technology of the Government of India, the National Science Centre and WUT ID-UB of Poland, German Bundesministerium f\"ur Bildung, Wissenschaft, Forschung and Technologie (BMBF), Helmholtz Association, Ministry of Education, Culture, Sports, Science, and Technology (MEXT), Japan Society for the Promotion of Science (JSPS), and Agencia Nacional de Investigacion y Desarrollo de Chile (ANID), Chile.
\end{acknowledgments}

%\clearpage
\appendix
\section{Femtoscopic parameters extracted from fits to the three-dimensional correlation functions}
\label{app:tables}

\begin{table*}[htbp]
\centering
\setlength{\tabcolsep}{4pt}
\renewcommand{\arraystretch}{1.15}
\begin{tabular}{|c|c|c|c|c|c|c|c|c|}
\hline
$\langle\phi_{k}\rangle$ & $\lambda$ & $R^{2}_{\mathrm{out}}\,(\mathrm{fm}^2)$ & $R^{2}_{\mathrm{side}}\,(\mathrm{fm}^2)$ & $R^{2}_{\mathrm{long}}\,(\mathrm{fm}^2)$ & $R^{2}_{\mathrm{out-side}}\,(\mathrm{fm}^2)$ & $R^{2}_{\mathrm{out-long}}\,(\mathrm{fm}^2)$ & $R^{2}_{\mathrm{side-long}}\,(\mathrm{fm}^2)$ & $\chi^{2}/\mathrm{NDF}$ \\
\hline
$0^{\circ}$ & $0.56 \pm 0.00$ & $32.96 \pm 0.39$ & $24.79 \pm 0.23$ & $28.99 \pm 0.31$ & $-0.00 \pm 0.21$ & $0.84 \pm 0.24$ & $0.04 \pm 0.18$ & $1.03$ \\
$45^{\circ}$ & $0.56 \pm 0.00$ & $32.91 \pm 0.38$ & $24.88 \pm 0.24$ & $28.67 \pm 0.31$ & $0.36 \pm 0.21$ & $0.46 \pm 0.24$ & $-0.56 \pm 0.18$ & $1.02$ \\
$90^{\circ}$ & $0.55 \pm 0.00$ & $32.27 \pm 0.39$ & $24.38 \pm 0.23$ & $28.86 \pm 0.32$ & $-0.00 \pm 0.21$ & $0.29 \pm 0.24$ & $-1.02 \pm 0.18$ & $1.02$ \\
$135^{\circ}$ & $0.55 \pm 0.00$ & $31.89 \pm 0.38$ & $24.96 \pm 0.24$ & $28.22 \pm 0.31$ & $-0.20 \pm 0.21$ & $-0.90 \pm 0.24$ & $-0.49 \pm 0.18$ & $1.03$ \\
$180^{\circ}$ & $0.54 \pm 0.00$ & $31.10 \pm 0.36$ & $24.57 \pm 0.23$ & $28.28 \pm 0.31$ & $0.19 \pm 0.20$ & $-1.10 \pm 0.24$ & $0.04 \pm 0.18$ & $1.03$ \\
$225^{\circ}$ & $0.56 \pm 0.00$ & $33.14 \pm 0.40$ & $24.80 \pm 0.24$ & $29.30 \pm 0.32$ & $0.46 \pm 0.21$ & $-0.95 \pm 0.25$ & $0.79 \pm 0.18$ & $1.03$ \\
$270^{\circ}$ & $0.55 \pm 0.00$ & $32.79 \pm 0.39$ & $24.35 \pm 0.23$ & $28.81 \pm 0.32$ & $0.14 \pm 0.21$ & $0.04 \pm 0.24$ & $1.52 \pm 0.18$ & $1.03$ \\
$315^{\circ}$ & $0.56 \pm 0.00$ & $32.88 \pm 0.39$ & $25.20 \pm 0.24$ & $28.15 \pm 0.30$ & $-0.33 \pm 0.21$ & $0.74 \pm 0.24$ & $0.84 \pm 0.18$ & $1.03$ \\
\hline
\end{tabular}
\caption{Extracted femtoscopic parameters from fits to the three-dimensional correlation functions obtained using Eq.~\ref{eq:expCF}, for all eight azimuthal angles of pion pairs with transverse pair momentum $k_{T} \in [0.15, 0.20]~\mathrm{GeV}/c$, in 0--10\% central collisions. The fit was performed using Eq.~\ref{eq:cfmodified}.}
\label{tab:fit_params_charge0_cent3_kt0}
\end{table*}

\begin{table*}[htbp]
\centering
\setlength{\tabcolsep}{4pt}
\renewcommand{\arraystretch}{1.15}
\begin{tabular}{|c|c|c|c|c|c|c|c|c|}
\hline
$\langle\phi_{k}\rangle$ & $\lambda$ & $R^{2}_{\mathrm{out}}\,(\mathrm{fm}^2)$ & $R^{2}_{\mathrm{side}}\,(\mathrm{fm}^2)$ & $R^{2}_{\mathrm{long}}\,(\mathrm{fm}^2)$ & $R^{2}_{\mathrm{out-side}}\,(\mathrm{fm}^2)$ & $R^{2}_{\mathrm{out-long}}\,(\mathrm{fm}^2)$ & $R^{2}_{\mathrm{side-long}}\,(\mathrm{fm}^2)$ & $\chi^{2}/\mathrm{NDF}$ \\
\hline
$0^{\circ}$ & $0.54 \pm 0.00$ & $29.79 \pm 0.23$ & $22.49 \pm 0.17$ & $23.10 \pm 0.20$ & $0.20 \pm 0.14$ & $0.50 \pm 0.15$ & $-0.02 \pm 0.12$ & $1.02$ \\
$45^{\circ}$ & $0.53 \pm 0.00$ & $30.16 \pm 0.24$ & $22.28 \pm 0.17$ & $23.12 \pm 0.20$ & $0.13 \pm 0.14$ & $0.37 \pm 0.15$ & $-0.50 \pm 0.12$ & $1.02$ \\
$90^{\circ}$ & $0.53 \pm 0.00$ & $30.83 \pm 0.25$ & $21.66 \pm 0.17$ & $22.68 \pm 0.20$ & $0.22 \pm 0.14$ & $0.02 \pm 0.15$ & $-0.87 \pm 0.12$ & $1.02$ \\
$135^{\circ}$ & $0.54 \pm 0.00$ & $30.22 \pm 0.24$ & $22.10 \pm 0.17$ & $22.58 \pm 0.19$ & $0.02 \pm 0.14$ & $-0.31 \pm 0.15$ & $-0.50 \pm 0.12$ & $1.02$ \\
$180^{\circ}$ & $0.53 \pm 0.00$ & $29.80 \pm 0.24$ & $22.27 \pm 0.17$ & $22.76 \pm 0.20$ & $0.01 \pm 0.14$ & $-0.54 \pm 0.15$ & $-0.07 \pm 0.12$ & $1.02$ \\
$225^{\circ}$ & $0.53 \pm 0.00$ & $29.79 \pm 0.23$ & $22.22 \pm 0.17$ & $22.87 \pm 0.20$ & $0.58 \pm 0.14$ & $-0.63 \pm 0.15$ & $0.47 \pm 0.12$ & $1.02$ \\
$270^{\circ}$ & $0.54 \pm 0.00$ & $30.61 \pm 0.24$ & $21.80 \pm 0.17$ & $23.10 \pm 0.20$ & $0.15 \pm 0.14$ & $-0.36 \pm 0.15$ & $0.84 \pm 0.12$ & $1.01$ \\
$315^{\circ}$ & $0.53 \pm 0.00$ & $30.02 \pm 0.24$ & $21.89 \pm 0.17$ & $22.81 \pm 0.20$ & $-0.16 \pm 0.14$ & $0.49 \pm 0.15$ & $0.67 \pm 0.12$ & $1.02$ \\
\hline
\end{tabular}
\caption{Extracted femtoscopic parameters from fits to the three-dimensional correlation functions obtained using Eq.~\ref{eq:expCF}, for all eight azimuthal angles of pion pairs with transverse pair momentum $k_{T} \in [0.20, 0.25]~\mathrm{GeV}/c$, in 0--10\% central collisions. The fit was performed using Eq.~\ref{eq:cfmodified}.}
\label{tab:fit_params_charge0_cent3_kt1}
\end{table*}

\begin{table*}[htbp]
\centering
\setlength{\tabcolsep}{4pt}
\renewcommand{\arraystretch}{1.15}
\begin{tabular}{|c|c|c|c|c|c|c|c|c|}
\hline
$\langle\phi_{k}\rangle$ & $\lambda$ & $R^{2}_{\mathrm{out}}\,(\mathrm{fm}^2)$ & $R^{2}_{\mathrm{side}}\,(\mathrm{fm}^2)$ & $R^{2}_{\mathrm{long}}\,(\mathrm{fm}^2)$ & $R^{2}_{\mathrm{out-side}}\,(\mathrm{fm}^2)$ & $R^{2}_{\mathrm{out-long}}\,(\mathrm{fm}^2)$ & $R^{2}_{\mathrm{side-long}}\,(\mathrm{fm}^2)$ & $\chi^{2}/\mathrm{NDF}$ \\
\hline
$0^{\circ}$ & $0.54 \pm 0.00$ & $25.96 \pm 0.22$ & $20.13 \pm 0.17$ & $18.28 \pm 0.17$ & $0.25 \pm 0.13$ & $0.57 \pm 0.13$ & $-0.10 \pm 0.11$ & $1.02$ \\
$45^{\circ}$ & $0.54 \pm 0.00$ & $26.08 \pm 0.22$ & $19.67 \pm 0.16$ & $17.92 \pm 0.17$ & $0.36 \pm 0.13$ & $0.37 \pm 0.13$ & $-0.28 \pm 0.10$ & $1.02$ \\
$90^{\circ}$ & $0.53 \pm 0.00$ & $26.32 \pm 0.23$ & $19.26 \pm 0.16$ & $18.01 \pm 0.17$ & $0.20 \pm 0.13$ & $-0.01 \pm 0.13$ & $-0.80 \pm 0.11$ & $1.02$ \\
$135^{\circ}$ & $0.54 \pm 0.00$ & $26.06 \pm 0.22$ & $19.69 \pm 0.16$ & $18.54 \pm 0.17$ & $-0.32 \pm 0.13$ & $-0.47 \pm 0.13$ & $-0.32 \pm 0.11$ & $1.02$ \\
$180^{\circ}$ & $0.54 \pm 0.00$ & $25.81 \pm 0.22$ & $20.09 \pm 0.17$ & $18.32 \pm 0.17$ & $0.06 \pm 0.13$ & $-0.69 \pm 0.13$ & $-0.02 \pm 0.11$ & $1.01$ \\
$225^{\circ}$ & $0.54 \pm 0.00$ & $26.20 \pm 0.22$ & $19.60 \pm 0.16$ & $18.27 \pm 0.17$ & $0.24 \pm 0.13$ & $-0.16 \pm 0.13$ & $0.49 \pm 0.11$ & $1.01$ \\
$270^{\circ}$ & $0.54 \pm 0.00$ & $26.66 \pm 0.23$ & $19.45 \pm 0.16$ & $18.27 \pm 0.17$ & $-0.18 \pm 0.13$ & $0.04 \pm 0.13$ & $0.52 \pm 0.10$ & $1.02$ \\
$315^{\circ}$ & $0.54 \pm 0.00$ & $25.86 \pm 0.22$ & $19.81 \pm 0.16$ & $18.34 \pm 0.17$ & $-0.05 \pm 0.13$ & $0.20 \pm 0.13$ & $0.38 \pm 0.11$ & $1.02$ \\
\hline
\end{tabular}
\caption{Extracted femtoscopic parameters from fits to the three-dimensional correlation functions obtained using Eq.~\ref{eq:expCF}, for all eight azimuthal angles of pion pairs with transverse pair momentum $k_{T} \in [0.25, 0.30]~\mathrm{GeV}/c$, in 0--10\% central collisions. The fit was performed using Eq.~\ref{eq:cfmodified}.}
\label{tab:fit_params_charge0_cent3_kt2}
\end{table*}

\begin{table*}[htbp]
\centering
\setlength{\tabcolsep}{4pt}
\renewcommand{\arraystretch}{1.15}
\begin{tabular}{|c|c|c|c|c|c|c|c|c|}
\hline
$\langle\phi_{k}\rangle$ & $\lambda$ & $R^{2}_{\mathrm{out}}\,(\mathrm{fm}^2)$ & $R^{2}_{\mathrm{side}}\,(\mathrm{fm}^2)$ & $R^{2}_{\mathrm{long}}\,(\mathrm{fm}^2)$ & $R^{2}_{\mathrm{out-side}}\,(\mathrm{fm}^2)$ & $R^{2}_{\mathrm{out-long}}\,(\mathrm{fm}^2)$ & $R^{2}_{\mathrm{side-long}}\,(\mathrm{fm}^2)$ & $\chi^{2}/\mathrm{NDF}$ \\
\hline
$0^{\circ}$ & $0.56 \pm 0.00$ & $22.50 \pm 0.22$ & $18.06 \pm 0.17$ & $15.19 \pm 0.16$ & $0.27 \pm 0.13$ & $0.32 \pm 0.13$ & $0.19 \pm 0.10$ & $1.02$ \\
$45^{\circ}$ & $0.56 \pm 0.00$ & $22.69 \pm 0.22$ & $18.13 \pm 0.17$ & $14.92 \pm 0.16$ & $0.32 \pm 0.13$ & $0.07 \pm 0.13$ & $-0.33 \pm 0.10$ & $1.02$ \\
$90^{\circ}$ & $0.56 \pm 0.00$ & $23.00 \pm 0.23$ & $17.86 \pm 0.17$ & $14.92 \pm 0.16$ & $0.06 \pm 0.13$ & $0.14 \pm 0.13$ & $-0.45 \pm 0.10$ & $1.02$ \\
$135^{\circ}$ & $0.56 \pm 0.00$ & $22.59 \pm 0.22$ & $18.04 \pm 0.17$ & $14.99 \pm 0.16$ & $-0.20 \pm 0.13$ & $-0.44 \pm 0.13$ & $-0.42 \pm 0.10$ & $1.01$ \\
$180^{\circ}$ & $0.56 \pm 0.00$ & $21.91 \pm 0.21$ & $18.20 \pm 0.17$ & $15.01 \pm 0.16$ & $-0.08 \pm 0.13$ & $-0.29 \pm 0.12$ & $-0.05 \pm 0.10$ & $1.01$ \\
$225^{\circ}$ & $0.57 \pm 0.00$ & $22.98 \pm 0.22$ & $18.01 \pm 0.17$ & $15.40 \pm 0.16$ & $0.43 \pm 0.13$ & $-0.35 \pm 0.13$ & $0.33 \pm 0.10$ & $1.01$ \\
$270^{\circ}$ & $0.56 \pm 0.00$ & $22.80 \pm 0.22$ & $17.42 \pm 0.16$ & $15.01 \pm 0.16$ & $-0.20 \pm 0.13$ & $0.04 \pm 0.13$ & $0.50 \pm 0.10$ & $1.01$ \\
$315^{\circ}$ & $0.56 \pm 0.00$ & $22.40 \pm 0.22$ & $17.84 \pm 0.17$ & $15.03 \pm 0.16$ & $-0.12 \pm 0.13$ & $0.25 \pm 0.12$ & $0.41 \pm 0.10$ & $1.01$ \\
\hline
\end{tabular}
\caption{Extracted femtoscopic parameters from fits to the three-dimensional correlation functions obtained using Eq.~\ref{eq:expCF}, for all eight azimuthal angles of pion pairs with transverse pair momentum $k_{T} \in [0.30, 0.35]~\mathrm{GeV}/c$, in 0--10\% central collisions. The fit was performed using Eq.~\ref{eq:cfmodified}.}
\label{tab:fit_params_charge0_cent3_kt3}
\end{table*}

\begin{table*}[htbp]
\centering
\setlength{\tabcolsep}{4pt}
\renewcommand{\arraystretch}{1.15}
\begin{tabular}{|c|c|c|c|c|c|c|c|c|}
\hline
$\langle\phi_{k}\rangle$ & $\lambda$ & $R^{2}_{\mathrm{out}}\,(\mathrm{fm}^2)$ & $R^{2}_{\mathrm{side}}\,(\mathrm{fm}^2)$ & $R^{2}_{\mathrm{long}}\,(\mathrm{fm}^2)$ & $R^{2}_{\mathrm{out-side}}\,(\mathrm{fm}^2)$ & $R^{2}_{\mathrm{out-long}}\,(\mathrm{fm}^2)$ & $R^{2}_{\mathrm{side-long}}\,(\mathrm{fm}^2)$ & $\chi^{2}/\mathrm{NDF}$ \\
\hline
$0^{\circ}$ & $0.57 \pm 0.00$ & $19.31 \pm 0.17$ & $15.84 \pm 0.14$ & $11.43 \pm 0.11$ & $0.10 \pm 0.10$ & $0.17 \pm 0.09$ & $-0.02 \pm 0.08$ & $1.01$ \\
$45^{\circ}$ & $0.56 \pm 0.00$ & $19.09 \pm 0.18$ & $15.61 \pm 0.14$ & $11.36 \pm 0.11$ & $0.38 \pm 0.10$ & $0.23 \pm 0.09$ & $-0.20 \pm 0.08$ & $1.01$ \\
$90^{\circ}$ & $0.57 \pm 0.00$ & $19.34 \pm 0.18$ & $15.73 \pm 0.14$ & $11.54 \pm 0.11$ & $0.07 \pm 0.10$ & $-0.01 \pm 0.09$ & $-0.31 \pm 0.08$ & $1.02$ \\
$135^{\circ}$ & $0.57 \pm 0.00$ & $18.95 \pm 0.18$ & $15.79 \pm 0.14$ & $11.75 \pm 0.12$ & $0.01 \pm 0.10$ & $-0.24 \pm 0.09$ & $-0.36 \pm 0.08$ & $1.01$ \\
$180^{\circ}$ & $0.57 \pm 0.00$ & $19.11 \pm 0.18$ & $15.70 \pm 0.14$ & $11.64 \pm 0.11$ & $0.11 \pm 0.10$ & $-0.32 \pm 0.09$ & $-0.02 \pm 0.08$ & $1.01$ \\
$225^{\circ}$ & $0.57 \pm 0.00$ & $19.37 \pm 0.18$ & $15.85 \pm 0.14$ & $11.71 \pm 0.11$ & $0.26 \pm 0.10$ & $-0.21 \pm 0.09$ & $0.31 \pm 0.08$ & $1.01$ \\
$270^{\circ}$ & $0.58 \pm 0.00$ & $19.47 \pm 0.18$ & $15.46 \pm 0.14$ & $11.97 \pm 0.12$ & $-0.13 \pm 0.10$ & $-0.02 \pm 0.10$ & $0.36 \pm 0.08$ & $1.01$ \\
$315^{\circ}$ & $0.57 \pm 0.00$ & $19.46 \pm 0.18$ & $15.55 \pm 0.14$ & $11.55 \pm 0.11$ & $-0.17 \pm 0.10$ & $0.23 \pm 0.09$ & $0.18 \pm 0.08$ & $1.01$ \\
\hline
\end{tabular}
\caption{Extracted femtoscopic parameters from fits to the three-dimensional correlation functions obtained using Eq.~\ref{eq:expCF}, for all eight azimuthal angles of pion pairs with transverse pair momentum $k_{T} \in [0.35, 0.45]~\mathrm{GeV}/c$, in 0--10\% central collisions. The fit was performed using Eq.~\ref{eq:cfmodified}.}
\label{tab:fit_params_charge0_cent3_kt4}
\end{table*}

\begin{table*}[htbp]
\centering
\setlength{\tabcolsep}{4pt}
\renewcommand{\arraystretch}{1.15}
\begin{tabular}{|c|c|c|c|c|c|c|c|c|}
\hline
$\langle\phi_{k}\rangle$ & $\lambda$ & $R^{2}_{\mathrm{out}}\,(\mathrm{fm}^2)$ & $R^{2}_{\mathrm{side}}\,(\mathrm{fm}^2)$ & $R^{2}_{\mathrm{long}}\,(\mathrm{fm}^2)$ & $R^{2}_{\mathrm{out-side}}\,(\mathrm{fm}^2)$ & $R^{2}_{\mathrm{out-long}}\,(\mathrm{fm}^2)$ & $R^{2}_{\mathrm{side-long}}\,(\mathrm{fm}^2)$ & $\chi^{2}/\mathrm{NDF}$ \\
\hline
$0^{\circ}$ & $0.60 \pm 0.01$ & $13.93 \pm 0.18$ & $13.07 \pm 0.15$ & $7.90 \pm 0.10$ & $-0.08 \pm 0.11$ & $0.10 \pm 0.09$ & $-0.10 \pm 0.08$ & $1.01$ \\
$45^{\circ}$ & $0.61 \pm 0.01$ & $14.33 \pm 0.19$ & $12.82 \pm 0.15$ & $7.77 \pm 0.10$ & $0.31 \pm 0.11$ & $0.17 \pm 0.09$ & $-0.33 \pm 0.07$ & $1.01$ \\
$90^{\circ}$ & $0.60 \pm 0.01$ & $14.53 \pm 0.19$ & $12.87 \pm 0.15$ & $7.73 \pm 0.10$ & $-0.14 \pm 0.11$ & $-0.04 \pm 0.09$ & $-0.31 \pm 0.08$ & $1.01$ \\
$135^{\circ}$ & $0.61 \pm 0.01$ & $14.44 \pm 0.19$ & $12.93 \pm 0.15$ & $7.78 \pm 0.10$ & $-0.25 \pm 0.11$ & $-0.04 \pm 0.09$ & $-0.14 \pm 0.08$ & $1.01$ \\
$180^{\circ}$ & $0.61 \pm 0.01$ & $14.71 \pm 0.19$ & $13.07 \pm 0.15$ & $7.92 \pm 0.11$ & $-0.01 \pm 0.11$ & $-0.22 \pm 0.09$ & $0.03 \pm 0.08$ & $1.01$ \\
$225^{\circ}$ & $0.60 \pm 0.01$ & $14.76 \pm 0.19$ & $12.87 \pm 0.15$ & $7.79 \pm 0.11$ & $0.08 \pm 0.11$ & $-0.17 \pm 0.09$ & $0.24 \pm 0.08$ & $1.02$ \\
$270^{\circ}$ & $0.62 \pm 0.01$ & $14.90 \pm 0.20$ & $12.93 \pm 0.15$ & $7.92 \pm 0.10$ & $-0.09 \pm 0.11$ & $-0.06 \pm 0.09$ & $0.35 \pm 0.08$ & $1.01$ \\
$315^{\circ}$ & $0.60 \pm 0.01$ & $14.47 \pm 0.19$ & $12.90 \pm 0.15$ & $7.81 \pm 0.10$ & $0.02 \pm 0.11$ & $0.20 \pm 0.09$ & $0.13 \pm 0.08$ & $1.01$ \\
\hline
\end{tabular}
\caption{Extracted femtoscopic parameters from fits to the three-dimensional correlation functions obtained using Eq.~\ref{eq:expCF}, for all eight azimuthal angles of pion pairs with transverse pair momentum $k_{T} \in [0.45, 0.80]~\mathrm{GeV}/c$, in 0--10\% central collisions. The fit was performed using Eq.~\ref{eq:cfmodified}.}
\label{tab:fit_params_charge0_cent3_kt5}
\end{table*}

\begin{table*}[htbp]
\centering
\setlength{\tabcolsep}{4pt}
\renewcommand{\arraystretch}{1.15}
\begin{tabular}{|c|c|c|c|c|c|c|c|c|}
\hline
$\langle\phi_{k}\rangle$ & $\lambda$ & $R^{2}_{\mathrm{out}}\,(\mathrm{fm}^2)$ & $R^{2}_{\mathrm{side}}\,(\mathrm{fm}^2)$ & $R^{2}_{\mathrm{long}}\,(\mathrm{fm}^2)$ & $R^{2}_{\mathrm{out-side}}\,(\mathrm{fm}^2)$ & $R^{2}_{\mathrm{out-long}}\,(\mathrm{fm}^2)$ & $R^{2}_{\mathrm{side-long}}\,(\mathrm{fm}^2)$ & $\chi^{2}/\mathrm{NDF}$ \\
\hline
$0^{\circ}$ & $0.57 \pm 0.00$ & $21.34 \pm 0.26$ & $18.14 \pm 0.16$ & $22.50 \pm 0.24$ & $0.18 \pm 0.14$ & $1.37 \pm 0.18$ & $-0.11 \pm 0.13$ & $1.02$ \\
$45^{\circ}$ & $0.56 \pm 0.00$ & $22.30 \pm 0.27$ & $16.79 \pm 0.15$ & $22.02 \pm 0.23$ & $1.23 \pm 0.14$ & $1.08 \pm 0.18$ & $-1.40 \pm 0.12$ & $1.02$ \\
$90^{\circ}$ & $0.56 \pm 0.00$ & $23.76 \pm 0.29$ & $16.05 \pm 0.15$ & $22.18 \pm 0.23$ & $0.14 \pm 0.14$ & $-0.33 \pm 0.18$ & $-1.79 \pm 0.12$ & $1.03$ \\
$135^{\circ}$ & $0.57 \pm 0.00$ & $22.71 \pm 0.28$ & $17.24 \pm 0.16$ & $22.45 \pm 0.24$ & $-0.95 \pm 0.14$ & $-1.16 \pm 0.18$ & $-1.12 \pm 0.13$ & $1.03$ \\
$180^{\circ}$ & $0.57 \pm 0.00$ & $20.99 \pm 0.26$ & $18.06 \pm 0.16$ & $22.22 \pm 0.24$ & $0.15 \pm 0.14$ & $-1.38 \pm 0.17$ & $0.28 \pm 0.13$ & $1.03$ \\
$225^{\circ}$ & $0.56 \pm 0.00$ & $22.05 \pm 0.28$ & $17.11 \pm 0.16$ & $22.13 \pm 0.23$ & $0.76 \pm 0.14$ & $-1.15 \pm 0.18$ & $1.32 \pm 0.12$ & $1.02$ \\
$270^{\circ}$ & $0.57 \pm 0.00$ & $23.43 \pm 0.28$ & $16.50 \pm 0.15$ & $22.44 \pm 0.24$ & $0.18 \pm 0.14$ & $0.20 \pm 0.18$ & $1.70 \pm 0.12$ & $1.02$ \\
$315^{\circ}$ & $0.56 \pm 0.00$ & $22.42 \pm 0.28$ & $17.03 \pm 0.16$ & $22.22 \pm 0.23$ & $-0.94 \pm 0.14$ & $1.18 \pm 0.18$ & $1.16 \pm 0.12$ & $1.03$ \\
\hline
\end{tabular}
\caption{Extracted femtoscopic parameters from fits to the three-dimensional correlation functions obtained using Eq.~\ref{eq:expCF}, for all eight azimuthal angles of pion pairs with transverse pair momentum $k_{T} \in [0.15, 0.20]~\mathrm{GeV}/c$, in 10--30\% central collisions. The fit was performed using Eq.~\ref{eq:cfmodified}.}
\label{tab:fit_params_charge0_cent2_kt0}
\end{table*}

\begin{table*}[htbp]
\centering
\setlength{\tabcolsep}{4pt}
\renewcommand{\arraystretch}{1.15}
\begin{tabular}{|c|c|c|c|c|c|c|c|c|}
\hline
$\langle\phi_{k}\rangle$ & $\lambda$ & $R^{2}_{\mathrm{out}}\,(\mathrm{fm}^2)$ & $R^{2}_{\mathrm{side}}\,(\mathrm{fm}^2)$ & $R^{2}_{\mathrm{long}}\,(\mathrm{fm}^2)$ & $R^{2}_{\mathrm{out-side}}\,(\mathrm{fm}^2)$ & $R^{2}_{\mathrm{out-long}}\,(\mathrm{fm}^2)$ & $R^{2}_{\mathrm{side-long}}\,(\mathrm{fm}^2)$ & $\chi^{2}/\mathrm{NDF}$ \\
\hline
$0^{\circ}$ & $0.54 \pm 0.00$ & $19.91 \pm 0.16$ & $16.52 \pm 0.12$ & $17.59 \pm 0.14$ & $0.10 \pm 0.09$ & $0.98 \pm 0.10$ & $0.04 \pm 0.09$ & $1.02$ \\
$45^{\circ}$ & $0.54 \pm 0.00$ & $21.55 \pm 0.17$ & $15.75 \pm 0.12$ & $17.90 \pm 0.15$ & $1.24 \pm 0.10$ & $0.73 \pm 0.11$ & $-0.96 \pm 0.09$ & $1.02$ \\
$90^{\circ}$ & $0.54 \pm 0.00$ & $22.59 \pm 0.18$ & $14.88 \pm 0.12$ & $18.27 \pm 0.15$ & $-0.13 \pm 0.10$ & $0.04 \pm 0.11$ & $-1.38 \pm 0.09$ & $1.01$ \\
$135^{\circ}$ & $0.54 \pm 0.00$ & $21.31 \pm 0.17$ & $16.04 \pm 0.12$ & $18.07 \pm 0.15$ & $-1.01 \pm 0.10$ & $-0.68 \pm 0.11$ & $-0.81 \pm 0.09$ & $1.02$ \\
$180^{\circ}$ & $0.53 \pm 0.00$ & $20.05 \pm 0.16$ & $16.49 \pm 0.12$ & $17.65 \pm 0.15$ & $0.07 \pm 0.10$ & $-1.08 \pm 0.11$ & $-0.00 \pm 0.09$ & $1.02$ \\
$225^{\circ}$ & $0.54 \pm 0.00$ & $21.37 \pm 0.17$ & $15.74 \pm 0.12$ & $18.25 \pm 0.15$ & $1.05 \pm 0.10$ & $-0.94 \pm 0.11$ & $0.94 \pm 0.09$ & $1.02$ \\
$270^{\circ}$ & $0.54 \pm 0.00$ & $22.62 \pm 0.18$ & $14.92 \pm 0.12$ & $18.37 \pm 0.16$ & $0.06 \pm 0.10$ & $-0.05 \pm 0.12$ & $1.42 \pm 0.09$ & $1.02$ \\
$315^{\circ}$ & $0.54 \pm 0.00$ & $21.38 \pm 0.17$ & $15.81 \pm 0.12$ & $18.10 \pm 0.15$ & $-0.85 \pm 0.10$ & $0.76 \pm 0.11$ & $0.86 \pm 0.09$ & $1.02$ \\
\hline
\end{tabular}
\caption{Extracted femtoscopic parameters from fits to the three-dimensional correlation functions obtained using Eq.~\ref{eq:expCF}, for all eight azimuthal angles of pion pairs with transverse pair momentum $k_{T} \in [0.20, 0.25]~\mathrm{GeV}/c$, in 10--30\% central collisions. The fit was performed using Eq.~\ref{eq:cfmodified}.}
\label{tab:fit_params_charge0_cent2_kt1}
\end{table*}

\begin{table*}[htbp]
\centering
\setlength{\tabcolsep}{4pt}
\renewcommand{\arraystretch}{1.15}
\begin{tabular}{|c|c|c|c|c|c|c|c|c|}
\hline
$\langle\phi_{k}\rangle$ & $\lambda$ & $R^{2}_{\mathrm{out}}\,(\mathrm{fm}^2)$ & $R^{2}_{\mathrm{side}}\,(\mathrm{fm}^2)$ & $R^{2}_{\mathrm{long}}\,(\mathrm{fm}^2)$ & $R^{2}_{\mathrm{out-side}}\,(\mathrm{fm}^2)$ & $R^{2}_{\mathrm{out-long}}\,(\mathrm{fm}^2)$ & $R^{2}_{\mathrm{side-long}}\,(\mathrm{fm}^2)$ & $\chi^{2}/\mathrm{NDF}$ \\
\hline
$0^{\circ}$ & $0.54 \pm 0.00$ & $17.64 \pm 0.15$ & $14.89 \pm 0.12$ & $14.02 \pm 0.12$ & $0.08 \pm 0.09$ & $0.79 \pm 0.09$ & $-0.01 \pm 0.08$ & $1.02$ \\
$45^{\circ}$ & $0.55 \pm 0.00$ & $18.70 \pm 0.16$ & $14.24 \pm 0.12$ & $14.27 \pm 0.13$ & $1.09 \pm 0.09$ & $0.40 \pm 0.10$ & $-0.77 \pm 0.08$ & $1.01$ \\
$90^{\circ}$ & $0.56 \pm 0.00$ & $20.45 \pm 0.16$ & $13.34 \pm 0.11$ & $14.56 \pm 0.13$ & $0.07 \pm 0.09$ & $-0.06 \pm 0.10$ & $-0.98 \pm 0.08$ & $1.02$ \\
$135^{\circ}$ & $0.55 \pm 0.00$ & $18.95 \pm 0.15$ & $14.22 \pm 0.12$ & $14.03 \pm 0.13$ & $-1.10 \pm 0.09$ & $-0.32 \pm 0.10$ & $-0.70 \pm 0.08$ & $1.02$ \\
$180^{\circ}$ & $0.55 \pm 0.00$ & $17.92 \pm 0.15$ & $14.96 \pm 0.12$ & $14.09 \pm 0.13$ & $0.27 \pm 0.09$ & $-0.68 \pm 0.09$ & $0.08 \pm 0.08$ & $1.02$ \\
$225^{\circ}$ & $0.55 \pm 0.00$ & $19.09 \pm 0.16$ & $14.26 \pm 0.12$ & $14.13 \pm 0.13$ & $1.18 \pm 0.09$ & $-0.68 \pm 0.10$ & $0.72 \pm 0.08$ & $1.02$ \\
$270^{\circ}$ & $0.55 \pm 0.00$ & $20.53 \pm 0.17$ & $13.31 \pm 0.11$ & $14.51 \pm 0.13$ & $-0.07 \pm 0.09$ & $-0.12 \pm 0.10$ & $0.88 \pm 0.08$ & $1.02$ \\
$315^{\circ}$ & $0.55 \pm 0.00$ & $19.02 \pm 0.16$ & $14.37 \pm 0.12$ & $14.22 \pm 0.13$ & $-0.86 \pm 0.09$ & $0.73 \pm 0.10$ & $0.46 \pm 0.08$ & $1.01$ \\
\hline
\end{tabular}
\caption{Extracted femtoscopic parameters from fits to the three-dimensional correlation functions obtained using Eq.~\ref{eq:expCF}, for all eight azimuthal angles of pion pairs with transverse pair momentum $k_{T} \in [0.25, 0.30]~\mathrm{GeV}/c$, in 10--30\% central collisions. The fit was performed using Eq.~\ref{eq:cfmodified}.}
\label{tab:fit_params_charge0_cent2_kt2}
\end{table*}

\begin{table*}[htbp]
\centering
\setlength{\tabcolsep}{4pt}
\renewcommand{\arraystretch}{1.15}
\begin{tabular}{|c|c|c|c|c|c|c|c|c|}
\hline
$\langle\phi_{k}\rangle$ & $\lambda$ & $R^{2}_{\mathrm{out}}\,(\mathrm{fm}^2)$ & $R^{2}_{\mathrm{side}}\,(\mathrm{fm}^2)$ & $R^{2}_{\mathrm{long}}\,(\mathrm{fm}^2)$ & $R^{2}_{\mathrm{out-side}}\,(\mathrm{fm}^2)$ & $R^{2}_{\mathrm{out-long}}\,(\mathrm{fm}^2)$ & $R^{2}_{\mathrm{side-long}}\,(\mathrm{fm}^2)$ & $\chi^{2}/\mathrm{NDF}$ \\
\hline
$0^{\circ}$ & $0.56 \pm 0.00$ & $15.83 \pm 0.15$ & $13.81 \pm 0.13$ & $11.52 \pm 0.12$ & $0.07 \pm 0.09$ & $0.41 \pm 0.09$ & $-0.01 \pm 0.08$ & $1.02$ \\
$45^{\circ}$ & $0.57 \pm 0.00$ & $16.39 \pm 0.16$ & $12.73 \pm 0.12$ & $11.93 \pm 0.12$ & $0.83 \pm 0.09$ & $0.24 \pm 0.09$ & $-0.65 \pm 0.08$ & $1.01$ \\
$90^{\circ}$ & $0.57 \pm 0.00$ & $17.62 \pm 0.17$ & $12.15 \pm 0.12$ & $11.74 \pm 0.12$ & $0.03 \pm 0.09$ & $0.01 \pm 0.10$ & $-0.57 \pm 0.08$ & $1.01$ \\
$135^{\circ}$ & $0.56 \pm 0.00$ & $16.38 \pm 0.16$ & $12.94 \pm 0.12$ & $11.46 \pm 0.12$ & $-0.83 \pm 0.09$ & $-0.12 \pm 0.09$ & $-0.39 \pm 0.08$ & $1.02$ \\
$180^{\circ}$ & $0.56 \pm 0.00$ & $15.82 \pm 0.15$ & $13.64 \pm 0.13$ & $11.41 \pm 0.12$ & $0.22 \pm 0.09$ & $-0.51 \pm 0.09$ & $-0.13 \pm 0.08$ & $1.02$ \\
$225^{\circ}$ & $0.57 \pm 0.00$ & $16.82 \pm 0.16$ & $12.71 \pm 0.12$ & $11.74 \pm 0.12$ & $0.81 \pm 0.09$ & $-0.58 \pm 0.10$ & $0.58 \pm 0.08$ & $1.02$ \\
$270^{\circ}$ & $0.56 \pm 0.00$ & $17.35 \pm 0.17$ & $12.19 \pm 0.12$ & $11.49 \pm 0.12$ & $-0.06 \pm 0.10$ & $-0.06 \pm 0.10$ & $0.89 \pm 0.08$ & $1.01$ \\
$315^{\circ}$ & $0.56 \pm 0.00$ & $16.40 \pm 0.16$ & $12.79 \pm 0.12$ & $11.65 \pm 0.12$ & $-1.05 \pm 0.09$ & $0.47 \pm 0.09$ & $0.44 \pm 0.08$ & $1.01$ \\
\hline
\end{tabular}
\caption{Extracted femtoscopic parameters from fits to the three-dimensional correlation functions obtained using Eq.~\ref{eq:expCF}, for all eight azimuthal angles of pion pairs with transverse pair momentum $k_{T} \in [0.30, 0.35]~\mathrm{GeV}/c$, in 10--30\% central collisions. The fit was performed using Eq.~\ref{eq:cfmodified}.}
\label{tab:fit_params_charge0_cent2_kt3}
\end{table*}

\begin{table*}[htbp]
\centering
\setlength{\tabcolsep}{4pt}
\renewcommand{\arraystretch}{1.15}
\begin{tabular}{|c|c|c|c|c|c|c|c|c|}
\hline
$\langle\phi_{k}\rangle$ & $\lambda$ & $R^{2}_{\mathrm{out}}\,(\mathrm{fm}^2)$ & $R^{2}_{\mathrm{side}}\,(\mathrm{fm}^2)$ & $R^{2}_{\mathrm{long}}\,(\mathrm{fm}^2)$ & $R^{2}_{\mathrm{out-side}}\,(\mathrm{fm}^2)$ & $R^{2}_{\mathrm{out-long}}\,(\mathrm{fm}^2)$ & $R^{2}_{\mathrm{side-long}}\,(\mathrm{fm}^2)$ & $\chi^{2}/\mathrm{NDF}$ \\
\hline
$0^{\circ}$ & $0.58 \pm 0.00$ & $13.06 \pm 0.12$ & $11.98 \pm 0.10$ & $8.76 \pm 0.08$ & $0.12 \pm 0.07$ & $0.46 \pm 0.06$ & $0.06 \pm 0.06$ & $1.02$ \\
$45^{\circ}$ & $0.58 \pm 0.00$ & $14.16 \pm 0.13$ & $11.22 \pm 0.10$ & $9.18 \pm 0.09$ & $0.83 \pm 0.07$ & $0.31 \pm 0.07$ & $-0.38 \pm 0.06$ & $1.02$ \\
$90^{\circ}$ & $0.58 \pm 0.00$ & $14.93 \pm 0.14$ & $10.74 \pm 0.10$ & $9.18 \pm 0.09$ & $-0.13 \pm 0.08$ & $0.05 \pm 0.07$ & $-0.53 \pm 0.06$ & $1.01$ \\
$135^{\circ}$ & $0.58 \pm 0.00$ & $13.83 \pm 0.13$ & $11.59 \pm 0.10$ & $8.94 \pm 0.09$ & $-0.79 \pm 0.07$ & $-0.10 \pm 0.07$ & $-0.40 \pm 0.06$ & $1.02$ \\
$180^{\circ}$ & $0.58 \pm 0.00$ & $13.06 \pm 0.12$ & $12.08 \pm 0.10$ & $9.03 \pm 0.09$ & $0.05 \pm 0.07$ & $-0.33 \pm 0.07$ & $-0.02 \pm 0.06$ & $1.01$ \\
$225^{\circ}$ & $0.58 \pm 0.00$ & $14.15 \pm 0.13$ & $11.44 \pm 0.10$ & $8.96 \pm 0.09$ & $0.88 \pm 0.07$ & $-0.31 \pm 0.07$ & $0.45 \pm 0.06$ & $1.02$ \\
$270^{\circ}$ & $0.58 \pm 0.00$ & $14.76 \pm 0.14$ & $10.72 \pm 0.10$ & $9.04 \pm 0.09$ & $-0.05 \pm 0.08$ & $-0.02 \pm 0.07$ & $0.55 \pm 0.06$ & $1.01$ \\
$315^{\circ}$ & $0.58 \pm 0.00$ & $13.99 \pm 0.13$ & $11.65 \pm 0.10$ & $8.95 \pm 0.09$ & $-0.85 \pm 0.07$ & $0.25 \pm 0.07$ & $0.49 \pm 0.06$ & $1.02$ \\
\hline
\end{tabular}
\caption{Extracted femtoscopic parameters from fits to the three-dimensional correlation functions obtained using Eq.~\ref{eq:expCF}, for all eight azimuthal angles of pion pairs with transverse pair momentum $k_{T} \in [0.35, 0.45]~\mathrm{GeV}/c$, in 10--30\% central collisions. The fit was performed using Eq.~\ref{eq:cfmodified}.}
\label{tab:fit_params_charge0_cent2_kt4}
\end{table*}

\begin{table*}[htbp]
\centering
\setlength{\tabcolsep}{4pt}
\renewcommand{\arraystretch}{1.15}
\begin{tabular}{|c|c|c|c|c|c|c|c|c|}
\hline
$\langle\phi_{k}\rangle$ & $\lambda$ & $R^{2}_{\mathrm{out}}\,(\mathrm{fm}^2)$ & $R^{2}_{\mathrm{side}}\,(\mathrm{fm}^2)$ & $R^{2}_{\mathrm{long}}\,(\mathrm{fm}^2)$ & $R^{2}_{\mathrm{out-side}}\,(\mathrm{fm}^2)$ & $R^{2}_{\mathrm{out-long}}\,(\mathrm{fm}^2)$ & $R^{2}_{\mathrm{side-long}}\,(\mathrm{fm}^2)$ & $\chi^{2}/\mathrm{NDF}$ \\
\hline
$0^{\circ}$ & $0.63 \pm 0.01$ & $9.98 \pm 0.12$ & $10.11 \pm 0.11$ & $5.87 \pm 0.08$ & $0.02 \pm 0.08$ & $0.24 \pm 0.06$ & $0.10 \pm 0.06$ & $1.01$ \\
$45^{\circ}$ & $0.64 \pm 0.01$ & $10.91 \pm 0.14$ & $9.66 \pm 0.11$ & $6.17 \pm 0.08$ & $0.79 \pm 0.08$ & $0.17 \pm 0.07$ & $-0.36 \pm 0.06$ & $1.02$ \\
$90^{\circ}$ & $0.63 \pm 0.01$ & $11.89 \pm 0.16$ & $8.74 \pm 0.11$ & $6.06 \pm 0.08$ & $-0.02 \pm 0.08$ & $0.01 \pm 0.07$ & $-0.42 \pm 0.06$ & $1.01$ \\
$135^{\circ}$ & $0.62 \pm 0.01$ & $10.76 \pm 0.14$ & $9.53 \pm 0.11$ & $5.82 \pm 0.08$ & $-0.61 \pm 0.08$ & $-0.19 \pm 0.07$ & $-0.27 \pm 0.06$ & $1.02$ \\
$180^{\circ}$ & $0.63 \pm 0.01$ & $9.99 \pm 0.12$ & $10.18 \pm 0.11$ & $5.93 \pm 0.08$ & $0.11 \pm 0.08$ & $-0.27 \pm 0.06$ & $-0.00 \pm 0.06$ & $1.02$ \\
$225^{\circ}$ & $0.62 \pm 0.01$ & $10.71 \pm 0.14$ & $9.34 \pm 0.11$ & $5.99 \pm 0.08$ & $0.67 \pm 0.08$ & $-0.08 \pm 0.07$ & $0.31 \pm 0.06$ & $1.02$ \\
$270^{\circ}$ & $0.64 \pm 0.01$ & $11.70 \pm 0.15$ & $8.90 \pm 0.11$ & $6.16 \pm 0.08$ & $-0.11 \pm 0.08$ & $0.04 \pm 0.07$ & $0.39 \pm 0.06$ & $1.01$ \\
$315^{\circ}$ & $0.63 \pm 0.01$ & $10.59 \pm 0.13$ & $9.68 \pm 0.11$ & $6.09 \pm 0.08$ & $-0.67 \pm 0.08$ & $0.17 \pm 0.07$ & $0.30 \pm 0.06$ & $1.01$ \\
\hline
\end{tabular}
\caption{Extracted femtoscopic parameters from fits to the three-dimensional correlation functions obtained using Eq.~\ref{eq:expCF}, for all eight azimuthal angles of pion pairs with transverse pair momentum $k_{T} \in [0.45, 0.80]~\mathrm{GeV}/c$, in 10--30\% central collisions. The fit was performed using Eq.~\ref{eq:cfmodified}.}
\label{tab:fit_params_charge0_cent2_kt5}
\end{table*}

\begin{table*}[htbp]
\centering
\setlength{\tabcolsep}{4pt}
\renewcommand{\arraystretch}{1.15}
\begin{tabular}{|c|c|c|c|c|c|c|c|c|}
\hline
$\langle\phi_{k}\rangle$ & $\lambda$ & $R^{2}_{\mathrm{out}}\,(\mathrm{fm}^2)$ & $R^{2}_{\mathrm{side}}\,(\mathrm{fm}^2)$ & $R^{2}_{\mathrm{long}}\,(\mathrm{fm}^2)$ & $R^{2}_{\mathrm{out-side}}\,(\mathrm{fm}^2)$ & $R^{2}_{\mathrm{out-long}}\,(\mathrm{fm}^2)$ & $R^{2}_{\mathrm{side-long}}\,(\mathrm{fm}^2)$ & $\chi^{2}/\mathrm{NDF}$ \\
\hline
$0^{\circ}$ & $0.59 \pm 0.01$ & $13.10 \pm 0.29$ & $11.11 \pm 0.16$ & $15.79 \pm 0.26$ & $-0.05 \pm 0.15$ & $1.46 \pm 0.19$ & $0.10 \pm 0.13$ & $1.03$ \\
$45^{\circ}$ & $0.58 \pm 0.01$ & $13.86 \pm 0.30$ & $10.20 \pm 0.15$ & $15.39 \pm 0.25$ & $0.99 \pm 0.14$ & $0.82 \pm 0.19$ & $-0.80 \pm 0.13$ & $1.03$ \\
$90^{\circ}$ & $0.59 \pm 0.01$ & $14.94 \pm 0.32$ & $9.66 \pm 0.14$ & $15.57 \pm 0.25$ & $0.07 \pm 0.14$ & $0.18 \pm 0.20$ & $-1.14 \pm 0.12$ & $1.03$ \\
$135^{\circ}$ & $0.58 \pm 0.01$ & $13.91 \pm 0.31$ & $10.27 \pm 0.15$ & $14.98 \pm 0.25$ & $-1.01 \pm 0.14$ & $-0.85 \pm 0.19$ & $-0.59 \pm 0.12$ & $1.04$ \\
$180^{\circ}$ & $0.57 \pm 0.01$ & $12.61 \pm 0.27$ & $11.03 \pm 0.16$ & $14.80 \pm 0.25$ & $0.34 \pm 0.14$ & $-1.20 \pm 0.18$ & $-0.23 \pm 0.13$ & $1.04$ \\
$225^{\circ}$ & $0.58 \pm 0.01$ & $13.43 \pm 0.29$ & $10.10 \pm 0.14$ & $15.08 \pm 0.25$ & $1.53 \pm 0.14$ & $-0.26 \pm 0.18$ & $0.74 \pm 0.12$ & $1.03$ \\
$270^{\circ}$ & $0.60 \pm 0.01$ & $14.52 \pm 0.30$ & $9.42 \pm 0.13$ & $15.47 \pm 0.25$ & $0.03 \pm 0.13$ & $0.26 \pm 0.19$ & $1.23 \pm 0.12$ & $1.04$ \\
$315^{\circ}$ & $0.58 \pm 0.01$ & $13.62 \pm 0.27$ & $10.37 \pm 0.15$ & $14.97 \pm 0.25$ & $-0.78 \pm 0.14$ & $0.41 \pm 0.19$ & $0.53 \pm 0.12$ & $1.03$ \\
\hline
\end{tabular}
\caption{Extracted femtoscopic parameters from fits to the three-dimensional correlation functions obtained using Eq.~\ref{eq:expCF}, for all eight azimuthal angles of pion pairs with transverse pair momentum $k_{T} \in [0.15, 0.20]~\mathrm{GeV}/c$, in 30--50\% central collisions. The fit was performed using Eq.~\ref{eq:cfmodified}.}
\label{tab:fit_params_charge0_cent1_kt0}
\end{table*}

\begin{table*}[htbp]
\centering
\setlength{\tabcolsep}{4pt}
\renewcommand{\arraystretch}{1.15}
\begin{tabular}{|c|c|c|c|c|c|c|c|c|}
\hline
$\langle\phi_{k}\rangle$ & $\lambda$ & $R^{2}_{\mathrm{out}}\,(\mathrm{fm}^2)$ & $R^{2}_{\mathrm{side}}\,(\mathrm{fm}^2)$ & $R^{2}_{\mathrm{long}}\,(\mathrm{fm}^2)$ & $R^{2}_{\mathrm{out-side}}\,(\mathrm{fm}^2)$ & $R^{2}_{\mathrm{out-long}}\,(\mathrm{fm}^2)$ & $R^{2}_{\mathrm{side-long}}\,(\mathrm{fm}^2)$ & $\chi^{2}/\mathrm{NDF}$ \\
\hline
$0^{\circ}$ & $0.54 \pm 0.00$ & $12.26 \pm 0.16$ & $10.75 \pm 0.13$ & $12.16 \pm 0.16$ & $0.13 \pm 0.10$ & $0.74 \pm 0.11$ & $-0.15 \pm 0.09$ & $1.03$ \\
$45^{\circ}$ & $0.55 \pm 0.00$ & $13.68 \pm 0.17$ & $9.70 \pm 0.12$ & $11.98 \pm 0.16$ & $1.23 \pm 0.10$ & $0.52 \pm 0.11$ & $-0.62 \pm 0.09$ & $1.02$ \\
$90^{\circ}$ & $0.55 \pm 0.00$ & $14.76 \pm 0.18$ & $8.77 \pm 0.11$ & $12.51 \pm 0.16$ & $-0.04 \pm 0.10$ & $-0.04 \pm 0.12$ & $-0.89 \pm 0.09$ & $1.02$ \\
$135^{\circ}$ & $0.55 \pm 0.00$ & $13.44 \pm 0.17$ & $9.84 \pm 0.12$ & $12.34 \pm 0.16$ & $-1.12 \pm 0.09$ & $-0.69 \pm 0.11$ & $-0.62 \pm 0.09$ & $1.02$ \\
$180^{\circ}$ & $0.54 \pm 0.00$ & $12.41 \pm 0.17$ & $10.50 \pm 0.12$ & $12.00 \pm 0.15$ & $0.17 \pm 0.09$ & $-0.96 \pm 0.11$ & $-0.01 \pm 0.09$ & $1.02$ \\
$225^{\circ}$ & $0.56 \pm 0.00$ & $13.68 \pm 0.17$ & $9.68 \pm 0.12$ & $12.60 \pm 0.16$ & $1.23 \pm 0.10$ & $-0.67 \pm 0.12$ & $0.67 \pm 0.09$ & $1.04$ \\
$270^{\circ}$ & $0.54 \pm 0.00$ & $14.66 \pm 0.19$ & $8.72 \pm 0.11$ & $12.15 \pm 0.16$ & $-0.16 \pm 0.10$ & $-0.13 \pm 0.12$ & $0.83 \pm 0.08$ & $1.02$ \\
$315^{\circ}$ & $0.56 \pm 0.00$ & $13.64 \pm 0.17$ & $9.74 \pm 0.11$ & $12.05 \pm 0.15$ & $-1.08 \pm 0.09$ & $0.49 \pm 0.11$ & $0.56 \pm 0.09$ & $1.03$ \\
\hline
\end{tabular}
\caption{Extracted femtoscopic parameters from fits to the three-dimensional correlation functions obtained using Eq.~\ref{eq:expCF}, for all eight azimuthal angles of pion pairs with transverse pair momentum $k_{T} \in [0.20, 0.25]~\mathrm{GeV}/c$, in 30--50\% central collisions. The fit was performed using Eq.~\ref{eq:cfmodified}.}
\label{tab:fit_params_charge0_cent1_kt1}
\end{table*}

\begin{table*}[htbp]
\centering
\setlength{\tabcolsep}{4pt}
\renewcommand{\arraystretch}{1.15}
\begin{tabular}{|c|c|c|c|c|c|c|c|c|}
\hline
$\langle\phi_{k}\rangle$ & $\lambda$ & $R^{2}_{\mathrm{out}}\,(\mathrm{fm}^2)$ & $R^{2}_{\mathrm{side}}\,(\mathrm{fm}^2)$ & $R^{2}_{\mathrm{long}}\,(\mathrm{fm}^2)$ & $R^{2}_{\mathrm{out-side}}\,(\mathrm{fm}^2)$ & $R^{2}_{\mathrm{out-long}}\,(\mathrm{fm}^2)$ & $R^{2}_{\mathrm{side-long}}\,(\mathrm{fm}^2)$ & $\chi^{2}/\mathrm{NDF}$ \\
\hline
$0^{\circ}$ & $0.56 \pm 0.01$ & $11.37 \pm 0.15$ & $9.64 \pm 0.12$ & $9.29 \pm 0.13$ & $0.09 \pm 0.09$ & $0.53 \pm 0.09$ & $-0.06 \pm 0.08$ & $1.02$ \\
$45^{\circ}$ & $0.55 \pm 0.01$ & $12.50 \pm 0.16$ & $8.65 \pm 0.11$ & $9.46 \pm 0.13$ & $1.16 \pm 0.09$ & $0.32 \pm 0.10$ & $-0.25 \pm 0.08$ & $1.01$ \\
$90^{\circ}$ & $0.57 \pm 0.01$ & $13.74 \pm 0.18$ & $7.94 \pm 0.10$ & $9.61 \pm 0.14$ & $0.02 \pm 0.09$ & $0.10 \pm 0.11$ & $-0.74 \pm 0.08$ & $1.01$ \\
$135^{\circ}$ & $0.56 \pm 0.01$ & $12.46 \pm 0.16$ & $8.67 \pm 0.11$ & $9.59 \pm 0.13$ & $-1.07 \pm 0.09$ & $-0.51 \pm 0.10$ & $-0.34 \pm 0.08$ & $1.02$ \\
$180^{\circ}$ & $0.56 \pm 0.01$ & $11.22 \pm 0.13$ & $9.76 \pm 0.12$ & $9.49 \pm 0.13$ & $0.18 \pm 0.09$ & $-0.53 \pm 0.09$ & $0.01 \pm 0.08$ & $1.01$ \\
$225^{\circ}$ & $0.56 \pm 0.01$ & $12.44 \pm 0.17$ & $8.73 \pm 0.11$ & $9.62 \pm 0.14$ & $1.07 \pm 0.09$ & $-0.32 \pm 0.10$ & $0.30 \pm 0.08$ & $1.02$ \\
$270^{\circ}$ & $0.57 \pm 0.01$ & $13.69 \pm 0.18$ & $8.07 \pm 0.11$ & $9.79 \pm 0.14$ & $-0.13 \pm 0.09$ & $0.01 \pm 0.11$ & $0.54 \pm 0.08$ & $1.01$ \\
$315^{\circ}$ & $0.56 \pm 0.01$ & $12.26 \pm 0.16$ & $8.89 \pm 0.11$ & $9.33 \pm 0.13$ & $-1.17 \pm 0.09$ & $0.43 \pm 0.10$ & $0.39 \pm 0.08$ & $1.02$ \\
\hline
\end{tabular}
\caption{Extracted femtoscopic parameters from fits to the three-dimensional correlation functions obtained using Eq.~\ref{eq:expCF}, for all eight azimuthal angles of pion pairs with transverse pair momentum $k_{T} \in [0.25, 0.30]~\mathrm{GeV}/c$, in 30--50\% central collisions. The fit was performed using Eq.~\ref{eq:cfmodified}.}
\label{tab:fit_params_charge0_cent1_kt2}
\end{table*}

\begin{table*}[htbp]
\centering
\setlength{\tabcolsep}{4pt}
\renewcommand{\arraystretch}{1.15}
\begin{tabular}{|c|c|c|c|c|c|c|c|c|}
\hline
$\langle\phi_{k}\rangle$ & $\lambda$ & $R^{2}_{\mathrm{out}}\,(\mathrm{fm}^2)$ & $R^{2}_{\mathrm{side}}\,(\mathrm{fm}^2)$ & $R^{2}_{\mathrm{long}}\,(\mathrm{fm}^2)$ & $R^{2}_{\mathrm{out-side}}\,(\mathrm{fm}^2)$ & $R^{2}_{\mathrm{out-long}}\,(\mathrm{fm}^2)$ & $R^{2}_{\mathrm{side-long}}\,(\mathrm{fm}^2)$ & $\chi^{2}/\mathrm{NDF}$ \\
\hline
$0^{\circ}$ & $0.58 \pm 0.01$ & $9.96 \pm 0.15$ & $8.99 \pm 0.13$ & $7.66 \pm 0.12$ & $0.08 \pm 0.09$ & $0.33 \pm 0.09$ & $0.00 \pm 0.08$ & $1.02$ \\
$45^{\circ}$ & $0.58 \pm 0.01$ & $11.22 \pm 0.17$ & $8.24 \pm 0.12$ & $7.55 \pm 0.12$ & $1.14 \pm 0.10$ & $0.26 \pm 0.10$ & $-0.31 \pm 0.08$ & $1.02$ \\
$90^{\circ}$ & $0.58 \pm 0.01$ & $12.38 \pm 0.19$ & $7.38 \pm 0.11$ & $7.72 \pm 0.13$ & $0.11 \pm 0.10$ & $-0.04 \pm 0.11$ & $-0.48 \pm 0.08$ & $1.02$ \\
$135^{\circ}$ & $0.58 \pm 0.01$ & $11.19 \pm 0.17$ & $8.15 \pm 0.12$ & $7.63 \pm 0.12$ & $-1.05 \pm 0.10$ & $-0.36 \pm 0.10$ & $-0.40 \pm 0.08$ & $1.02$ \\
$180^{\circ}$ & $0.58 \pm 0.01$ & $9.99 \pm 0.15$ & $9.12 \pm 0.13$ & $7.45 \pm 0.12$ & $0.19 \pm 0.09$ & $-0.41 \pm 0.09$ & $0.09 \pm 0.08$ & $1.02$ \\
$225^{\circ}$ & $0.57 \pm 0.01$ & $11.23 \pm 0.17$ & $8.10 \pm 0.12$ & $7.55 \pm 0.13$ & $1.21 \pm 0.10$ & $-0.16 \pm 0.10$ & $0.43 \pm 0.08$ & $1.02$ \\
$270^{\circ}$ & $0.58 \pm 0.01$ & $11.92 \pm 0.18$ & $7.42 \pm 0.11$ & $7.58 \pm 0.13$ & $-0.15 \pm 0.10$ & $-0.24 \pm 0.10$ & $0.52 \pm 0.08$ & $1.02$ \\
$315^{\circ}$ & $0.58 \pm 0.01$ & $10.86 \pm 0.16$ & $8.40 \pm 0.13$ & $7.59 \pm 0.12$ & $-1.26 \pm 0.10$ & $0.37 \pm 0.10$ & $0.23 \pm 0.08$ & $1.01$ \\
\hline
\end{tabular}
\caption{Extracted femtoscopic parameters from fits to the three-dimensional correlation functions obtained using Eq.~\ref{eq:expCF}, for all eight azimuthal angles of pion pairs with transverse pair momentum $k_{T} \in [0.30, 0.35]~\mathrm{GeV}/c$, in 30--50\% central collisions. The fit was performed using Eq.~\ref{eq:cfmodified}.}
\label{tab:fit_params_charge0_cent1_kt3}
\end{table*}

\begin{table*}[htbp]
\centering
\setlength{\tabcolsep}{4pt}
\renewcommand{\arraystretch}{1.15}
\begin{tabular}{|c|c|c|c|c|c|c|c|c|}
\hline
$\langle\phi_{k}\rangle$ & $\lambda$ & $R^{2}_{\mathrm{out}}\,(\mathrm{fm}^2)$ & $R^{2}_{\mathrm{side}}\,(\mathrm{fm}^2)$ & $R^{2}_{\mathrm{long}}\,(\mathrm{fm}^2)$ & $R^{2}_{\mathrm{out-side}}\,(\mathrm{fm}^2)$ & $R^{2}_{\mathrm{out-long}}\,(\mathrm{fm}^2)$ & $R^{2}_{\mathrm{side-long}}\,(\mathrm{fm}^2)$ & $\chi^{2}/\mathrm{NDF}$ \\
\hline
$0^{\circ}$ & $0.59 \pm 0.01$ & $8.26 \pm 0.11$ & $8.14 \pm 0.11$ & $5.62 \pm 0.08$ & $-0.03 \pm 0.07$ & $0.20 \pm 0.07$ & $-0.02 \pm 0.06$ & $1.02$ \\
$45^{\circ}$ & $0.60 \pm 0.01$ & $9.46 \pm 0.13$ & $7.18 \pm 0.10$ & $5.82 \pm 0.09$ & $1.07 \pm 0.08$ & $0.13 \pm 0.07$ & $-0.14 \pm 0.06$ & $1.01$ \\
$90^{\circ}$ & $0.60 \pm 0.01$ & $10.42 \pm 0.15$ & $6.51 \pm 0.09$ & $6.14 \pm 0.10$ & $-0.19 \pm 0.08$ & $0.05 \pm 0.08$ & $-0.24 \pm 0.06$ & $1.01$ \\
$135^{\circ}$ & $0.59 \pm 0.01$ & $9.28 \pm 0.13$ & $7.34 \pm 0.10$ & $5.85 \pm 0.09$ & $-1.03 \pm 0.08$ & $-0.22 \pm 0.07$ & $-0.27 \pm 0.06$ & $1.02$ \\
$180^{\circ}$ & $0.60 \pm 0.01$ & $8.39 \pm 0.12$ & $8.42 \pm 0.11$ & $5.78 \pm 0.09$ & $-0.02 \pm 0.08$ & $-0.27 \pm 0.07$ & $-0.08 \pm 0.06$ & $1.01$ \\
$225^{\circ}$ & $0.60 \pm 0.01$ & $9.54 \pm 0.13$ & $7.42 \pm 0.10$ & $6.04 \pm 0.09$ & $1.18 \pm 0.08$ & $-0.19 \pm 0.08$ & $0.21 \pm 0.06$ & $1.02$ \\
$270^{\circ}$ & $0.60 \pm 0.01$ & $10.42 \pm 0.15$ & $6.51 \pm 0.09$ & $6.10 \pm 0.10$ & $-0.05 \pm 0.08$ & $0.11 \pm 0.08$ & $0.30 \pm 0.06$ & $1.02$ \\
$315^{\circ}$ & $0.61 \pm 0.01$ & $9.49 \pm 0.13$ & $7.53 \pm 0.10$ & $6.12 \pm 0.09$ & $-0.92 \pm 0.08$ & $0.10 \pm 0.07$ & $0.27 \pm 0.06$ & $1.01$ \\
\hline
\end{tabular}
\caption{Extracted femtoscopic parameters from fits to the three-dimensional correlation functions obtained using Eq.~\ref{eq:expCF}, for all eight azimuthal angles of pion pairs with transverse pair momentum $k_{T} \in [0.35, 0.45]~\mathrm{GeV}/c$, in 30--50\% central collisions. The fit was performed using Eq.~\ref{eq:cfmodified}.}
\label{tab:fit_params_charge0_cent1_kt4}
\end{table*}

\begin{table*}[htbp]
\centering
\setlength{\tabcolsep}{4pt}
\renewcommand{\arraystretch}{1.15}
\begin{tabular}{|c|c|c|c|c|c|c|c|c|}
\hline
$\langle\phi_{k}\rangle$ & $\lambda$ & $R^{2}_{\mathrm{out}}\,(\mathrm{fm}^2)$ & $R^{2}_{\mathrm{side}}\,(\mathrm{fm}^2)$ & $R^{2}_{\mathrm{long}}\,(\mathrm{fm}^2)$ & $R^{2}_{\mathrm{out-side}}\,(\mathrm{fm}^2)$ & $R^{2}_{\mathrm{out-long}}\,(\mathrm{fm}^2)$ & $R^{2}_{\mathrm{side-long}}\,(\mathrm{fm}^2)$ & $\chi^{2}/\mathrm{NDF}$ \\
\hline
$0^{\circ}$ & $0.68 \pm 0.01$ & $6.70 \pm 0.13$ & $7.45 \pm 0.14$ & $4.00 \pm 0.08$ & $0.10 \pm 0.09$ & $0.02 \pm 0.07$ & $-0.03 \pm 0.06$ & $1.03$ \\
$45^{\circ}$ & $0.67 \pm 0.01$ & $7.45 \pm 0.15$ & $6.42 \pm 0.12$ & $4.15 \pm 0.09$ & $0.89 \pm 0.09$ & $0.10 \pm 0.07$ & $-0.16 \pm 0.06$ & $1.04$ \\
$90^{\circ}$ & $0.67 \pm 0.01$ & $8.32 \pm 0.17$ & $5.51 \pm 0.11$ & $4.18 \pm 0.09$ & $0.01 \pm 0.09$ & $0.04 \pm 0.08$ & $-0.20 \pm 0.06$ & $1.03$ \\
$135^{\circ}$ & $0.66 \pm 0.01$ & $7.31 \pm 0.15$ & $6.65 \pm 0.13$ & $3.89 \pm 0.08$ & $-0.76 \pm 0.09$ & $-0.07 \pm 0.07$ & $-0.14 \pm 0.06$ & $1.04$ \\
$180^{\circ}$ & $0.67 \pm 0.01$ & $6.66 \pm 0.13$ & $7.24 \pm 0.13$ & $3.95 \pm 0.08$ & $0.06 \pm 0.08$ & $-0.09 \pm 0.07$ & $-0.03 \pm 0.06$ & $1.03$ \\
$225^{\circ}$ & $0.66 \pm 0.01$ & $7.48 \pm 0.15$ & $6.33 \pm 0.12$ & $3.96 \pm 0.08$ & $0.94 \pm 0.09$ & $-0.13 \pm 0.07$ & $0.10 \pm 0.06$ & $1.04$ \\
$270^{\circ}$ & $0.68 \pm 0.01$ & $8.66 \pm 0.18$ & $5.51 \pm 0.11$ & $4.24 \pm 0.09$ & $-0.06 \pm 0.09$ & $-0.06 \pm 0.08$ & $0.25 \pm 0.06$ & $1.04$ \\
$315^{\circ}$ & $0.70 \pm 0.01$ & $7.79 \pm 0.16$ & $6.53 \pm 0.12$ & $4.27 \pm 0.09$ & $-1.00 \pm 0.09$ & $0.07 \pm 0.07$ & $0.16 \pm 0.06$ & $1.03$ \\
\hline
\end{tabular}
\caption{Extracted femtoscopic parameters from fits to the three-dimensional correlation functions obtained using Eq.~\ref{eq:expCF}, for all eight azimuthal angles of pion pairs with transverse pair momentum $k_{T} \in [0.45, 0.80]~\mathrm{GeV}/c$, in 30--50\% central collisions. The fit was performed using Eq.~\ref{eq:cfmodified}.}
\label{tab:fit_params_charge0_cent1_kt5}
\end{table*}

\begin{table*}[htbp]
\centering
\setlength{\tabcolsep}{4pt}
\renewcommand{\arraystretch}{1.15}
\begin{tabular}{|c|c|c|c|c|c|c|c|c|}
\hline
$\langle\phi_{k}\rangle$ & $\lambda$ & $R^{2}_{\mathrm{out}}\,(\mathrm{fm}^2)$ & $R^{2}_{\mathrm{side}}\,(\mathrm{fm}^2)$ & $R^{2}_{\mathrm{long}}\,(\mathrm{fm}^2)$ & $R^{2}_{\mathrm{out-side}}\,(\mathrm{fm}^2)$ & $R^{2}_{\mathrm{out-long}}\,(\mathrm{fm}^2)$ & $R^{2}_{\mathrm{side-long}}\,(\mathrm{fm}^2)$ & $\chi^{2}/\mathrm{NDF}$ \\
\hline
$0^{\circ}$ & $0.58 \pm 0.01$ & $6.87 \pm 0.37$ & $5.30 \pm 0.14$ & $9.71 \pm 0.29$ & $-0.10 \pm 0.15$ & $0.63 \pm 0.23$ & $-0.08 \pm 0.13$ & $1.10$ \\
$45^{\circ}$ & $0.60 \pm 0.01$ & $6.40 \pm 0.35$ & $4.98 \pm 0.13$ & $9.89 \pm 0.28$ & $0.16 \pm 0.14$ & $0.01 \pm 0.22$ & $-0.27 \pm 0.12$ & $1.09$ \\
$90^{\circ}$ & $0.61 \pm 0.01$ & $7.60 \pm 0.37$ & $4.72 \pm 0.12$ & $10.05 \pm 0.27$ & $0.17 \pm 0.14$ & $0.21 \pm 0.22$ & $-0.36 \pm 0.11$ & $1.08$ \\
$135^{\circ}$ & $0.60 \pm 0.01$ & $6.97 \pm 0.37$ & $5.14 \pm 0.14$ & $9.72 \pm 0.29$ & $-0.34 \pm 0.14$ & $-0.08 \pm 0.22$ & $-0.16 \pm 0.12$ & $1.08$ \\
$180^{\circ}$ & $0.60 \pm 0.01$ & $6.35 \pm 0.35$ & $5.54 \pm 0.14$ & $9.04 \pm 0.27$ & $0.17 \pm 0.15$ & $-0.49 \pm 0.21$ & $-0.27 \pm 0.12$ & $1.08$ \\
$225^{\circ}$ & $0.59 \pm 0.01$ & $6.76 \pm 0.36$ & $4.96 \pm 0.13$ & $9.50 \pm 0.27$ & $0.35 \pm 0.14$ & $-0.19 \pm 0.21$ & $0.02 \pm 0.12$ & $1.09$ \\
$270^{\circ}$ & $0.59 \pm 0.01$ & $7.16 \pm 0.37$ & $4.71 \pm 0.13$ & $9.94 \pm 0.29$ & $0.06 \pm 0.14$ & $0.03 \pm 0.22$ & $0.45 \pm 0.12$ & $1.08$ \\
$315^{\circ}$ & $0.60 \pm 0.01$ & $6.40 \pm 0.36$ & $5.11 \pm 0.14$ & $10.17 \pm 0.29$ & $-0.36 \pm 0.14$ & $0.33 \pm 0.22$ & $0.51 \pm 0.12$ & $1.10$ \\
\hline
\end{tabular}
\caption{Extracted femtoscopic parameters from fits to the three-dimensional correlation functions obtained using Eq.~\ref{eq:expCF}, for all eight azimuthal angles of pion pairs with transverse pair momentum $k_{T} \in [0.15, 0.20]~\mathrm{GeV}/c$, in 50--80\% central collisions. The fit was performed using Eq.~\ref{eq:cfmodified}.}
\label{tab:fit_params_charge0_cent0_kt0}
\end{table*}

\begin{table*}[htbp]
\centering
\setlength{\tabcolsep}{4pt}
\renewcommand{\arraystretch}{1.15}
\begin{tabular}{|c|c|c|c|c|c|c|c|c|}
\hline
$\langle\phi_{k}\rangle$ & $\lambda$ & $R^{2}_{\mathrm{out}}\,(\mathrm{fm}^2)$ & $R^{2}_{\mathrm{side}}\,(\mathrm{fm}^2)$ & $R^{2}_{\mathrm{long}}\,(\mathrm{fm}^2)$ & $R^{2}_{\mathrm{out-side}}\,(\mathrm{fm}^2)$ & $R^{2}_{\mathrm{out-long}}\,(\mathrm{fm}^2)$ & $R^{2}_{\mathrm{side-long}}\,(\mathrm{fm}^2)$ & $\chi^{2}/\mathrm{NDF}$ \\
\hline
$0^{\circ}$ & $0.55 \pm 0.01$ & $6.85 \pm 0.17$ & $5.57 \pm 0.13$ & $7.54 \pm 0.17$ & $-0.04 \pm 0.10$ & $0.38 \pm 0.11$ & $-0.10 \pm 0.09$ & $1.07$ \\
$45^{\circ}$ & $0.55 \pm 0.01$ & $7.26 \pm 0.19$ & $5.24 \pm 0.12$ & $7.84 \pm 0.18$ & $0.46 \pm 0.10$ & $0.23 \pm 0.12$ & $0.04 \pm 0.09$ & $1.07$ \\
$90^{\circ}$ & $0.54 \pm 0.01$ & $7.51 \pm 0.19$ & $4.65 \pm 0.11$ & $7.83 \pm 0.18$ & $0.05 \pm 0.09$ & $0.13 \pm 0.12$ & $-0.12 \pm 0.09$ & $1.07$ \\
$135^{\circ}$ & $0.55 \pm 0.01$ & $7.45 \pm 0.19$ & $5.16 \pm 0.12$ & $7.59 \pm 0.18$ & $-0.60 \pm 0.10$ & $-0.40 \pm 0.12$ & $-0.10 \pm 0.09$ & $1.07$ \\
$180^{\circ}$ & $0.56 \pm 0.01$ & $6.82 \pm 0.17$ & $5.59 \pm 0.12$ & $7.41 \pm 0.17$ & $0.07 \pm 0.09$ & $-0.26 \pm 0.11$ & $-0.21 \pm 0.09$ & $1.08$ \\
$225^{\circ}$ & $0.54 \pm 0.01$ & $7.39 \pm 0.19$ & $5.18 \pm 0.12$ & $7.33 \pm 0.17$ & $0.48 \pm 0.10$ & $-0.39 \pm 0.12$ & $0.23 \pm 0.09$ & $1.08$ \\
$270^{\circ}$ & $0.56 \pm 0.01$ & $7.99 \pm 0.20$ & $4.80 \pm 0.11$ & $8.13 \pm 0.19$ & $0.08 \pm 0.10$ & $0.03 \pm 0.13$ & $0.36 \pm 0.09$ & $1.08$ \\
$315^{\circ}$ & $0.55 \pm 0.01$ & $7.36 \pm 0.19$ & $5.19 \pm 0.12$ & $7.35 \pm 0.17$ & $-0.42 \pm 0.09$ & $0.09 \pm 0.12$ & $0.13 \pm 0.09$ & $1.08$ \\
\hline
\end{tabular}
\caption{Extracted femtoscopic parameters from fits to the three-dimensional correlation functions obtained using Eq.~\ref{eq:expCF}, for all eight azimuthal angles of pion pairs with transverse pair momentum $k_{T} \in [0.20, 0.25]~\mathrm{GeV}/c$, in 50--80\% central collisions. The fit was performed using Eq.~\ref{eq:cfmodified}.}
\label{tab:fit_params_charge0_cent0_kt1}
\end{table*}

\begin{table*}[htbp]
\centering
\setlength{\tabcolsep}{4pt}
\renewcommand{\arraystretch}{1.15}
\begin{tabular}{|c|c|c|c|c|c|c|c|c|}
\hline
$\langle\phi_{k}\rangle$ & $\lambda$ & $R^{2}_{\mathrm{out}}\,(\mathrm{fm}^2)$ & $R^{2}_{\mathrm{side}}\,(\mathrm{fm}^2)$ & $R^{2}_{\mathrm{long}}\,(\mathrm{fm}^2)$ & $R^{2}_{\mathrm{out-side}}\,(\mathrm{fm}^2)$ & $R^{2}_{\mathrm{out-long}}\,(\mathrm{fm}^2)$ & $R^{2}_{\mathrm{side-long}}\,(\mathrm{fm}^2)$ & $\chi^{2}/\mathrm{NDF}$ \\
\hline
$0^{\circ}$ & $0.58 \pm 0.01$ & $6.93 \pm 0.17$ & $5.45 \pm 0.13$ & $6.04 \pm 0.15$ & $0.05 \pm 0.10$ & $0.28 \pm 0.11$ & $0.02 \pm 0.09$ & $1.06$ \\
$45^{\circ}$ & $0.58 \pm 0.01$ & $7.56 \pm 0.19$ & $4.78 \pm 0.11$ & $6.00 \pm 0.15$ & $0.45 \pm 0.10$ & $0.12 \pm 0.11$ & $-0.20 \pm 0.09$ & $1.05$ \\
$90^{\circ}$ & $0.59 \pm 0.01$ & $8.02 \pm 0.19$ & $4.54 \pm 0.11$ & $6.21 \pm 0.16$ & $-0.06 \pm 0.10$ & $-0.12 \pm 0.12$ & $-0.02 \pm 0.08$ & $1.07$ \\
$135^{\circ}$ & $0.58 \pm 0.01$ & $7.40 \pm 0.19$ & $4.74 \pm 0.12$ & $6.14 \pm 0.16$ & $-0.50 \pm 0.09$ & $0.05 \pm 0.11$ & $-0.14 \pm 0.08$ & $1.05$ \\
$180^{\circ}$ & $0.56 \pm 0.01$ & $6.62 \pm 0.17$ & $5.35 \pm 0.13$ & $5.86 \pm 0.15$ & $-0.01 \pm 0.10$ & $-0.27 \pm 0.10$ & $-0.08 \pm 0.09$ & $1.06$ \\
$225^{\circ}$ & $0.57 \pm 0.01$ & $7.22 \pm 0.18$ & $4.89 \pm 0.12$ & $6.16 \pm 0.16$ & $0.47 \pm 0.10$ & $-0.16 \pm 0.11$ & $-0.03 \pm 0.09$ & $1.06$ \\
$270^{\circ}$ & $0.56 \pm 0.01$ & $7.67 \pm 0.18$ & $4.33 \pm 0.11$ & $6.06 \pm 0.15$ & $-0.11 \pm 0.10$ & $-0.03 \pm 0.12$ & $0.09 \pm 0.08$ & $1.06$ \\
$315^{\circ}$ & $0.57 \pm 0.01$ & $7.28 \pm 0.18$ & $5.05 \pm 0.13$ & $6.08 \pm 0.15$ & $-0.58 \pm 0.10$ & $0.21 \pm 0.11$ & $0.07 \pm 0.09$ & $1.06$ \\
\hline
\end{tabular}
\caption{Extracted femtoscopic parameters from fits to the three-dimensional correlation functions obtained using Eq.~\ref{eq:expCF}, for all eight azimuthal angles of pion pairs with transverse pair momentum $k_{T} \in [0.25, 0.30]~\mathrm{GeV}/c$, in 50--80\% central collisions. The fit was performed using Eq.~\ref{eq:cfmodified}.}
\label{tab:fit_params_charge0_cent0_kt2}
\end{table*}

\begin{table*}[htbp]
\centering
\setlength{\tabcolsep}{4pt}
\renewcommand{\arraystretch}{1.15}
\begin{tabular}{|c|c|c|c|c|c|c|c|c|}
\hline
$\langle\phi_{k}\rangle$ & $\lambda$ & $R^{2}_{\mathrm{out}}\,(\mathrm{fm}^2)$ & $R^{2}_{\mathrm{side}}\,(\mathrm{fm}^2)$ & $R^{2}_{\mathrm{long}}\,(\mathrm{fm}^2)$ & $R^{2}_{\mathrm{out-side}}\,(\mathrm{fm}^2)$ & $R^{2}_{\mathrm{out-long}}\,(\mathrm{fm}^2)$ & $R^{2}_{\mathrm{side-long}}\,(\mathrm{fm}^2)$ & $\chi^{2}/\mathrm{NDF}$ \\
\hline
$0^{\circ}$ & $0.58 \pm 0.01$ & $5.86 \pm 0.16$ & $4.61 \pm 0.13$ & $4.69 \pm 0.14$ & $0.08 \pm 0.09$ & $0.21 \pm 0.10$ & $0.04 \pm 0.08$ & $1.08$ \\
$45^{\circ}$ & $0.58 \pm 0.01$ & $7.17 \pm 0.20$ & $4.43 \pm 0.13$ & $4.72 \pm 0.14$ & $0.56 \pm 0.11$ & $0.14 \pm 0.11$ & $0.06 \pm 0.09$ & $1.09$ \\
$90^{\circ}$ & $0.61 \pm 0.01$ & $6.95 \pm 0.19$ & $4.35 \pm 0.13$ & $5.14 \pm 0.15$ & $-0.17 \pm 0.10$ & $0.17 \pm 0.11$ & $-0.13 \pm 0.09$ & $1.09$ \\
$135^{\circ}$ & $0.60 \pm 0.01$ & $6.67 \pm 0.19$ & $4.60 \pm 0.13$ & $4.84 \pm 0.14$ & $-0.68 \pm 0.10$ & $-0.14 \pm 0.11$ & $-0.11 \pm 0.08$ & $1.08$ \\
$180^{\circ}$ & $0.60 \pm 0.01$ & $6.37 \pm 0.17$ & $4.92 \pm 0.14$ & $4.96 \pm 0.15$ & $0.22 \pm 0.11$ & $-0.09 \pm 0.11$ & $0.05 \pm 0.09$ & $1.08$ \\
$225^{\circ}$ & $0.57 \pm 0.01$ & $6.44 \pm 0.19$ & $4.50 \pm 0.13$ & $4.64 \pm 0.14$ & $0.45 \pm 0.10$ & $-0.15 \pm 0.11$ & $0.16 \pm 0.09$ & $1.08$ \\
$270^{\circ}$ & $0.59 \pm 0.01$ & $7.11 \pm 0.19$ & $4.04 \pm 0.12$ & $5.00 \pm 0.15$ & $-0.07 \pm 0.10$ & $0.18 \pm 0.12$ & $0.02 \pm 0.08$ & $1.08$ \\
$315^{\circ}$ & $0.61 \pm 0.01$ & $7.08 \pm 0.20$ & $4.60 \pm 0.13$ & $4.88 \pm 0.14$ & $-0.62 \pm 0.10$ & $0.03 \pm 0.11$ & $0.06 \pm 0.09$ & $1.08$ \\
\hline
\end{tabular}
\caption{Extracted femtoscopic parameters from fits to the three-dimensional correlation functions obtained using Eq.~\ref{eq:expCF}, for all eight azimuthal angles of pion pairs with transverse pair momentum $k_{T} \in [0.30, 0.35]~\mathrm{GeV}/c$, in 50--80\% central collisions. The fit was performed using Eq.~\ref{eq:cfmodified}.}
\label{tab:fit_params_charge0_cent0_kt3}
\end{table*}

\begin{table*}[htbp]
\centering
\setlength{\tabcolsep}{4pt}
\renewcommand{\arraystretch}{1.15}
\begin{tabular}{|c|c|c|c|c|c|c|c|c|}
\hline
$\langle\phi_{k}\rangle$ & $\lambda$ & $R^{2}_{\mathrm{out}}\,(\mathrm{fm}^2)$ & $R^{2}_{\mathrm{side}}\,(\mathrm{fm}^2)$ & $R^{2}_{\mathrm{long}}\,(\mathrm{fm}^2)$ & $R^{2}_{\mathrm{out-side}}\,(\mathrm{fm}^2)$ & $R^{2}_{\mathrm{out-long}}\,(\mathrm{fm}^2)$ & $R^{2}_{\mathrm{side-long}}\,(\mathrm{fm}^2)$ & $\chi^{2}/\mathrm{NDF}$ \\
\hline
$0^{\circ}$ & $0.62 \pm 0.01$ & $5.32 \pm 0.14$ & $4.55 \pm 0.12$ & $3.64 \pm 0.10$ & $0.07 \pm 0.08$ & $0.03 \pm 0.08$ & $-0.04 \pm 0.07$ & $1.08$ \\
$45^{\circ}$ & $0.64 \pm 0.01$ & $5.99 \pm 0.16$ & $4.49 \pm 0.12$ & $3.90 \pm 0.11$ & $0.32 \pm 0.09$ & $-0.00 \pm 0.09$ & $-0.03 \pm 0.07$ & $1.09$ \\
$90^{\circ}$ & $0.63 \pm 0.01$ & $6.53 \pm 0.17$ & $3.88 \pm 0.11$ & $3.70 \pm 0.11$ & $-0.00 \pm 0.09$ & $-0.15 \pm 0.09$ & $0.08 \pm 0.06$ & $1.10$ \\
$135^{\circ}$ & $0.60 \pm 0.01$ & $5.62 \pm 0.15$ & $4.20 \pm 0.11$ & $3.56 \pm 0.10$ & $-0.50 \pm 0.08$ & $-0.06 \pm 0.08$ & $-0.13 \pm 0.07$ & $1.09$ \\
$180^{\circ}$ & $0.62 \pm 0.01$ & $5.30 \pm 0.14$ & $4.82 \pm 0.13$ & $3.59 \pm 0.10$ & $0.06 \pm 0.08$ & $-0.12 \pm 0.08$ & $0.04 \pm 0.07$ & $1.09$ \\
$225^{\circ}$ & $0.61 \pm 0.01$ & $6.13 \pm 0.17$ & $4.19 \pm 0.12$ & $3.78 \pm 0.11$ & $0.41 \pm 0.09$ & $0.02 \pm 0.09$ & $0.03 \pm 0.07$ & $1.09$ \\
$270^{\circ}$ & $0.61 \pm 0.01$ & $6.20 \pm 0.17$ & $3.70 \pm 0.10$ & $3.81 \pm 0.11$ & $-0.06 \pm 0.08$ & $-0.06 \pm 0.09$ & $0.12 \pm 0.06$ & $1.09$ \\
$315^{\circ}$ & $0.63 \pm 0.01$ & $5.81 \pm 0.15$ & $4.20 \pm 0.11$ & $3.70 \pm 0.10$ & $-0.56 \pm 0.09$ & $0.01 \pm 0.08$ & $0.06 \pm 0.07$ & $1.09$ \\
\hline
\end{tabular}
\caption{Extracted femtoscopic parameters from fits to the three-dimensional correlation functions obtained using Eq.~\ref{eq:expCF}, for all eight azimuthal angles of pion pairs with transverse pair momentum $k_{T} \in [0.35, 0.45]~\mathrm{GeV}/c$, in 50--80\% central collisions. The fit was performed using Eq.~\ref{eq:cfmodified}.}
\label{tab:fit_params_charge0_cent0_kt4}
\end{table*}

\begin{table*}[htbp]
\centering
\setlength{\tabcolsep}{4pt}
\renewcommand{\arraystretch}{1.15}
\begin{tabular}{|c|c|c|c|c|c|c|c|c|}
\hline
$\langle\phi_{k}\rangle$ & $\lambda$ & $R^{2}_{\mathrm{out}}\,(\mathrm{fm}^2)$ & $R^{2}_{\mathrm{side}}\,(\mathrm{fm}^2)$ & $R^{2}_{\mathrm{long}}\,(\mathrm{fm}^2)$ & $R^{2}_{\mathrm{out-side}}\,(\mathrm{fm}^2)$ & $R^{2}_{\mathrm{out-long}}\,(\mathrm{fm}^2)$ & $R^{2}_{\mathrm{side-long}}\,(\mathrm{fm}^2)$ & $\chi^{2}/\mathrm{NDF}$ \\
\hline
$0^{\circ}$ & $0.69 \pm 0.02$ & $4.39 \pm 0.17$ & $4.23 \pm 0.15$ & $2.47 \pm 0.10$ & $0.05 \pm 0.10$ & $0.01 \pm 0.08$ & $0.13 \pm 0.07$ & $1.33$ \\
$45^{\circ}$ & $0.72 \pm 0.02$ & $4.88 \pm 0.19$ & $3.83 \pm 0.14$ & $2.65 \pm 0.11$ & $0.46 \pm 0.10$ & $-0.07 \pm 0.09$ & $0.09 \pm 0.07$ & $1.37$ \\
$90^{\circ}$ & $0.69 \pm 0.02$ & $5.54 \pm 0.22$ & $3.38 \pm 0.13$ & $2.51 \pm 0.10$ & $0.13 \pm 0.11$ & $0.03 \pm 0.10$ & $-0.08 \pm 0.07$ & $1.36$ \\
$135^{\circ}$ & $0.71 \pm 0.02$ & $4.85 \pm 0.19$ & $4.10 \pm 0.16$ & $2.48 \pm 0.10$ & $-0.43 \pm 0.11$ & $0.09 \pm 0.08$ & $-0.11 \pm 0.07$ & $1.33$ \\
$180^{\circ}$ & $0.71 \pm 0.02$ & $4.49 \pm 0.17$ & $4.24 \pm 0.15$ & $2.44 \pm 0.09$ & $-0.03 \pm 0.10$ & $0.10 \pm 0.08$ & $0.01 \pm 0.07$ & $1.32$ \\
$225^{\circ}$ & $0.68 \pm 0.02$ & $4.75 \pm 0.19$ & $3.89 \pm 0.16$ & $2.50 \pm 0.10$ & $0.53 \pm 0.10$ & $0.01 \pm 0.09$ & $-0.07 \pm 0.07$ & $1.34$ \\
$270^{\circ}$ & $0.69 \pm 0.02$ & $5.14 \pm 0.20$ & $3.26 \pm 0.13$ & $2.43 \pm 0.10$ & $0.13 \pm 0.10$ & $-0.05 \pm 0.09$ & $0.06 \pm 0.07$ & $1.36$ \\
$315^{\circ}$ & $0.72 \pm 0.02$ & $4.57 \pm 0.17$ & $4.02 \pm 0.14$ & $2.38 \pm 0.09$ & $-0.57 \pm 0.10$ & $0.12 \pm 0.08$ & $0.01 \pm 0.07$ & $1.35$ \\
\hline
\end{tabular}
\caption{Extracted femtoscopic parameters from fits to the three-dimensional correlation functions obtained using Eq.~\ref{eq:expCF}, for all eight azimuthal angles of pion pairs with transverse pair momentum $k_{T} \in [0.45, 0.80]~\mathrm{GeV}/c$, in 50--80\% central collisions. The fit was performed using Eq.~\ref{eq:cfmodified}.}
\label{tab:fit_params_charge0_cent0_kt5}
\end{table*}

\clearpage

\bibliography{apssamp}

\end{document}